\documentclass[letterpaper,12pt]{report} 

\usepackage[dvips]{graphicx,color} 
\usepackage{array,hhline,multicol,multirow,dcolumn} 
\usepackage{rotating} 
\usepackage{epsfig,graphics,rotate,color}
\usepackage{lineno}

\newcommand{\nubar}{\bar{\nu}}
\newcommand{\numu}{\nu_{\mu}}
\newcommand{\numubar}{\bar{\nu}_{\mu}}
\newcommand{\nue}{\nu_e}
\newcommand{\nuebar}{\bar{\nu}_e}

\newcommand{\nux}{\nu_x}

\newcommand{\gtwid}{\mathrel{\raise.3ex\hbox{$>$\kern-.75em\lower1ex\hbox{$\sim$}}}}
\newcommand{\ltwid}{\mathrel{\raise.3ex\hbox{$<$\kern-.75em\lower1ex\hbox{$\sim$}}}}

\begin{document}


\begin{titlepage}

\vspace{1.5in}

\centerline{\Huge Bringing the SciBar Detector}
\vspace{.1in}
\centerline{\Huge to the Booster Neutrino Beam} 

\vspace{.4in}

\centerline{\today}

\vspace{.4in}

\begin{center}
A.~A.~Aguilar-Arevalo$^1$, 
J.~Alcaraz$^4$, 
S.~Andringa$^4$,
S.~J.~Brice$^2$,
B.~C.~Brown$^2$,
L.~Bugel$^{10}$,
J.~Catala$^6$, 
A.~Cervera$^6$,
J.~M.~Conrad$^1$,
E.~Couce$^6$, 
U.~Dore$^{11}$, 
X.~Espinal$^4$,
D.~A.~Finley$^2$,
J.~J.~Gomez-Cadenas$^6$, 
Y.~Hayato$^5$,
K.~Hiraide$^7$,
T.~Ishii$^3$, 
G.~Jover$^4$,
T.~Kobilarcik$^2$,
Y.~Kurimoto$^7$, 
Y.~Kurosawa$^7$,
W.~C.~Louis$^8$,
P.~F.~Loverre$^{11}$,
L.~Ludovici$^{11}$,
T.~Lux$^4$, 
J.~Martin-Albo$^6$, 
C.~Mariani$^{11}$, 
K.~B.~M.~Mahn$^1$,
K.~Matsuoka$^7$,
W.~Metcalf$^9$,
J.~Monroe$^1$,
T.~Nakaya$^{7}$, 
F.~Nova$^4$,
P.~Novella$^6$, 
A.~Y.~Rodriguez$^4$, 
F.~Sanchez$^4$, 
M.~H.~Shaevitz$^1$,
M.~Sorel$^6$, 
R.~Stefanski$^2$, 
M.~Taguchi$^7$,
H.~Tanaka$^7$, 
A.~Tornero$^6$, 
R.~Van~de~Water$^8$,
M.~O.~Wascko$^{9}$, 
M.~Wilking$^{12}$, 
M.~Yokoyama$^7$, 
G.~P.~Zeller$^1$,
E.~D.~Zimmerman$^{12}$
\end{center}

\vspace{.2in}

{\it
\begin{center}
  $^1$Columbia University\\
  $^2$Fermi National Accelerator Laboratory\\
  $^3$High Energy Accelerator Research Organization~(KEK)\\
  $^4$Institut de F\`{\i}sica d'Altes Energies~(IFAE), Universitat Aut\`onoma de Barcelona\\
  $^5$Institute for Cosmic Ray Research (ICRR)\\
  $^6$Instituto de F\`{i}sica Corpuscular, Universidad de Val\`{e}ncia\\
  $^7$Kyoto University\\
  $^8$Los Alamos National Laboratory\\
  $^9$Louisiana State University\\
  $^{10}$Stratton Mountain School\\
  $^{11}$Universita' degli Studi di Roma "La Sapienza" and Instituto Nazionale di Fisica Nucleare~(INFN)\\
  $^{12}$University of Colorado, Boulder\\
\end{center}
}

\end{titlepage}
\clearpage

\noindent
{\bf \huge Executive Summary}\\
\label{execsum}

This document presents the physics case for bringing SciBar, the fully
active, finely segmented tracking detector at KEK, to the FNAL Booster
Neutrino Beam (BNB) line. This unique opportunity arose with the
termination of K2K beam operations in 2005. At that time, the SciBar
detector became available for use in other neutrino beam lines,
including the BNB, which has been providing neutrinos to the
MiniBooNE experiment since late 2002.

The physics that can be done with SciBar/BNB can be put into three
categories, each involving several measurements. First are neutrino
cross section measurements which are interesting in their own right,
including analyses of multi-particle final states, with unprecedented
statistics.  Second are measurements of processes that represent the
signal and primary background channels for the upcoming T2K
experiment. Third are measurements which improve existing or planned
MiniBooNE analyses and the understanding of the BNB, both in neutrino
and antineutrino mode.

For each of these proposed measurements, the SciBar/BNB combination
presents a unique opportunity or will significantly improve upon
current or near-future experiments for several reasons. First, the
fine granularity of SciBar allows detailed reconstruction of final
states not possible with the MiniBooNE detector.  Additionally, the
BNB neutrino energy spectrum is a close match to the expected T2K
energy spectrum in a region where cross sections are expected to vary
dramatically with energy.  As a result, the SciBar/BNB combination
will provide cross-section measurements in an energy range
complementary to MINER$\nu$A and complete our knowledge of neutrino
cross sections over the entire energy range of interest to the
upcoming off-axis experiments.

SciBar and BNB have both been built and operated with great success.
As a result, the cost of SciBar/BNB is far less than building a
detector from scratch and both systems are well understood with
existing detailed and calibrated Monte Carlo simulations. The
performance expectations assumed in this document are therefore
well-grounded in reality and carry little risk of not meeting
expectations.

This document includes a site optimization study with trade-offs
between the excavation costs associated with placing the detector at
different angles from the axis of the BNB and the physics which can be
performed with the neutrino flux expected at these locations. Table 1
provides a summary of the impact of placing SciBar at these locations
on the proposed measurements.  The overwhelming conclusion of this
study is that an on-axis location presents the best physics case and
offsets the additional costs due to excavation. The estimated cost of
the detector enclosure at the desired on-axis location is \$505K.

This proposal requests an extension of the BNB run through the end of
FY2007, one year past its currently approved run, regardless of the
outcome of the MiniBooNE $\nue$ appearance search.  Our schedules show
that SciBar would be operational in the BNB within 9 months of
initiation of the project, allowing ample time to achieve our physics
goals in FY2007.  In this document, we assume a total data set of
$2.0\times10^{20}$~POT, with $0.5\times10^{20}$~POT in neutrino mode
and $1.5\times10^{20}$~POT in antineutrino mode.

\newpage

\begin{center}

{\bf SciBooNE Collaboration}

\vspace{0.4in}

J.~M.~Conrad,
K.~B.~M.~Mahn,
M.~H.~Shaevitz,
G.~P.~Zeller\\
{\it Columbia University, New York, NY}\\

\vspace{0.2in}

S.~J.~Brice,
B.~C.~Brown,
D.~A.~Finley,
T.~Kobilarcik,
R.~Stefanski\\
{\it Fermi National Accelerator Laboratory, Batavia, IL}\\

\vspace{0.2in}

T.~Ishii\\
{\it High Energy Accelerator Research Organization~(KEK), Tsukuba, Japan}\\

\vspace{0.2in}

J.~Alcaraz, 
S.~Andringa,
X.~Espinal,
G.~Jover, 
T.~Lux, 
F.~Nova, 
A.~Y.~Rodriguez, 
F.~Sanchez\\
{\it Institut de F\`{\i}sica d'Altes Energies~(IFAE), Universitat Aut\`onoma de Barcelona, Barcelona, Spain}\\

\vspace{0.2in}

Y.~Hayato\\
{\it Institute for Cosmic Ray Research (ICRR), Tokyo, Japan}\\

\vspace{0.2in}

J.~Catala, 
A.~Cervera, 
E.~Couce, 
J.~J.~Gomez-Cadenas, 
J.~Martin-Albo, 
P.~Novella, 
M.~Sorel, 
A.~Tornero\\
{\it Instituto de F\`{i}sica Corpuscular, Universidad de Val\`{e}ncia, Valencia, Spain}\\

\vspace{0.2in}

K.~Hiraide, 
Y.~Kurimoto, 
Y.~Kurosawa, 
T.~Nakaya$^*$,
K.~Matsuoka,
M.~Taguchi,
H.~Tanaka, 
M.~Yokoyama\\
{\it Kyoto University, Kyoto, Japan}\\

\vspace{0.2in}

W.~C.~Louis,
R.~Van~de~Water,\\
{\it Los Alamos National Laboratory, Los Alamos, NM}\\

\vspace{0.2in}

W.~Metcalf,
M.~O.~Wascko$^*$\\
{\it Louisiana State University, Baton Rouge, LA}\\

\vspace{0.2in}

L.~Bugel\\
{\it Stratton Mountain School, Stratton Mountain, VT}\\

\vspace{0.2in}

U.~Dore, 
P.~F.~Loverre, 
L.~Ludovici, 
C.~Mariani,\\
{\it Universita' degli Studi di Roma "La Sapienza" and Instituto Nazionale di Fisica Nucleare~(INFN), Rome, Italy}\\

\vspace{0.2in}

M.~Wilking,
E.~D.~Zimmerman\\ 
{\it University of Colorado, Boulder, CO}\\
\end{center}

$^*$Co-spokespersons

\begin{sidewaystable}
\newlength{\CW} \setlength{\CW}{20mm}
\centering
\renewcommand{\multirowsetup}{\centering}
\begin{tabular}{|p{15mm}|p{22mm}|p{25mm}|p{\CW}|p{\CW}|p{\CW}|p{\CW}|p{\CW}|p{\CW}|p{\CW}|} \cline{4-10}
  \multicolumn{3}{c|}{}        &  Location A (on-axis)   &  Location B  & Location C & Location D & Location H & Mini-BooNE alone & K2K, MINOS, MINER$\nu$A \\ \cline{1-10}
  \multicolumn{3}{|r|}{Distance from MB target}     &  100m  & 100m & 100m & 100m & 250m & 541m & \multicolumn{1}{|c|}{---} \\
  \multicolumn{3}{|r|}{Height above beam center}    &    0cm & 300cm& 500cm& 700cm& 300cm&   0cm& \multicolumn{1}{|c|}{---} \\
  \multicolumn{3}{|r|}{Total $\nu$ flux ($\times$10$^{-10}$cm$^{-2}$POT$^{-1}$)}&  350  &  250  &  180  &  140  &  40 & 5 & 160 (K2K)\\
  \multicolumn{3}{|r|}{Peak $\nu$ energy (GeV)}           & 0.6 & 0.45 & 0.35 & 0.25 & 0.6 & 0.6 & 1.2,3,7,12 \\
  \multicolumn{3}{|r|}{Enclosure cost}              & \$505k & \$431k & \$292k & \$219k & \$431k & \multicolumn{1}{|c|}{---} & \multicolumn{1}{|c|}{---} \\ \hline\hline

\multirow{3}{15mm}[-9mm]{Leverage MB} 

& WS BG spectrum            & $\nubar$:1.5$\times$10$^{20}$ POT & \Large{$\star\star\star$} & \Large{$\star$} & \Large{$\oslash$}  & \Large{$\oslash$}  & \Large{$\oslash$}  &   \Large{$\star$}    &  \Large{$\oslash$}  \\ \cline{2-10}

& $\numu$ Disappearance   & $\nu$:0.5$\times$10$^{20}$ POT & \Large{$\star\star\star$} & \Large{$\oslash$}  & \Large{$\oslash$}  & \Large{$\oslash$}  & \Large{$\oslash$}  & \Large{$\star\star\star$} &  \Large{$\oslash$}  \\ \cline{2-10}

& $\numubar$ Disappearance   & $\nubar$:1.5$\times$10$^{20}$ POT & \Large{$\star\star\star$} & \Large{$\oslash$}  & \Large{$\oslash$}  & \Large{$\oslash$}  & \Large{$\oslash$}  & \Large{$\star\star$} &  \Large{$\oslash$}  \\ \cline{2-10}

& Intrinsic $\nue$        & $\nu$:0.5$\times$10$^{20}$ POT &
 \Large{$\star\star\star$} & \Large{$\oslash$}  & \Large{$\oslash$}  & \Large{$\oslash$}  & \Large{$\oslash$}  & \Large{$\star\star\star$}  & \Large{$\oslash$} \\ \hline\hline

\multirow{3}{15mm}[-9mm]{Help T2K}

& $\numu$ CC$\pi^+$ systematics & $\nu$:0.5$\times$10$^{20}$ POT &
\Large{$\star\star\star$}& \Large{$\star$} & \Large{$\oslash$} & \Large{$\oslash$} & \Large{$\star$} & \Large{$\star\star$} & \Large{$\surd$} \\ \cline{2-10}

& $\numu$ NC$\pi^0$ systematics & $\nu$:0.5$\times$10$^{20}$ POT &
\Large{$\star\star\star$} & \Large{$\star$} & \Large{$\oslash$} & \Large{$\oslash$} & \Large{$\oslash$} & \Large{$\star\star$} & \Large{$\surd$} \\ \cline{2-10}

& anti-$\nu$ Measurements & $\bar\nu$:1.5$\times$10$^{20}$ POT &
\Large{$\star\star\star$} & \Large{$\oslash$} & \Large{$\oslash$} & \Large{$\oslash$} & \Large{$\oslash$} & \Large{$\star\star$} & \Large{?} \\ \hline\hline

\multirow{3}{15mm}[-9mm]{SciBar Physics} 

& Exclusive anti-$\nu$ $\pi$-p & $\bar\nu$:1.5$\times$10$^{20}$ POT &
\Large{$\star\star\star$} & \Large{$\star$} & \Large{$\oslash$} & \Large{$\oslash$} & \Large{$\oslash$} & \Large{$\oslash$} & \Large{?} \\ \cline{2-10}

& NC$\pi^0$ Energy Dependence  & $\nu$:0.5$\times$10$^{20}$ POT &
\Large{$\star\star\star$} & \Large{$\star$} & \Large{$\oslash$} & \Large{$\oslash$} & \Large{$\star$} & \Large{$\oslash$} & \Large{$\surd$} \\ \cline{2-10}

& $\Delta\,\rightarrow\, N\gamma$ & $\nu$:0.5$\times$10$^{20}$ POT  $\bar\nu$:1.5$\times$10$^{20}$ POT  & \Large{$\star\star\star$} & \Large{$\oslash$} & \Large{$\oslash$} & \Large{$\oslash$} & \Large{$\oslash$} & \Large{$\oslash$} & \Large{$\surd$} \\ \hline

\end{tabular}
\caption{\em Relative performance merit for each of the measurements
  at each of the detector locations.  The number of stars indicates
  the precision of the measurement, $\oslash$ indicates that the
  measurement is not possible at that location, and $\surd$ indicates
  that a measurement can be made, but not in the energy range of
  interest to MiniBooNE or T2K.  Please see the text for further details
  of each measurement.}
\label{tab:physics}
\end{sidewaystable}

\vfill\eject

\tableofcontents

\vfill\eject


\chapter{Introduction}
\label{intro}

The American Physical Society's Divisions of Nuclear Physics and Particles and Fields, together with the Divisions of Astrophysics and the Physics of Beams, have recently conducted a ``Study on the Physics of Neutrinos''. The resulting APS report~\cite{the-neutrino-matrix} stated:

\begin{quote}
{\it We recommend, as a high priority, a comprehensive U.S. program to complete our understanding of neutrino mixing, to determine the character of the neutrino mass spectrum, and to search for CP violation among neutrinos.}
\end{quote}

This document presents the physics case for installing the SciBar
detector of the K2K experiment in the BNB at Fermilab. K2K beam
operations were terminated in 2005.  SciBar then became available for
use in other neutrino beam lines, including BNB, which has been
providing neutrinos to the MiniBooNE detector since late 2002.

The physics that can be accomplished with this configuration directly
addresses the high priority recommendation of the APS study, and, more
specifically, addresses two special points also mentioned in the
report:

\begin{quote}
{\it Support for decisive resolution of the high-$\Delta$m$^2$ puzzle. This issue is currently addressed by a single experiment now running in a neutrino beam at Fermilab. Ultimately, a decisive resolution of the puzzle may require additional studies with beams of antineutrinos.}
\end{quote}

and

\begin{quote}
{\it The precise determination of neutrino cross sections is an essential ingredient in the interpretation of neutrino experiments and is, in addition, capable of revealing exotic and unexpected phenomena.}
\end{quote}

The marriage of SciBar and the BNB presents a low risk opportunity for
a broad physics reach.  Both are already built and have been operated
very successfully. This means that:

\begin{enumerate}
\item the cost of bringing SciBar to Fermilab is far smaller than
  building a new detector from scratch,
\item both systems are very well understood with detailed and
  calibrated Monte Carlo simulations---the predictions of performance
  in this document have already been demonstrated with real operation.
\end{enumerate}

The remainder of this introduction provides the information necessary
to follow the physics case outlined in the later chapters.  The BNB is
described in Section~\ref{sec:boo_beam} and the SciBar detector in
Section~\ref{sec:intro_det}. The specific locations where the SciBar
detector might be placed in the BNB are discussed in
Section~\ref{sec:intro_loc}, and the expected event rates at each
location are detailed in Section~\ref{sec:evt_rate}. The introduction
ends with a discussion of time constraints in Section~\ref{sec:time}.

Three distinct types of measurements become possible with SciBar in
the BNB.  First, there are ways that SciBar can leverage the existing
investment in the MiniBooNE detector.  Chapter~\ref{chap:leverage}
describes the ways in which SciBar can improve measurements using
MiniBooNE tank data.  Next, Chapter~\ref{chap:t2k} describes the
reasons why the K2K collaboration would like to place SciBar in the
BNB, and describes how a number of cross section measurements can be
made that are vital to T2K reaching their desired oscillation
sensitivity. The last class of measurements, in
Chapter~\ref{chap:sb_phys}, cover physics topics that can be addressed
by SciBar/BNB alone.

For each SciBar measurement, this document states:

\begin{enumerate}
\item why the measurement is interesting,
\item the expected statistics for the measurement, and whether the
  beamline needs to be in neutrino or antineutrino mode,
\item why the measurement cannot be done at all or as well by any
  other past, present, or near future experiment, and
\item how the different potential detector locations for SciBar in the
  Booster Neutrino Beam affect the measurement.
\end{enumerate}

Table~\ref{tab:physics} provides a handy summary of the potential for
success of each of the proposed measurements at each of the detector
locations considered.  The document concludes with discussion of
schedule and costs in Chapter~\ref{chap:cost}.

\section{Booster Neutrino Beam Description}
\label{sec:boo_beam}

To create the BNB, 8 GeV protons are extracted from the Booster and
steered to strike a 71~cm long, 1~cm diameter beryllium target. This
target sits at the upstream end of a magnetic focusing horn that is
pulsed with $\sim$170~kA to focus the mesons produced by the proton-Be
interactions.  Following the horn is a 50~m long decay pipe that gives
the pions a chance to decay and produce neutrinos, before the mesons
encounter an absorber and then dirt which serve to remove all but the
neutrinos from the beam.

The protons from the Booster arrive in batches of 84 bunches, each of
which is $\sim$4~ns wide with $\sim$19~ns peak-to-peak separation,
giving a length of $\sim$1.6~$\mu$s to the whole batch. The batches
are extracted at a maximum rate of 5~Hz, a limit set by the horn, and
each contains $\sim$$4.5 \times 10^{12}$ protons. This timing
structure is carried through to the neutrino beam, and provides a
tight constraint on cosmic backgrounds.

In its current mode of operation, the horn focuses $\pi^+$ and
defocuses $\pi^-$ thus producing a $\nu_\mu$ beam. By reversing the
polarity of the horn current, $\pi^-$ are focused and a predominantly
$\bar{\nu}_\mu$ beam is created. In addition there is an absorber that
can be lowered into the beam at 25~m. Though currently not in use, the
absorber would alter the beam spectrum and composition in ways that
may prove useful for background checks or to reduce the effects of
beam parallax on a nearby detector.

The pion and kaon production cross sections from p-Be interactions are
the most important input to the BNB neutrino flux prediction, and the
most uncertain. These cross sections are being measured very precisely
by the HARP experiment at CERN~\cite{harp}.  The collaboration has
released its first result, a precise measurement of the production
cross section of pions in proton-aluminum interactions at 12.9 GeV/c,
which is the K2K neutrino beam energy and target
material~\cite{harp_al}.  The beryllium analysis is expected to be
released within the next few months, and HARP anticipates
uncertainties of $\sim$5\% on the pion production cross sections, for
both $\pi^+$ and $\pi^-$.  This will allow a very precise prediction
of the neutrino and antineutrino fluxes in the BNB by the time
SciBooNE proposes to start its data run.

\subsection{Expectations for Proton Delivery}
\label{subsec:pot}

\begin{figure}[hb]
\begin{center}
\includegraphics[width=\textwidth]{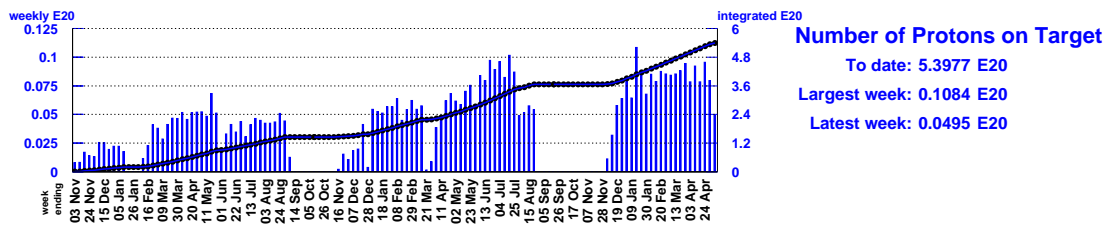}
\end{center}
\caption{\label{fig:pot_delivery} \em Proton delivery to the Booster
  Neutrino Beam target from the start of operations in late 2002 to
  present (May 2005). The histogram records the weekly proton rate and
  the curve shows the cumulative total.}
\end{figure}

The Booster Neutrino Beam saw first protons on target (POT) in
September of 2002 and Fig.~\ref{fig:pot_delivery} records the weekly
and cumulative proton delivery since then.

At present, the NuMI beam has started running and the Booster Neutrino
Beam has been receiving significantly fewer protons. The letter from
the Fermilab Director to MiniBooNE entitled ``Prospects for the
Booster Neutrino Beam,'' and dated August 6, 2004 states:

\begin{quote}
  {\em Collaborations proposing experiments to run in the Booster neutrino
    beam in FY2006 and beyond should plan their physics program on the
    basis of $1-2\times10^{20}$ POT per year. Proponents may want
    to discuss what additional physics could be done with somewhat 
    more protons, but they should understand that is beyond our present
    expectations for the beam.}
\end{quote}

In this document, we make the assumption that $2\times10^{20}$ POT
will be delivered to the BNB in one year. This assumption is
consistent with the Lab's current ``proton plan~\cite{protonplan}.''
Additionally, the improvements in proton delivery made since the
Director's letter and indicated in the latter portions of
Fig.~\ref{fig:pot_delivery} justify this optimistic assumption.
Because MiniBooNE is currently approved to run only through the end of
FY2006, this proposal is a request for an extension of the BNB run
through the end of FY2007, regardless of the result of the MiniBooNE
oscillation search in neutrino mode.  

Operating the BNB for one year in the NuMI era requires running the
Booster accelerator $\sim$2~Hz more than it would run without the
BNB~\cite{prebys}.  This cost increase has been estimated to be
approximately \$40 per hour~\cite{prebys}.  Additionally, the 8 GeV
line costs \$11 per hour to run~\cite{moore}.  Assuming a BNB run of
2.0$\times$10$^{20}$POT, this indicates a BNB run of
2.2$\times$10$^7$s, which amounts to an approximate incremental cost
increase of \$315,000.  Additionally, this added running has the
potential to increase the failure rates of components in the Linac and
Booster.  This impact has been studied and is expected to be
minimal~\cite{prebys}.

MiniBooNE will likely switch the polarity of the horn and begin
accumulating statistics in antineutrino mode before the end of 2005,
continuing until the next accelerator shutdown.  However, the decision
of how to run the BNB in 2006 hinges on whether or not MiniBooNE sees
a $\nu_e$ appearance oscillation signal; the MiniBooNE collaboration
has recently stated that this result will not be out before the end of
2005.  If MiniBooNE sees a signal then the case for installing SciBar
in the beam becomes very strong as it will provide a powerful check on
the $\nu_\mu$ spectrum and will reduce the uncertainty on the
intrinsic $\nu_e$ background by measuring it at a near location (see
Chapter~\ref{chap:leverage} for details).  If MiniBooNE does not see a
$\nu_e$ oscillation signal then the beamline will most likely switch
to antineutrino mode in FY2006. The physics justification for this
switch is laid out in \cite{fy06_loi}.  This document focuses on the
case where MiniBooNE does not see a $\nu_e$ appearance signal and the
ensuing data are taken primarily in antineutrino mode. In this
scenario, we assume that in one year $0.5 \times 10^{20}$ POT will be
delivered in neutrino mode and $1.5 \times 10^{20}$ POT in
antineutrino mode.

\section{SciBar Detector Description}
\label{sec:intro_det}

\begin{figure}[hb]
  \begin{center}
    \includegraphics[keepaspectratio=true,width=4.5in]{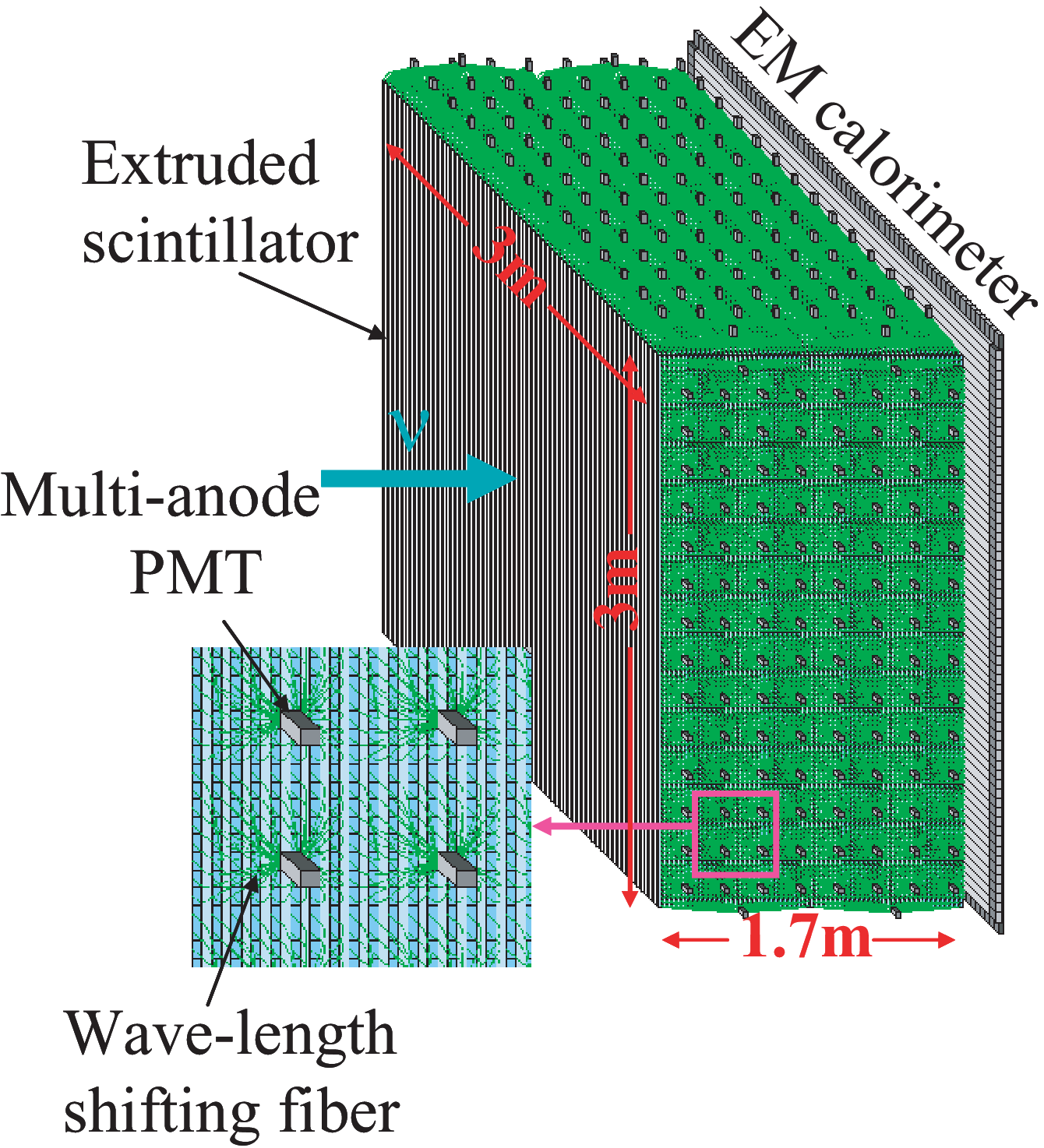}
  \end{center}
  \caption[Schematic view of SciBar]{\em Schematic view of SciBar.
    Extruded scintillator strips are arranged vertically and
    horizontally, with WLS fibers embedded in each strip.  Each WLS
    fiber is read out by a 64-channel MA-PMT.  An electromagnetic
    calorimeter sits immediately downstream of SciBar.}
  \label{fig:k2k_scibar.eps}
\end{figure}

\subsection{The K2K SciBar Detector}
SciBar \cite{Nitta:2004nt} is a fully active, finely segmented
tracking detector consisting of plastic scintillator bars. It was
constructed in summer 2003 as a new near detector for K2K, and
operated until late 2004.  The cost of SciBar was approximately \$2M,
not including contingencies or labor.

Figure~\ref{fig:k2k_scibar.eps} shows a schematic view of SciBar. The
tracker consists of 14,848 extruded scintillator strips, each
$1.3\times 2.5\times 300$ cm. The scintillators are arranged
vertically and horizontally to construct a $3\times 3\times 1.7$ m$^3$
volume with a total mass of 15 tons, and a fiducial mass of 9.38 tons.
Each strip is read out by a wavelength-shifting (WLS) fiber attached
to a 64-channel multi-anode PMT (MA-PMT). Charge and timing
information from each MA-PMT is recorded by custom
electronics \cite{Yoshida:2004mh}. The specification of each component
of SciBar is summarized in Table \ref{table:specification}.

\begin{table}[h]
 \caption{\em Specifications and measured performance merits of SciBar components.}
 \label{table:specification} 
 \begin{center}
  \begin{tabular}{lll}
    \hline \hline
     Structure    & Dimensions       & 3 m (horizontal), 3 m (vertical), 1.7m (thickness) \\
                  & Weight           & 15 tons \\
                  & Number of strips & 14,848 \\
                  & Number of PMTs   & 224 \\
    \hline
     Scintillator & Material            & Polystyrene with PPO(1\%) and POPOP(0.03\%) \\
                  & Size                & 2.5 $\times$ 1.3 $\times$ 300 cm$^2$ \\
                  & Coating             & 0.25 mm (TiO$_2$) \\
                  & Emission wavelength & 420 nm (peak) \\
    \hline
     Fiber        & Type                  & Kuraray Y11(200)MS \\    
                  & Diameter              & 1.5 mm \\
                  & Refractive index      & 1.59 (outer)/ 1.50 (middle)/ 1.42 (inner) \\
                  & Absorption wavelength & 430 nm (peak) \\
                  & Emission wavelength   & 476 nm (peak) \\
                  & Attenuation length    & 350 cm \\
    \hline
     PMT          & Model                   & Hamamatsu H8804 \\                      
                  & Cathode material        & Bialkali \\
                  & Anode                   & $8\times 8$ ($2\times 2$ mm$^2$/pixel) \\                  
                  & Quantum efficiency      & 12\% for 500 nm photons \\                  
                  & Typical gain            & $6\times 10^5$ at $\sim 800$ V \\                  
                  & Response linearity      & 200 PE at gain of $6\times 10^5$ \\
                  & Cross talk              & 4\% (adjacent pixel) \\               
     \hline
      DAQ         & VA/TA ASIC         & IDEAS VA32HDR11 and TA32CG \\                                                
                  & Shaping time       & 1.2 $\mu$sec (VA), 80 ns (TA) \\                  
                  & Noise              & 0.3 PE \\
                  & Response linearity & 5\% at 300 PE \\                 
                  & TDC resolution     & 0.78 ns \\             
                  & TDC full range     & 50 $\mu$sec \\                    
   \hline \hline
  \end{tabular}
 \end{center}
\end{table}

\clearpage

An electromagnetic calorimeter (EC) is installed downstream of SciBar.
The purpose of the EC is to measure the $\nu_e$ contamination in the
beam and the $\pi^0$ yield from neutrino interactions, particularly
for high momentum $\pi^0$s whose decay photons are boosted forward.
The EC consists of 32 (vertical) and 30 (horizontal) modules of the
so-called ``spaghetti calorimeter'' from the CHORUS experiment
\cite{Buontempo:1997ie}. Each module is made of 1~mm diameter
scintillating fibers embedded in the grooves of 1.9~mm thick lead
foils. The dimensions of each module are $4.0\times 8.2\times 262$
cm$^3$.  The light from each module is read out by two 1~'' PMTs on
both sides. The EC has a thickness of 11$X_0$ along the beam
direction, giving it a very high efficiency. The energy resolution of
the EC is $14\%/\sqrt{E_e \ {\rm [GeV]}}$.

\begin{figure}[t]
  \begin{center}
    \includegraphics[keepaspectratio=true,width=2.5in]{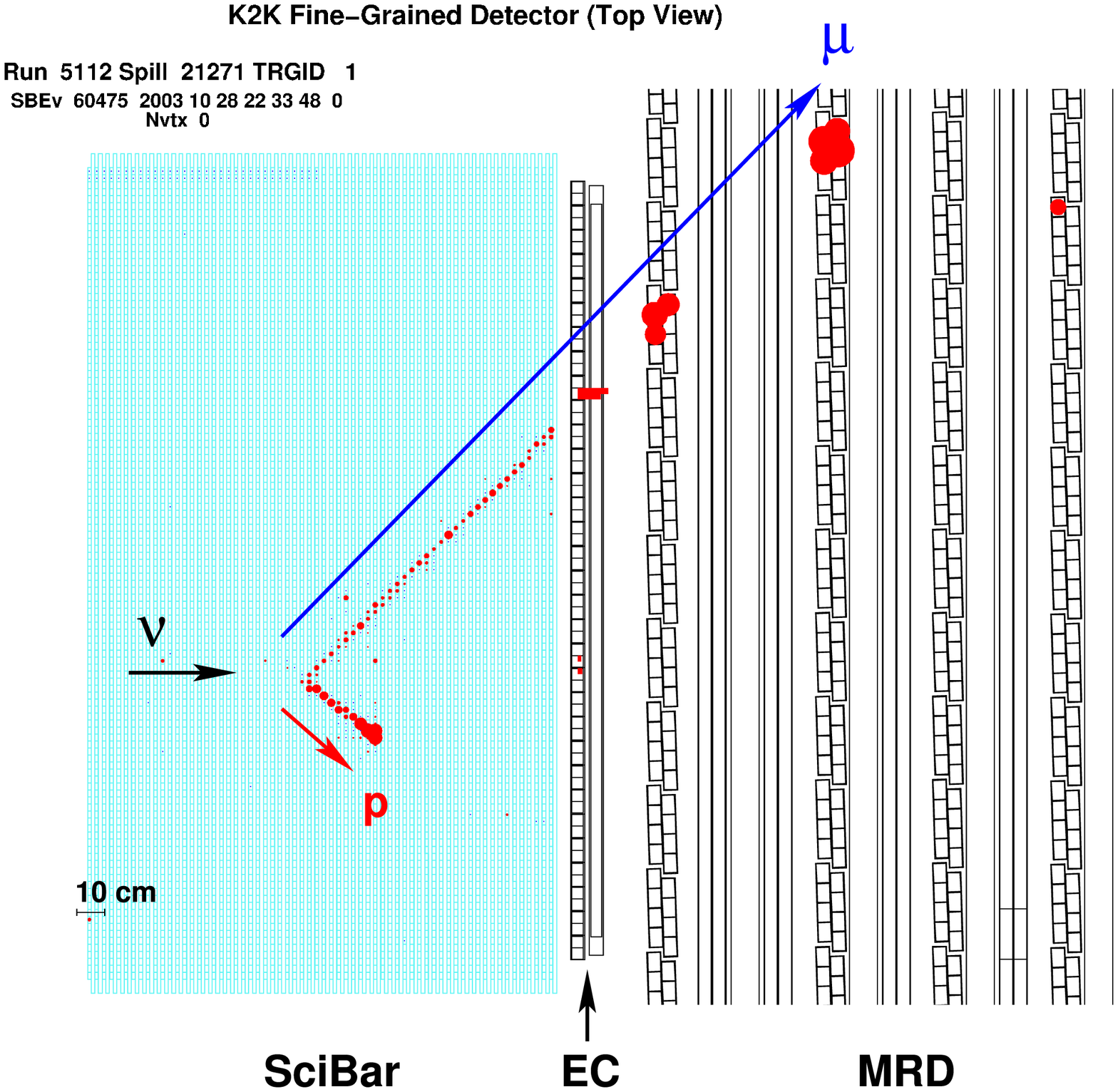}
    \includegraphics[keepaspectratio=true,width=2.5in]{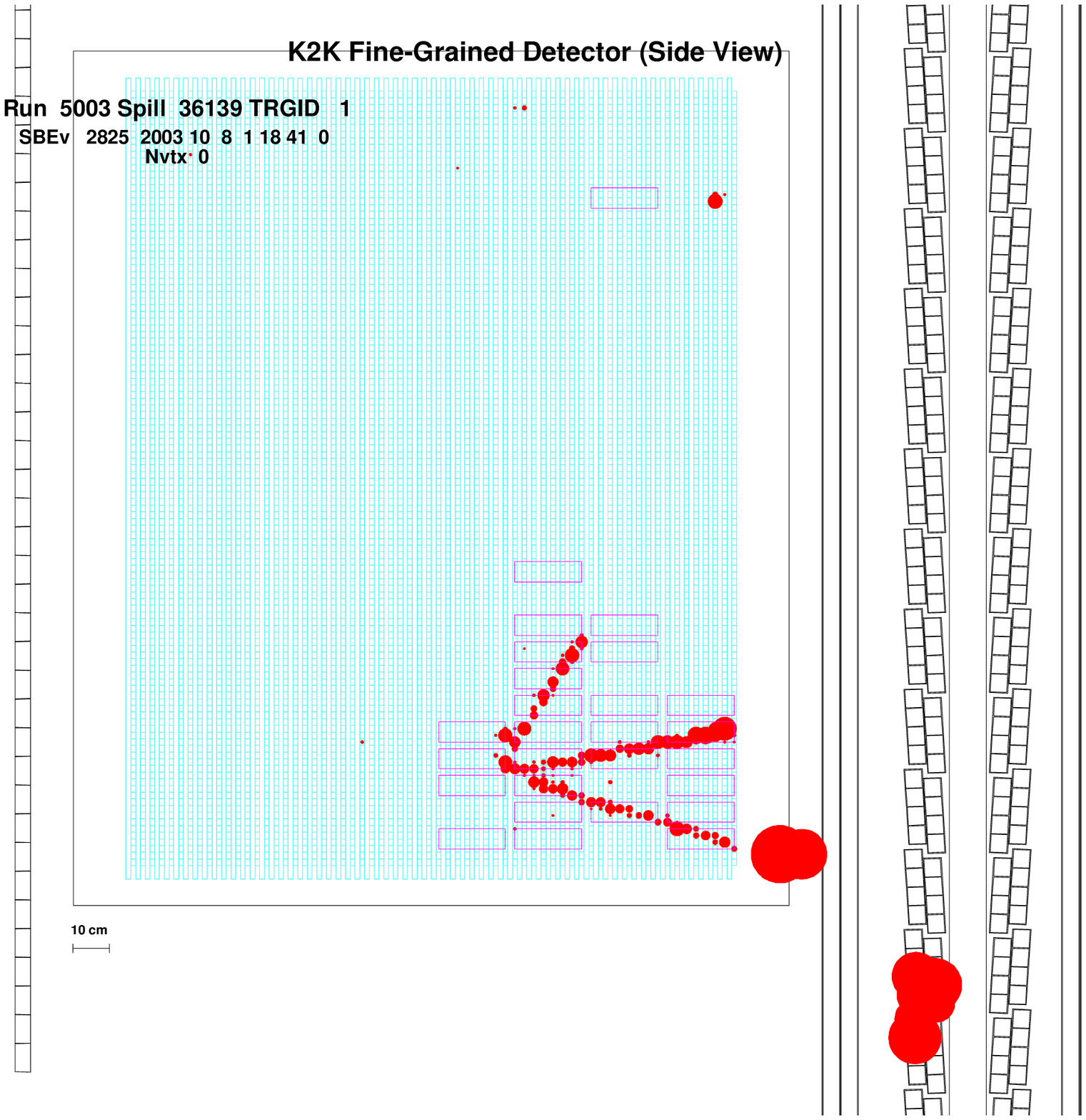}
  \end{center}
  \caption[Event display of a typical CC-QE candidate]
  {\em Event displays of typical $\numu$ interactions in SciBar at
    K2K. The left-hand panel shows a two track CC QE candidate event,
    and the right hand panel shows a three track CC1$\pi^+$ candidate.
    The red circles show the hit cells, and their areas are
    proportional to the recorded ADC counts.}
  \label{fig:cc_evt_displays}
\end{figure}

A muon range detector (MRD) \cite{Ishii:2001sj} is located downstream
of the EC. The MRD at KEK consists of 12 layers of iron plates
sandwiched between vertical and horizontal drift-tube layers. The
cross sectional size of a layer is approximately 7.6 m $\times$ 7.6 m.
The four upstream iron plates are 10 cm thick and the eight downstream
are 20 cm thick. The total iron thickness of 2.0 m covers up to 2.8
GeV muons.

Not including the MRD, the actual size of the SciBar detector's
experimental area at K2K is approximately 5.5 m wide and 2 m along the
beam direction; SciBar, the EC, and two electronics racks were
installed in that space.

\subsection{Detector Performance}

\begin{table}[h]
\centering
   \begin{tabular}{|c|c|} \hline
Particle type   &  Efficiency \\ \hline\hline
$\mu$           &  98.9\%  \\ 
$\pi^{\pm}$     &  98.1\%  \\
p               &  97.7\%  \\
e$^{\pm}$       &  94.5\% 
\\
\hline
   \end{tabular}
   \caption{\em SciBar Monte Carlo particle detection efficiencies.}
   \label{table:part_eff}
\end{table} 

The SciBar detector was operated at K2K from October 2003 to November
2004, for an accumulated data set of 0.21$\times$10$^{20}$ POT. During
that period, the number of dead channels was monitored and only six
dead channels were identified out of 14,336 channels.  Operationally,
SciBar performed very well, requiring only two detector accesses over
the duration of its neutrino beam run.

Light yield in SciBar was measured using cosmic ray data. The average
light yield is 18 photoelectrons (PE) for a 1.0 cm muon track at 40 cm
from the PMT along the fiber. The light yield is sufficient for track
finding and particle identification. The stability of the light yield
is also checked using cosmic ray data.  With PMT gain corrections, the
light yield was found to be stable at the 0.7\% level.

\begin{figure}[htbp]
  \begin{center}
    \includegraphics[keepaspectratio=true,height=45mm]{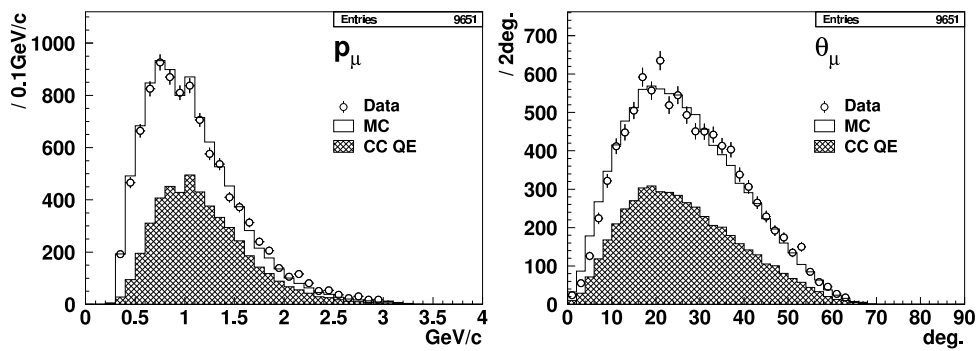}
  \end{center}
  \caption{\em Comparison of data (open circles) and Monte Carlo
    (histograms) $\numu$ charged current events in SciBar at K2K; show
    are muon momentum distributions (left) and angular distributions
    (right).  The MC distributions are normalized by entries.}
  \label{fig:muon_momentum.eps}
\end{figure}

Figure \ref{fig:cc_evt_displays} shows two event displays of actual
$\numu$ charged-current candidate events in SciBar. The first event
shown is a CC QE candidate, with two clear tracks, and the second
event shown is a CC1$\pi^+$ candidate with three clear tracks.  We can
clearly distinguish the muon/pion tracks from the proton tracks by
their energy depositions.  

Particle tracks are found in SciBar using the powerful {\em cellular
  automaton} track pattern recognition algorithm~\cite{cat}. The
minimum track length required to reconstruct a track is approximately
10 cm, which corresponds to 4-8 hits, depending on the angle of the
track with respect to the detector axes.  At 2.2 MeV deposited per cm
for a minimum ionizing particle, that corresponds roughly to a minimum
kinetic energy of 20-25 MeV for a particle to be detected.  The track
finding efficiency of a muon generated in a charged-current neutrino
interaction in SciBar is approximately 94\%, estimated using $\nu$
data.  Track finding efficiencies for various particles found using
the SciBar Monte Carlo are shown in Figure~\ref{table:part_eff}.

Figure \ref{fig:muon_momentum.eps} shows the distributions of muon
momentum ($p_\mu$) and angle with respect to the beam ($\theta_\mu$),
with the requirement that a track created in SciBar match a track (or
hits) found in the MRD.  The data and MC agree well except for the
forward ($\theta_\mu < 10$ degrees) region, which may point to new
physics, rather than a detector deficiency~\cite{k2k_coherent}.  The
energy resolution and angular resolution of the muons are 0.08 GeV and
1.6 degrees, respectively. The muon energy resolution is dominated by
the MRD resolution.  More detailed detector performance can be found
elsewhere~\cite{Maesaka:2005}.

\subsection{Modifications to Detector Configuration}

The detector configuration will be modified slightly for SciBooNE.
The detector complex will consist, as before, of three detectors:
SciBar, the EC and the MRD.  All SciBar components and most EC
components will be brought from KEK to Fermilab, and their
configuration will not change. In order to save costs, the MRD will
be assembled from detector components salvaged from past FNAL
experiments~\cite{e605}, rather than be shipped from Japan.

We have studied the effects of the changes in the MRD size and
acceptance on the physics potential of SciBooNE. We have found that
the size of the available iron plates, 3.5 m $\times$ 4 m and plate
thicknesses of 2.5 and 5 cm, does not significantly degrade the
sensitivity of the experiment.  We will use plastic scintillators for
the active detector elements instead of drift tubes.  Since 60~cm of
iron is sufficient to stop muons with kinetic energy of 1~GeV/c, we
will use only 12 planes of iron, each with thickness 5~cm.  Monte
Carlo studies indicate that this smaller MRD size reduces the
efficiency for SciBar-MRD track matching by only 10-20\%, depending on
the interaction type.

\begin{figure}[htbp]
  \begin{center}
    \includegraphics[keepaspectratio=true,width=5.0in]{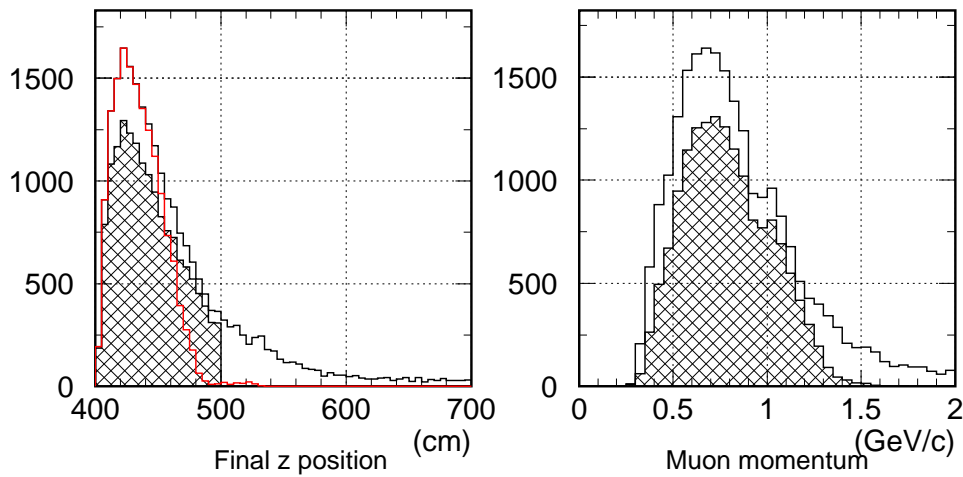}
  \end{center}
  \caption{\em Effect of smaller MRD size on the muon acceptance.
  Shown are the z-position (left) and momentum (right) of muons that
  stop in the MRD, for the K2K sized MRD (line) and the proposed new
  size MRD (cross-hatch).  The red histogram in the left hand panel
  shows the stopping position of all muons below 1~GeV/c; note that
  all muons below this value stop within the 60~cm depth of iron.}
  \label{fig:mrdsize}
\end{figure}

Figure~\ref{fig:mrdsize} shows the effect of the smaller MRD size on
the muon acceptance.  The figure compares the muon stopping position
along the beam direction of the K2K sized MRD (7.6~m$\times$7.6~m) and
the SciBooNE sized MRD (3.5~m$\times$4~m), as well as the momentum
distribution of stopping muons.  It can be seen that the new sized MRD
is sufficient to stop all muons with momentum below 1~GeV/c.

\clearpage

\section{Discussion of Specific Locations}
\label{sec:intro_loc}

In pursuing this project, we have explored potential detector sites
both on and off the beam axis.  In this section, we explore the
variations in flux and spectrum with detector location, with the goal
of selecting the detector location which best maximizes the physics
output.  We do this by comparing predicted event rates at the various
locations, based on current neutrino interaction cross sections and
the known efficiencies of the SciBar detector, and estimating the
measurements within reach based on those predicted event rates and
spectra.

\begin{figure}
\center
{\includegraphics[width=4.0in]{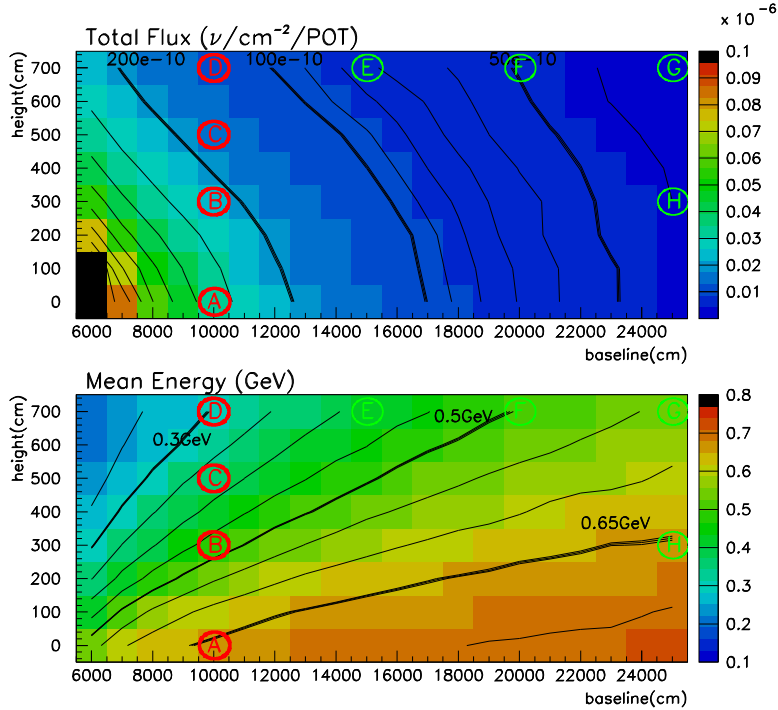}}
\vspace{-0.1in}
\caption{\em Total neutrino flux (top) and average neutrino energy
  (bottom) as a function of distance from the MiniBooNE target, in
  both longitudinal and vertical directions.  The flux is given in
  units of $\nu$/cm$^2$/POT, and the energy is given in units of GeV.
  The origin of beam coordinate system coincides with the neutrino
  production target, and is not shown in the plots.}
\label{fig:offaxis_fluxes}
\end{figure}

We begin with a general discussion of the Booster neutrino flux.
Figure~\ref{fig:offaxis_fluxes} shows the expected total flux and mean
energy of all neutrino species as a function of distance from the
target in the beam direction ($\hat{z}$) and the vertical direction
($\hat{y}$).  In the figure, the horizontal axis represents the
distance from the neutrino target in the beam direction ($\hat{z}$),
measured in cm, and the vertical axis represents the vertical
($\hat{y}$) distance from the beam axis measured in cm.

From Fig.~\ref{fig:offaxis_fluxes}(top), we see that there are
contours of constant flux, roughly ellipsoidal in shape with the major
axis aligned with the beam direction, emanating from the neutrino
target.  As an example, the flux line at z=125~m, y=0~m indicates that
we expect 200$\times$10$^{-10}\nu$/cm$^2$/POT at that location.
Following this flux line, we see that this same total flux is expected
at many more locations, for example at z=100~m, y=4~m and z=75~m,
y=7~m.

Following the contour lines of constant flux allows one to optimize
the detector with regard to total neutrino flux.  Alternatively, one
can optimize with regard to the energy spectrum.
Fig.~\ref{fig:offaxis_fluxes}(bottom) shows contours of constant mean
energy, for neutrinos less than 2~GeV\footnote{For this plot, the
  calculation of the neutrino mean energy was found using only
  neutrinos below 2~GeV, to remove the effect of the high energy
  tails.}; these contour lines appear to radiate from the neutrino
target position.  Following the previous example which examined a line
of constant flux, we now follow a line of constant mean energy.
Noting that at z=100~m, y=0~m the mean neutrino energy is
$\sim$0.65~GeV, we follow the $\sim$0.65~GeV line and find that at
z=250m, y=3m we expect the same mean energy.

In this discussion, we consider eight different detector locations:
four locations at z=100~m, ranging vertically from 0~m (on-axis) to
7~m (on the surface), and four on the surface, ranging from 100~m to
250~m from the proton target.  We also consider one location at
z=250~m, y=3~m.  As discussed in Section~\ref{sec:evt_rate}, several
locations were eliminated immediately because they would produce
extremely poor statistics.

Not surprisingly, we find that the on-axis location at a distance of
100~m from the neutrino target is the best choice, providing the
largest possible physics reach.

\begin{figure}
\center
{\includegraphics[width=4.0in]{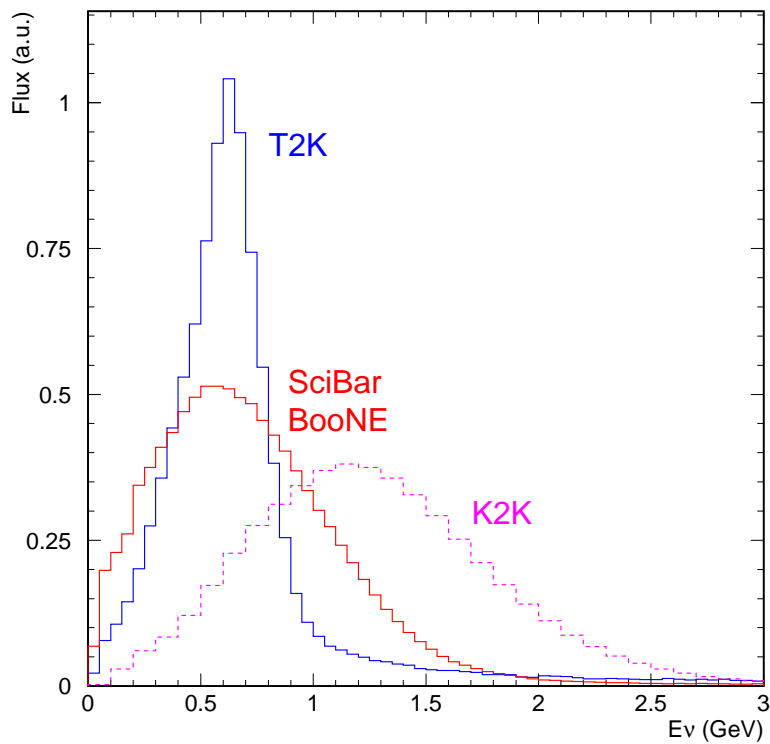}}
\vspace{-0.1in}
\caption{\em Comparison of the $\nu_{\mu}$ flux spectra at K2K, T2K,
  and the on-axis location at 100m.}
\label{fig:onaxis_comp}
\end{figure}

\subsubsection{Discussion of On-Axis Spectrum}

Figure~\ref{fig:onaxis_comp} shows a comparison of the $\nu_{\mu}$
flux spectra for K2K, T2K\, and this on-axis location.  This figure
indicates why the BNB is of direct interest to T2K: the energy peaks
of the two fluxes coincide and the entire range of the T2K energy flux
is encompassed within the flux peak of the BNB.  Thus, cross section
measurements made at FNAL will have direct relevance to neutrino
events at T2K.  Note also that the high energy tail of the T2K flux
extends much farther than the high energy tail of the BNB flux; this
high energy tail increases the uncertainty on cross section
measurements by increasing the number of misidentified inelastic
events.

\subsubsection{Discussion of Spectra at Off-axis Locations}

\begin{figure}[h]
\center
{\includegraphics[width=3.0in]{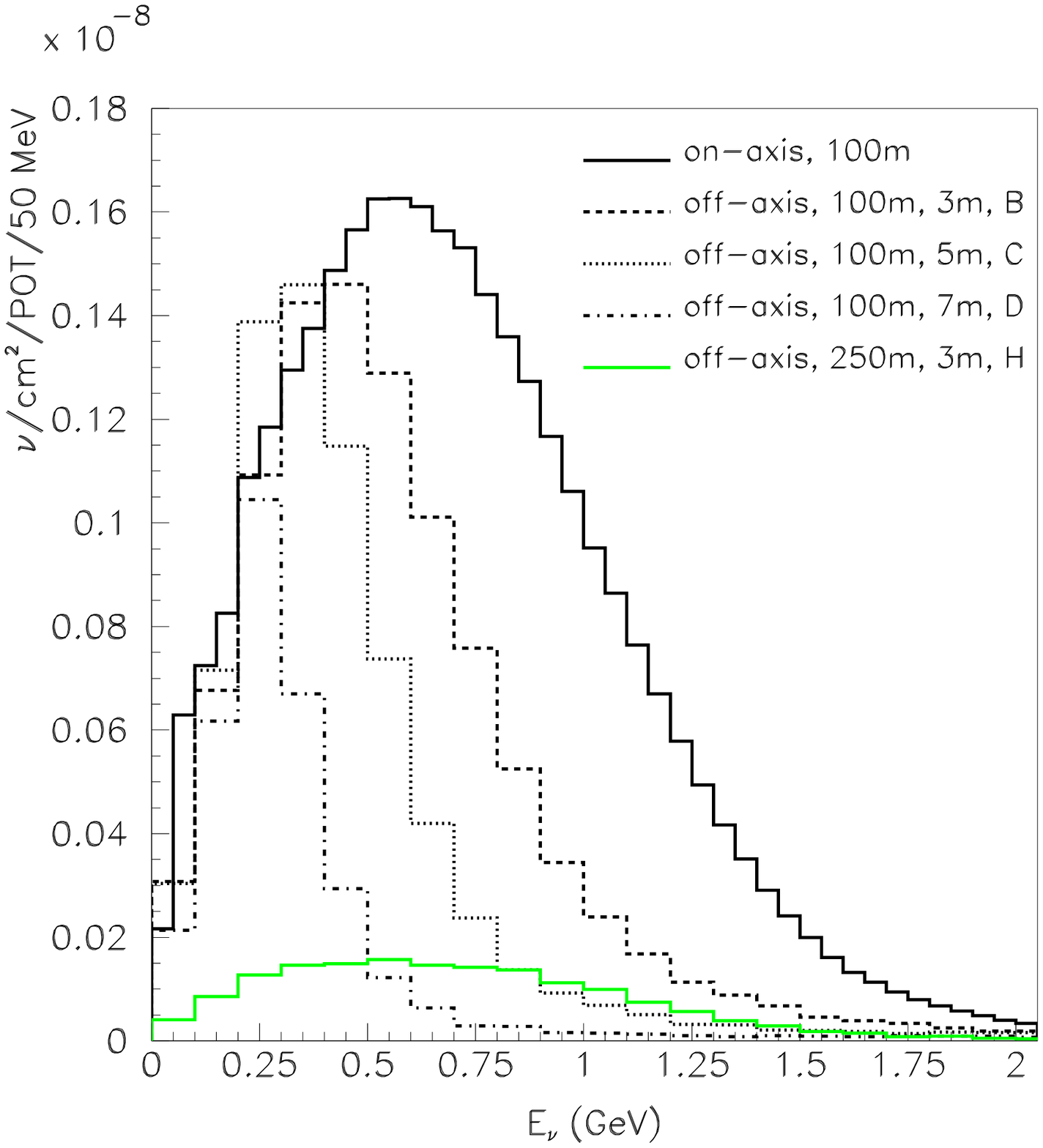}
\hspace{-0.3in}
\includegraphics[width=3.0in]{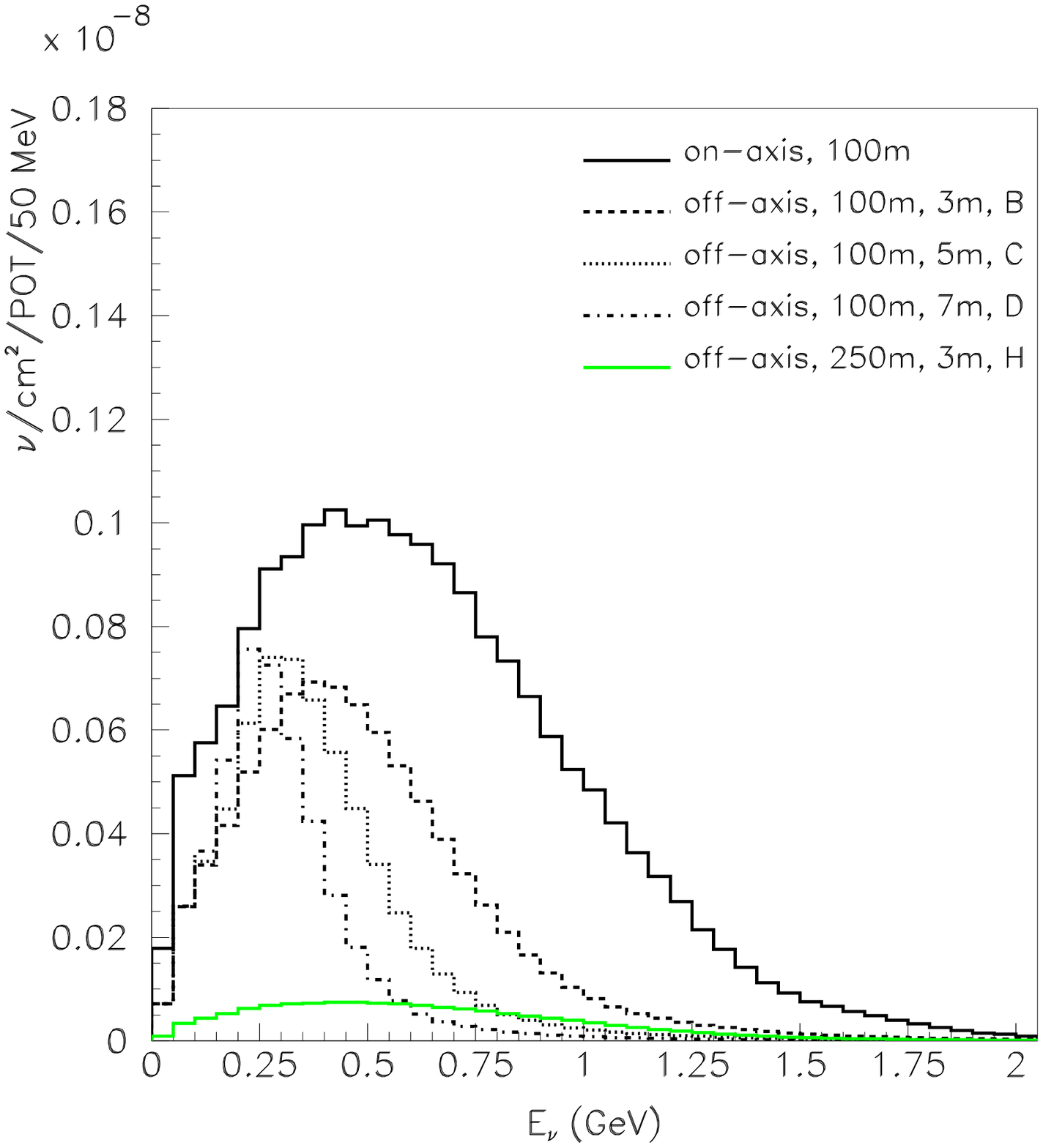}}
\vspace{-0.1in}
\caption{\em Comparison of neutrino (left) and antineutrino (right) mode 
  energy spectra for several different detector locations as indicated 
  in Figure~\ref{fig:offaxis_fluxes}.}
\label{fig:offaxis_comp}
\end{figure}

Figure~\ref{fig:offaxis_comp}(left) reveals in detail the effects of
going off-axis in the vertical direction.  The figure demonstrates
that at increasingly off-axis positions, the peak of the neutrino flux
moves to lower energy, and the overall flux decreases.  This behavior
was first seen in the discussion of Figure~\ref{fig:offaxis_fluxes}.
The off-axis behavior of the $\numubar$ flux expected for antineutrino
running mode is shown in Figure~\ref{fig:offaxis_comp}(right), and is
seen to exhibit the same behavior.

We have also considered several locations on the surface, at
increasing distance from the proton target.  These locations provide
different off-axis angles, but roughly equal costs because they all
involve the same excavation needs.  We have also selected a location,
at z=250~m, y=3~m, which gives a very similar energy spectrum to the
on-axis location at z=100~m.  However, all of these locations yield
event rates that are too low to make interesting measurements on the
time scales of this project.

\clearpage

\section{Expected Event Rates}
\label{sec:evt_rate}

Expected event rates in the SciBar detector for a variety of Booster beamline
locations were estimated using the NEUT Monte Carlo simulation which has been 
demonstrated to perform well in modeling SciBar data taken at K2K~\cite{neut}.
This section presents the number of events anticipated for various neutrino 
reactions and detector sites assuming a 9.38 ton fiducial $CH$ target and 
a total of $2\times10^{20}$ POT ($0.5 \times 10^{20}$ POT in neutrino mode, and 
$1.5 \times 10^{20}$ in antineutrino mode). 

%
%
\begin{table}[h]
\centering
   \begin{tabular}{|l|c|} \hline
Reaction   &  \# $\numu$ events
 \\ \hline\hline
CC QE                    & 31,720 \\ \hline
CC resonant 1$\pi^+$     & 14,108 \\ \hline
NC elastic               & 13,751 \\ \hline
CC multi-$\pi$           & 5,279  \\ \hline
NC resonant $1\pi^0$     & 3,723  \\ \hline
CC resonant 1$\pi^0$     & 3,106  \\ \hline
NC resonant $1\pi^{\pm}$ & 2,372  \\ \hline
NC multi-$\pi$           & 1,723  \\ \hline
CC coherent $1\pi^+$     & 1,432  \\ \hline
NC coherent $1\pi^0$     &   746  \\ \hline\hline
total                    & 77,960 
\\
\hline
   \end{tabular}
   \caption{\em Total number of $\numu$ events expected in neutrino mode
            assuming 9.38 ton fiducial volume, $0.5 \times 10^{20}$ POT,
          and on-axis z=100m SciBar location. $\numubar$ events have 
          been omitted from this table as they contribute $<2\%$ 
          to the total event rate.}
   \label{table:nu-rates-onaxis}
\end{table}

\begin{figure}[h]
\center
{\includegraphics[width=4.0in]{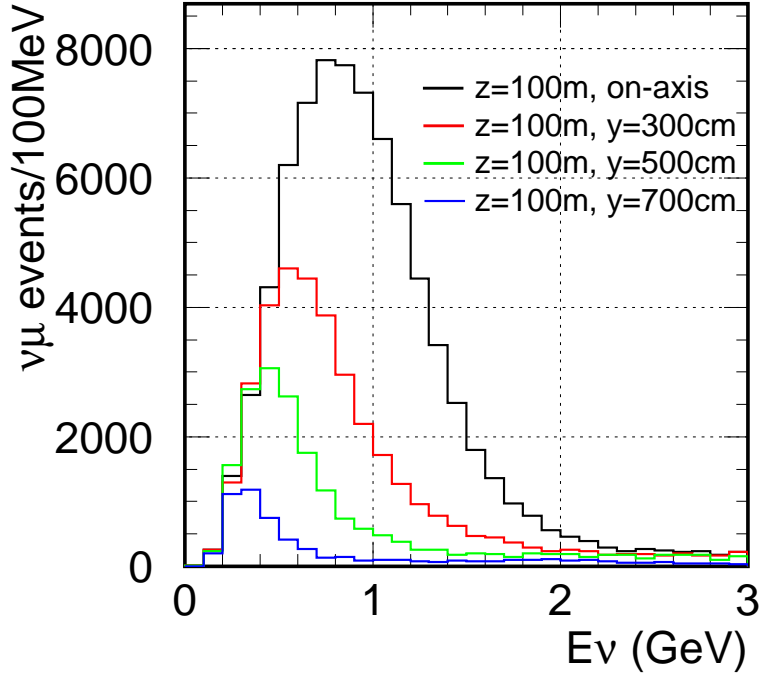}}
\vspace{-0.2in}
\caption{\em Expected energy distributions neutrino events (cross
  section weighted) for various SciBar on-axis and off-axis detector
  locations at z=100m. This plot assumes 0.5$\times$10$^{20}$ POT in
  neutrino mode, and 9.38 ton fiducial mass.}
\label{fig:offaxis-nu-rate-dist}
\end{figure}

\begin{table}[h]
\centering
   \begin{tabular}{|l|c|c|} \hline
Reaction   &  \# $\numubar$ events (RS) & \# $\numu$ events (WS)
 \\ \hline\hline
CC QE                    & 18,623 & 7,884  \\ \hline
NC elastic               &  7,563 & 3,516  \\ \hline
CC resonant 1$\pi^-$     &  4,494 &     0  \\ \hline
CC resonant 1$\pi^+$     &      0 & 4,481  \\ \hline
CC coherent $1\pi^-$     &  2,150 &     0  \\ \hline
CC coherent $1\pi^+$     &      0 &   377  \\ \hline
NC resonant $1\pi^0$     &  2,150 & 1,115  \\ \hline
CC multi-$\pi$           &  1,635 & 2,760 \\ \hline
NC resonant $1\pi^{\pm}$ &  1,227 &  735  \\ \hline
CC resonant 1$\pi^0$     &  1,127 &  960  \\ \hline
NC coherent $1\pi^0$     &  1,109 &  207  \\ \hline
NC multi-$\pi$           &    710 &  891  \\ \hline\hline
total                    & 40,685 & 22,925
\\
\hline
   \end{tabular}
   \caption{\em Total number of $\numubar$ and $\numu$ events expected in antineutrino
     mode assuming 9.38 ton fiducial volume, $1.5 \times 10^{20}$ POT and
     on-axis z=100m SciBar detector location.  Note that WS events make up
     30\% of the CC QE sample, but 36\% of the total events. }
   \label{table:nubar-rates-onaxis}
\end{table}

\begin{table}[h]
\centering
   \begin{tabular}{|c||r|r|r|r|r|r|r|r|} \hline
   & on-axis & B & C & D & E & F & G & H \\ 
   & z=100m  & z=100m & z=100m & z=100m & z=150m
   & z=200m  & z=250m & z=250m \\
   & d=0m    & d=3m   & d=5m  & d=7m  & d=7m & d=7m & d=7m & d=3m
 \\ \hline\hline
$<E_\nu>$   & 0.92 & 0.76 & 0.64  & 0.60 & 0.60 & 0.61  & 0.61  & 0.94 \\ \hline
$\# \numu$     & 78,397 & 37,230 & 19,357 & 6,001 & 3,791 & 2,807 & 2,200 & 8,112 \\ \hline
$\# \numubar$  &  1,138 &    636 &  467   &   176 &   113 &    88 &    67 &   109 \\ \hline
$\# \nue$     &    669 &    415  & 268    &  128  &  68  &   46  &    39 &    61  \\ \hline
$\#$ CC $\numu$ & 55,983 & 26,244 & 13,530 & 4,103 & 2,588 & 1,932 & 1,513 & 5,807 \\ \hline
$\#$ MRD   & 18,500 & 7,000  & 2,970  & 850   &  520  & 390  & 310  & 1,970  \\ \hline
\hline
   \end{tabular}
   \caption{\em Number of events expected in neutrino mode
     assuming 9.38 ton and $0.5 \times 10^{20}$ POT for the various 
     SciBar detector locations as identified in 
     Figure~\ref{fig:offaxis_fluxes}.  The first row reports the
     mean neutrino energy of the events (cross section 
     weighted) in GeV. The last row
     indicates the number of events with a matching track in the MRD.}
   \label{table:nu-rates-locations}
\end{table} 

\subsection{On-Axis}

The largest number of events are expected for the on-axis detector location
at 100m. Tables~\ref{table:nu-rates-onaxis} and \ref{table:nubar-rates-onaxis}
present these anticipated rates for on-axis running in both neutrino and
antineutrino configurations. Because wrong-sign backgrounds are non-negligible
in antineutrino running, the neutrino rates in this mode are explicitly 
provided (Table~\ref{table:nubar-rates-onaxis}). As can be  seen from 
both tables, the most copious interactions in the Booster beamline are CC QE.
A total of $\sim80,000$ interactions are expected in the full on-axis
neutrino exposure ($0.5 \times 10^{20}$ POT) and a total of $\sim60,000$
for on-axis antineutrino running ($1.5 \times 10^{20}$ POT).


\subsection{Off-Axis}

Table~\ref{table:nu-rates-locations} shows the number of neutrino events 
expected for the variety of off-axis SciBar detector locations that were 
considered (Figure~\ref{fig:offaxis_fluxes}). The expected energy 
distributions of events at these sites are shown in 
Figure~\ref{fig:offaxis-nu-rate-dist}. In general, the collected event samples
decrease and the energy spectra become softer as one moves off-axis. The event
rate decreases by a factor two in moving 3m vertically from the beam axis at 
z=100m (site B), and is down by a factor $\sim13$ at the surface (site D).

\clearpage

\section{Non-Neutrino Backgrounds}
\label{sec:intro_bg}

\begin{figure}
  \begin{center}
    \includegraphics[width=4.0in]{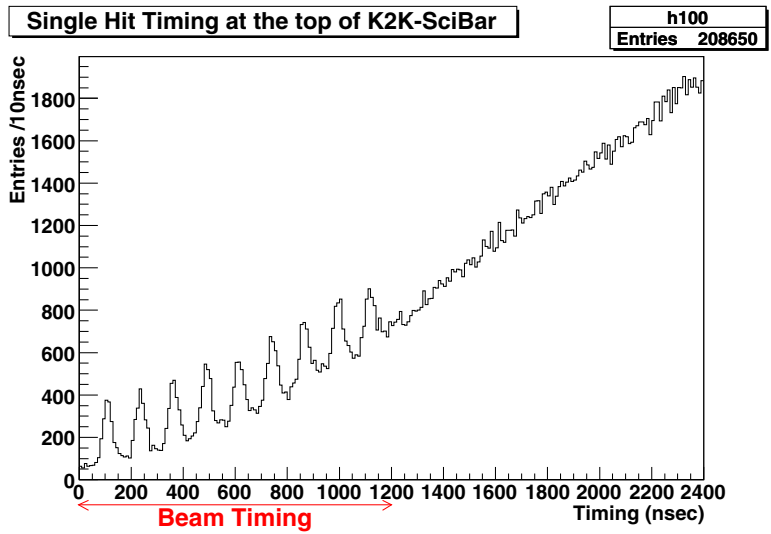}
    \caption{\em Skyshine events in SciBar at KEK.  Shown is the fine
      time structure of a single strip's hits in SciBar, during the
      K2K neutrino data runs.}
  \end{center}
  \label{fig:k2kskyshine}
\end{figure}

We anticipate background activity in the detector caused by sources
other than neutrino interactions in the fiducial volume.  They fall
into two broad categories: beam related and beam unrelated
backgrounds, described below.

\subsubsection{Beam Related Backgrounds}
\label{sec:bg_beam}

\begin{table}
  \begin{center}
    \begin{tabular}{|l|c|c|c|c|}
      \hline
                  & \multicolumn{2}{|c|}{60~m}& \multicolumn{2}{|c|}{90~m}\\ \cline{2-5}
                  & beam-on  & beam off & beam-on & beam-off \\ \hline \hline
      \# spills   & 25,589   & 10,072   & 33,441  & 10,233   \\
      singles (1) &     16   &      0   &     14  &      0   \\
      singles (2) &     37   &      0   &     20  &      1   \\
      coincidences&      5   &      0   &      4  &      0   \\
      \hline \hline
    \end{tabular}
  \caption{\em BNB skyshine test results.}
  \label{table:bnb-skyshine1}
  \end{center}
\end{table}

The two most significant beam related backgrounds are dirt neutrinos
and neutron skyshine.  Dirt neutrinos interact in the earth around
the detector hall, sending energetic particles into the detector, and
skyshine is the flux of neutrons from the decay pipe or beam dump
that are initially projected into the air but are scattered back
toward the ground and interact in the detector.  Experience with
MiniBooNE indicates that dirt neutrinos form a negligible background
for charged current events.  The expected effect on neutral current
analyses is also small due primarily to the lack of a high energy tail
in the BNB flux.

\begin{figure}
  \begin{center}
    \includegraphics[width=4in]{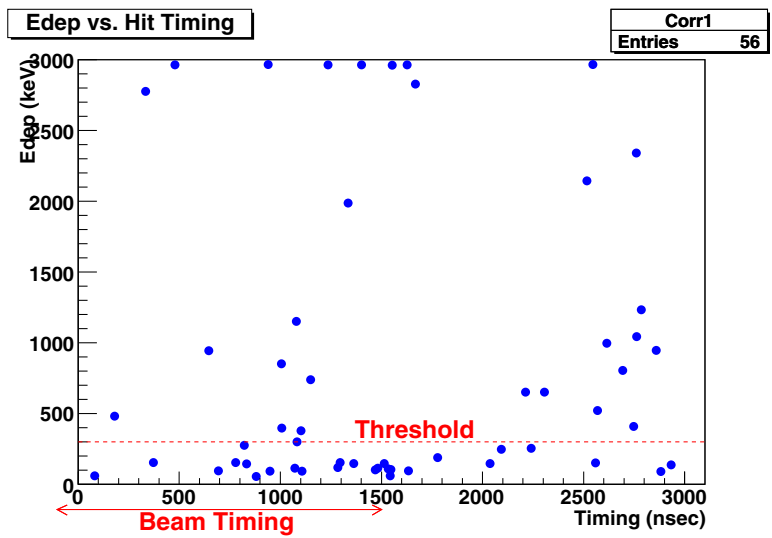}
    \caption{\em Skyshine test results at BNB: Energy deposited in one
      of the scintillation counters versus time with respect to the
      start of the beam time window.  This plot is from the beam-on
      run at 90~m.}
    \end{center}
  \label{fig:bnbskyshine}
\end{figure}

\subsubsection{Neutron Skyshine}

Neutron skyshine has been observed around particle accelerators for
many years, particularly in relation to spurious signals seen in
neutrino detectors~\cite{msu-skyshine,bnl-skyshine}.  Recent analysis
of the K2K SciBar data revealed a skyshine signature, which is
illustrated in Figure\ref{fig:k2kskyshine}.  To understand if this
background would be seen in the BNB, we performed a simple beam
related background test in July, 2005.

\begin{table}
  \begin{center}
    \begin{tabular}{|l|c|c|c|c|c|}
      \hline
                     & BNB 60 m & BNB 90 m & BNB 100 m      &  SciBar  & ground level\\
                     &          &          &(extrapolated~\cite{msu-skyshine})  &  at KEK  & at KEK\\ \hline \hline
      POT/spill      & \multicolumn{3}{|c|}{4.0-4.5$\times$10$^{20}$} & \multicolumn{2}{|c|}{5.0$\times$10$^{20}$} \\ \hline
      single hits    & 48$\pm$15                & 26$\pm$9                 & 21$\pm$10    &   2.3      &   18.0 \\
      coincidence & 8.0$\pm$3.6             & 4.9$\pm$2.5             & 4.2$\pm$2.9  &   N/A     &    N/A\\
      \hline \hline
    \end{tabular}
    \caption{\em Comparison of BNB skyshine test and K2K SciBar
      skyshine rate.  The rates (events/g/spill) in the BNB test were
      scaled up to the mass (15~ton) of SciBar.  The errors on the BNB
      skyshine rates come from the statistics of the counts in
      Table~\ref{table:bnb-skyshine1}.}
  \label{table:bnb-skyshine2}
  \end{center}
\end{table}

The test was performed by collecting hits above $\sim$300~keV from two
plastic scintillator paddles placed on the ground at distances
$\sim$60~m and $\sim$90~m from the neutrino target.  The threshold was
set around 300~keV because this is the energy deposit required to
create a signal in a SciBar scintillator bar.
Figure~\ref{fig:bnbskyshine} show the distribution of energy deposited
versus time relative to the start of the beam window for one of the
scintillator paddles at 90~m during the beam-on run.  Beam-on and
beam-off data were collected over the course of two days, with the
results summarized in Table~\ref{table:bnb-skyshine1}.  There is a
clear excess of hits with the beam on as compared to off.  Scaling the
observed rates from the masses of the two scintillators up to the
15~ton mass of SciBar indicates that the skyshine rates
(events/ton/beam~spill) in the BNB are comparable to the ground level
skyshine neutron rates seen above the SciBar near detector hall at
KEK, as seen in Table~\ref{table:bnb-skyshine2}.  This indicates that
additional shielding will not be necessary.

\subsubsection{Beam Unrelated Backgrounds}
\label{sec:bg_cosmic}

Cosmic rays are the main beam unrelated backgrounds.  The cosmic ray
rate can be cut down quite effectively with beam timing cuts, due to
the very low duty factor of the BNB.  Approximately 0.2\% of beam
neutrino events will be contaminated by a cosmic muon, but these
characteristic events can be vetoed easily.  However, the $\sim$1~kHz
rate of cosmic muons is actually useful, since it serves as a
calibration data sample for strip efficiency and track reconstruction
studies.

The average rate of cosmic ray neutrons above 50 MeV during periods of
normal solar activity at sea level and $\sim 40^{\circ}$ geomagnetic
latitude is approximately $9\times10^{-3}$sec$^{-1}$cm$^{-2}$, and the
momentum spectrum of cosmic ray-induced neutrons falls very steeply
with energy~\cite{bartol,neutrons}.  We therefore expect a cosmic
neutron rate of $\sim$2~Hz in SciBar, for neutrons above 100 MeV.
These will be a background for neutral current analyses.  These events
will be very hard to veto, since the neutrons sneak in unseen before
interacting with protons and masquerade as neutral current neutrino
events.  Therefore, we assume we will not veto any of these events.
The accidental coincidence rate should be $<3\times10^{-6}$.  Thus, we
expect to see $\sim$100-200 of these background events, depending on
Booster performance, which is consistent with previous predictions of
cosmic neutron rates at similar latitudes~\cite{nova}.  Moreover,
these cosmic background rates can be measured exactly with beam-off
data.

\section{External Time Constraints}
\label{sec:time}

There is a time constraint that affects when SciBar can operate in the
BNB: the SciBar detector will be needed back in Japan for insertion
into the T2K beamline sometime in 2008 or 2009.  While this deadline
is uncertain, it does set the upper limit on the duration of a
possible SciBar run in the BNB.

In the following three sections describing the physics that could be
done by SciBar/BNB it is assumed that the detector would be exposed to
2$\times$10$^{20}$ POT in one year of running. The current schedule
presented in Chapter~\ref{chap:cost} assumes that SciBar will be
installed and begin commissioning in the fall of 2006.

\chapter{SciBar Physics}
\label{chap:sb_phys}

The fine segmentation of the SciBar detector enables low energy cross
section measurements that can not be performed elsewhere.  Three such
opportunities are described here.  Two of these would be the first
measurements with antineutrinos, the third would be a first in
neutrinos as well.  All require the multi-track reconstruction
capabilities of SciBar.

We focus on these three measurements because the relevant analysis
techniques already exist or are in development at SciBar.  However,
tables~\ref{table:nu-rates-onaxis} and \ref{table:nubar-rates-onaxis}
show that a number of other cross sections are accessible at SciBar on
the BNB with statistics competitive or superior to previous or current
measurements in this energy range.

\section{Exclusive $\pi$-p Antineutrino Measurements}
\label{sec:pip}

%

Both K2K and MiniBooNE will provide direct measurements of the
inclusive neutrino NC $1\pi^0$ cross section at low energy. K2K has
already published an $11\%$ measurement of the NC 1$\pi^0$/total CC
ratio in their 1~kton water Cherenkov detector~\cite{k2k-pi0-paper}.
MiniBooNE is expected to have results soon from their neutrino mode
running. However, what is lacking in Cherenkov-ring based detection is
the ability to identify the final state nucleons in the event (most,
if not all, of the nucleons are below Cherenkov threshold). Because of
this, such detectors cannot provide separate measurements of the
contributing resonant cross sections, and hence, cannot separate
$\numu \, p \rightarrow \numu \, p \, \pi^0$ ($\Delta^+$) versus
$\numu \, n \rightarrow \numu \, n  \, \pi^0$ ($\Delta^0$) reactions.\\

\begin{figure}[h]
\center
{\includegraphics[width=4.0in]{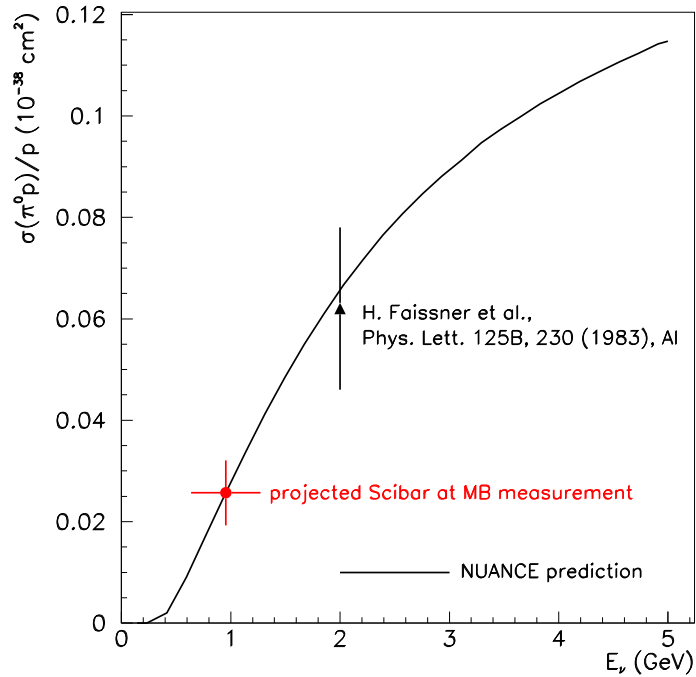}}
\vspace{-0.1in}
\caption{\em Experimental measurement~\cite{faissner} of the per
  nucleon cross section for the antineutrino resonant reaction,
  $\numubar \, p \rightarrow \numubar \, p \, \pi^0$.  Also plotted is
  the prediction from the NUANCE Monte Carlo~\cite{nuance} (which has
  not been corrected for an aluminum target). The expected measurement
  from SciBar/BNB, plotted at the Monte Carlo predicted central value,
  includes both statistical and systematic uncertainties.}
\label{fig:nc-nubar-pi0-measurements}
\end{figure}

K2K, with their currently collected near detector data, will make
a separate measurement of the  $\numu \, p \rightarrow \numu \, p  \, \pi^0$
cross section in SciBar at their mean beam energy. This result will be further
discussed in the next section. In contrast, MiniBooNE cannot measure
such an exclusive final state, but has plans to measure the inclusive 
$\numubar$ $1\pi^0$ cross section in an antineutrino exposure~\cite{fy06_loi}.
This leaves the exclusive $\numubar \, p \rightarrow \numubar \, p  \, \pi^0$ 
cross section unmeasured. Figure~\ref{fig:nc-nubar-pi0-measurements} shows 
the current available data on this particular reaction, a single measurement
on aluminum at $\sim2$ GeV appearing as a footnote~\cite{faissner}.\\

%
%
%

SciBar/BNB expects $\sim 1,100$ $\numubar \, p \rightarrow \numubar \,
p \, \pi^0$ interactions in antineutrino mode running for an on-axis
detector location (Table~\ref{table:nubar-rates-onaxis}).  Using this
sample, the experiment can make a $25\%$ measurement of this exclusive
channel. Such a measurement would be the first of its kind in the 1
GeV energy range (Figure~\ref{fig:nc-nubar-pi0-measurements}).
The statistics in the other locations would be prohibitively small.

\section{Energy Dependence of NC $1\pi^0$ Cross Section}
\label{sec:cohpion}

%
%
%

\begin{figure}[h]
\center
{\includegraphics[width=4.0in]{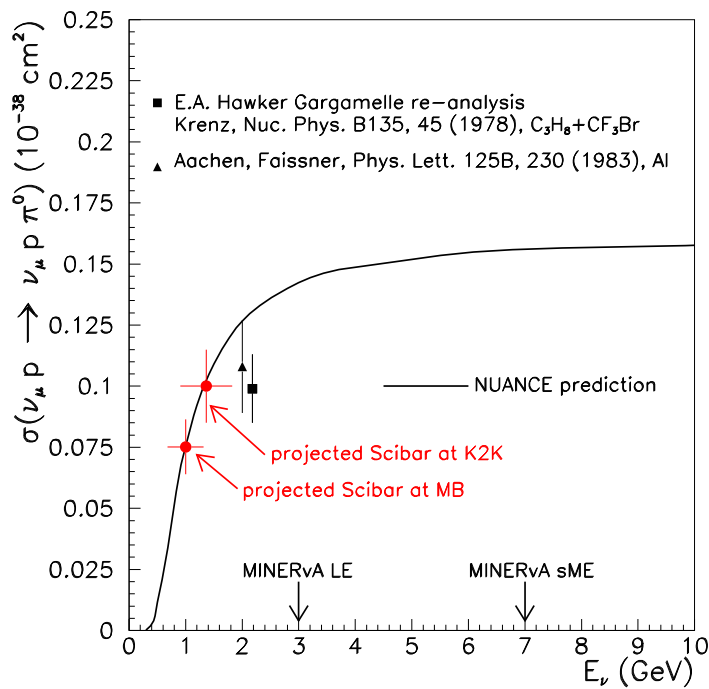}}
\vspace{-0.1in}
\caption{\em Experimental measurements~\cite{faissner,ggm} of the per
  nucleon cross section for the neutrino resonant reaction, $\numu \,
  p \rightarrow \numu \, p \, \pi^0$.  Also plotted is the prediction
  from the NUANCE Monte Carlo~\cite{nuance} (which has not been
  corrected for either the aluminum or propane-freon target data). The
  projected measurements from SciBar at both K2K and the BNB, plotted
  at the Monte Carlo predicted central value, include both statistical
  and systematic uncertainties.}
\label{fig:nc-nu-pi0-measurements}
\end{figure}

Because of the uncollected energy carried away by the final state neutrino 
in NC interactions, experiments are forced to report flux-averaged NC 
$1\pi^0$ cross sections at a single energy point. 
Figure~\ref{fig:nc-nu-pi0-measurements} shows two such published measurements 
that were both made near 2 GeV. \\

Given that future $\nue$ appearance experiments rely on precise
knowledge of their NC $1\pi^0$ backgrounds at low energy, and given
the sharp turn-on of this cross section in this energy region, one
would like to have solid experimental confirmation of the energy
dependence of the NC $1\pi^0$ cross section. SciBar can uniquely
provide such a measure in combining a NC $1\pi^0$ cross section
measurement made \emph{in situ} in the higher energy KEK beam with a
measurement made with the same detector in the Booster neutrino
beamline at Fermilab. With the 850 $\numu \, p \rightarrow \numu \, p
\, \pi^0$ events already collected with the SciBar detector at K2K, we
estimate that a $\sim15\%$ cross section measurement can be made at
the higher energy point. With the expected sample of $\sim 1,900$ such
interactions for the on-axis SciBar location at MiniBooNE (assuming
$0.5 \times 10^{20}$ POT), a $15\%$ cross section measurement can be
obtained at the lower energy point
(Figure~\ref{fig:nc-nu-pi0-measurements}). \\

The dual measurements at 1.3 GeV and 800 MeV would provide the first
mapping of this cross section in the region where it is varying most
rapidly.  Moreover, performing these measurements in the same
detector, with the same reconstruction, systematics, and model
assumptions, will provide an unprecedentedly powerful constraint.
Additionally, such information could be combined with NC $1\pi^0$
cross section measurements made at higher energy using the LE (3 GeV),
sME (7 GeV), and sHE (12 GeV) beam configurations at
MINER$\nu$A~\cite{minerva} to completely map out the NC $1\pi^0$ cross
section across the entire energy range.

It may be possible to further bin the SciBar NC cross section measurements
in energy by fully reconstructing the final state proton and $\pi^0$ in
the event. As an example, such a binned NC measurement has been reported in 
the past for the $\numu \, n \rightarrow \numu p \, \pi^-$ channel by a 
previous bubble chamber experiment at Argonne~\cite{anl-nc-p-pim}. So while it
may be possible to map out the energy dependence more finely than as
presented in Figure~\ref{fig:nc-nu-pi0-measurements}, this requires 
further detailed study.

\section{Radiative Delta Decay}
\label{sec:rad-delta}

The $\Delta$ resonance, which is produced in both CC and NC channels,
can decay radiatively ($\Delta\,\rightarrow\, N\gamma$) with a
branching fraction of 0.56\%.  Misidentification of neutral current
radiative $\Delta$ decay events are an important background for any
$\nue$ appearance search.  Distinguishing these events from $\nue$
interactions requires precise tracking, which is unavailable in large
Cherenkov detectors.  This radiative decay branching fraction has a
7\% uncertainty~\cite{pdg}, which seriously exacerbates the effects on
$\nue$ appearance oscillation analyses. Also, radiative decay modes
have only been observed via photo-production in the past, so a direct
observation of this decay mode would be the first observation of such
in a neutrino-induced (or antineutrino-induced) interaction.

With the tracking capabilities of SciBar, we can search for both a
proton and a detached photon vertex consistent with the $\Delta$ mass.
We expect ~60 $\Delta$ radiative decays (NC+CC) in the 9.38 ton
fiducial volume of SciBar during the 0.5$\times$10$^{20}$ POT neutrino
mode run, and ~30 $\Delta$ radiative decays (NC+CC, $\nu$ and
$\nubar$) in the 1.5$\times$10$^{20}$ POT antineutrino mode run.  In
the best possible case, MiniBooNE hopes to constrain $\Delta$
production using the CC1$\pi^+$ data sample to 20\%, but cannot
constrain the radiative decay branching ratio.  This results in a 25\%
systematic error on misidentified radiative $\Delta$ decays.  With the
conservative assumption of 50\% detection efficiency, we expect ~45
such events in SciBar in one year, which allows a determination of the
radiative decay branching fraction with an uncertainty of $\sim$15\%.
As mentioned above, this would be the first observation of these
events in neutrino-induced interactions.  Improvements in the
efficiency of detecting this decay mode could produce a result
competitive with the photoproduction BR measurement uncertainty.

Again, the statistics in the off-axis locations would be prohibitively small, 
leaving the on-axis location as the only viable choice for this analysis.

\chapter{Measurements that Help T2K}
\label{chap:t2k}

T2K~\cite{t2k} is a next-generation long baseline neutrino oscillation
experiment at the J-PARC facility~\cite{jparc} in Tokai, Japan. T2K is
an approved and funded experiment, currently under construction and
aiming to begin beam commissioning in 2009. T2K uses
Super-Kamiokande~\cite{sk} as a far detector with a neutrino flight
distance of 295~km to detect an intense neutrino beam with a peak
energy of 750~MeV; this gives sensitivity to the neutrino oscillation
maximum for $\Delta m^2_{23}=2.5 \times 10^{-3}$~$\rm eV^2$. The two
main physics goals of T2K are (1) a precise measurement of neutrino
oscillation parameters in $\nu_\mu \to \nu_X$ disappearance: $\delta
(\Delta m^2_{23}) \sim 10^{-4}$~$\rm eV^2$ and $\delta
(\sin^22\theta_{23}) \sim 0.01$, and (2) a sensitive search for the
unmeasured mixing angle $\theta_{13}$ in $\nu_\mu \to \nu_e$
appearance: $\sin^22\theta_{13} \ge 0.008$ at the 90\% C.L., depending
on the values of the other oscillation parameters.

Given the good match between the MiniBooNE neutrino spectrum and that
expected by T2K as shown in Figure~\ref{fig:onaxis_comp}, there are a
variety of cross-section measurements that can be made by SciBar/BNB
that would improve T2K.  We consider three such measurements.
The neutrino energies at K2K, MINOS, and MINER$\nu$A are higher and
these experiments have limited statistics in the range useful to T2K.
We note the cases in which the SciBar measurements are superior to
those made using MiniBooNE tank data alone.

The T2K collaboration is interested in having these measurements made
with SciBooNE for several reasons.  One reason is that they hope
to use the T2K near detectors to constrain their neutrino flux, which
requires accurate cross section measurements independent of their
data.  Such measurements do not currently exist and no other
experiment besides SciBooNE is capable of making them to the
required precision.  The HARP pion production cross section
measurements will give unprecedented precision to the neutrino flux
prediction in the BNB, which will allow more accurate neutrino-nucleus
cross section measurements below 1~GeV than has ever been possible
before.  The accuracy of the SciBooNE cross section measurements will
allow T2K to use their near detector event rate measurements to
extract the neutrino flux in the JPARC beam soon after it becomes
operational.

Furthermore, understanding the effects of the nuclear environment on
the neutrino interaction cross section is crucial to the success of
T2K.  Although T2K will primarily need to understand the cross
sections on oxygen, an understanding of neutrino-carbon interaction
will illuminate some of the nuclear effects.  Also, the simple fact
that the K2K collaboration is offering a \$2M detector for use at FNAL
is ample evidence of their enthusiasm for these measurements and
commitment to getting them done.

\section{$\numu$ CC$\pi^+$}
\label{sec:numupip} 

\begin{figure}[t]
\begin{center}
\includegraphics[width=0.49\textwidth]{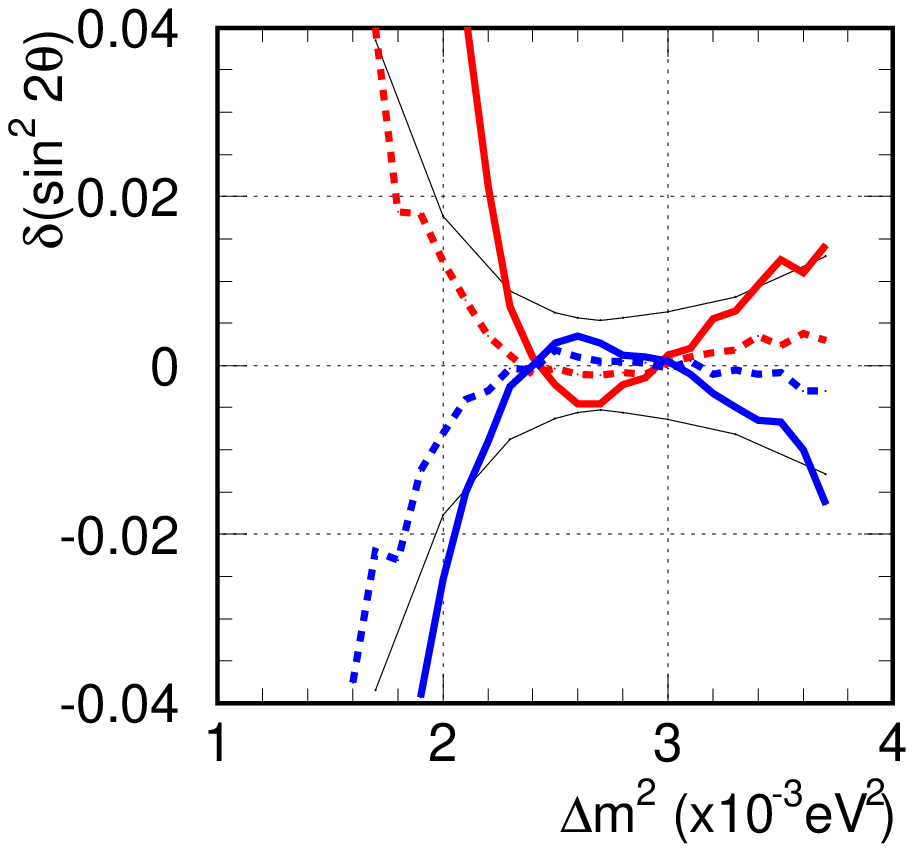}
\includegraphics[width=0.49\textwidth]{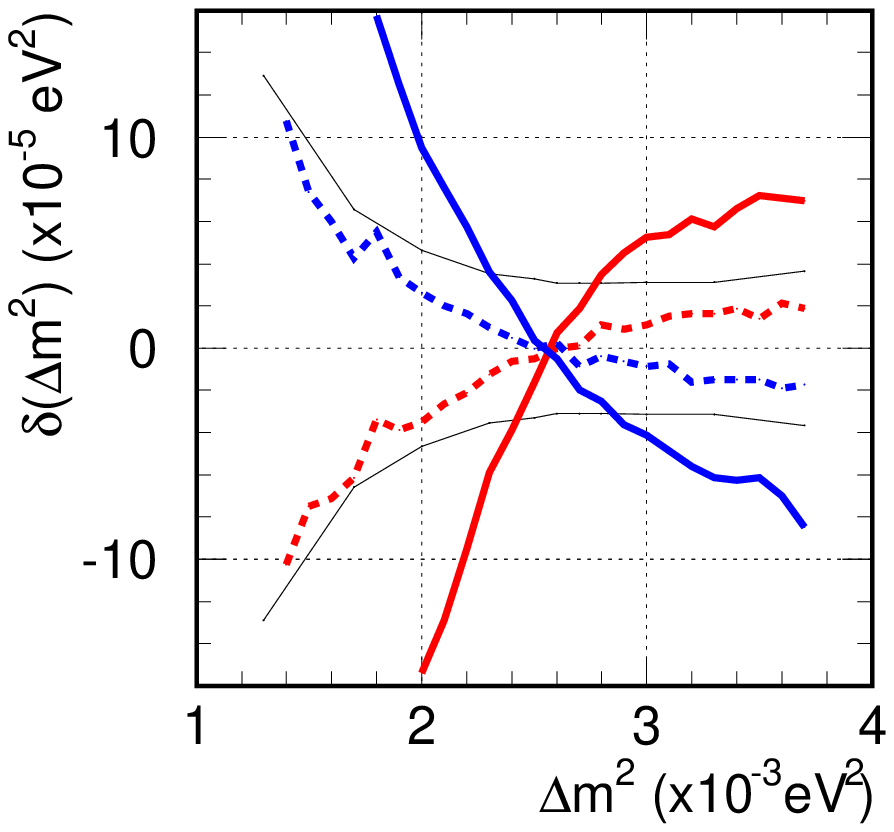}
\end{center}
\caption{\label{fig:numupip_level} \em The shift in the measurement of
  the atmospheric oscillation parameters as a function of true $\Delta
  m^2$ when an error of 20\% (solid) and 5\% (dashed) is assumed in
  predicting the the non-QE/QE ratio. The effect of shifting the
  background upward is shown by the blue line , and downward by the
  red line. The thin black line shows the irreducible uncertainty
  from statistics alone.}
\end{figure}

In T2K, the near maximal value of $\theta_{23}$ will cause a large
distortion in the $\nu_\mu$ spectrum that will be measured with
$\nu_\mu$ CC QE interactions. T2K will use this to measure
$\theta_{23}$ accurately. The background to this channel (referred to
generically as non-QE events) is dominated by single pion charged
current events (CC$\pi^+$), coming from either a $\Delta$ resonance or
by coherent production from the entire nucleus, in which the pion is
not observed so that the final state looks like a CC QE interaction.
To estimate the effect of this background, one needs only to
understand the CC non-QE/CC QE ratio as a function of energy.
Figure~\ref{fig:numupip_level} shows the effect on the oscillation
parameter measurements of making a 20\% mistake or a 5\% mistake in
predicting this background. This figure makes it clear that the
CC$\pi^+$ cross-section at these energies needs to be known to 5\% to
keep any resulting error on the oscillation parameters within
statistical uncertainties.

\begin{figure}[t]
\begin{center}
\includegraphics[width=3.0in]{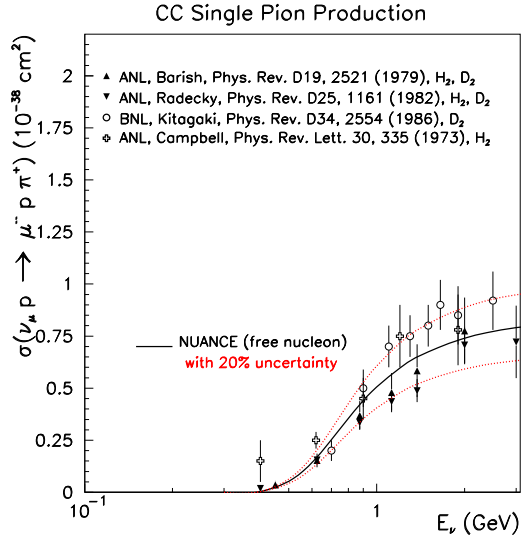}
\end{center}
\caption{\label{fig:current_ccpip} \em The current measurements of the
  $\numu\,p\,\rightarrow\,\mu^-p\pi^+$ (CC1$\pi^+$) cross section.
  Also shown is a 20\% uncertainty band around the predicted
  cross-section. Note that there are no data for any target heavier
  than deuterium below 1~GeV.}
\end{figure}

Fig.~\ref{fig:current_ccpip} shows the current state of knowledge of
the CC1$\pi^+$ interaction cross section in the 1~GeV range. This plot
shows that the current uncertainty on the CC1$\pi^+$ cross-section on
bare protons (deuterium is almost bare) is $\sim$20\%.  For carbon and
oxygen targets there are no data below 4.7 GeV; hence, the uncertainty
increases to 25-30\%, as nuclear model uncertainties become important.

Clearly, additional measurements are needed to get the uncertainty on
the CC1$\pi^+$ cross-section down to the desired 5\% level. As shown
in Table~\ref{table:nu-rates-onaxis} the expected number of CC1$\pi^+$
interactions in SciBar is over 14,000 assuming $0.5 \times 10^{20}$
POT. With cut efficiencies, we still expect $<$5\% statistical
uncertainty {\em per energy bin}.  SciBar's superior final state
resolution allows a more accurate reconstruction of the neutrino
energy and momentum transferred than is possible with large Cherenkov
calorimeter detectors. SciBar's ability to separate the final state
pion and muon from the protons that may be emitted means that, for a
subset of the events, it can actually reconstruct the invariant mass
of the resonant state.  This is allows a direct constraint on Delta
resonance production in carbon, which is a concern as the radiative
decay channel provides a non-negligible background to $\nue$ appearance
searches(see Section~\ref{sec:rad-delta}).

Since the neutrino energy can be reconstructed for CC1$\pi^+$
interactions, K2K, MINOS, and MINER$\nu$A could, in principle, measure
the cross-section despite having higher energy neutrino spectra. That
being said, at these low energies these experiments will suffer from
larger feed down from inelastic backgrounds. Some details on how well
K2K might be able to do can be found in \cite{k2k_coherent}. For
MINER$\nu$A, 1 GeV is about as low as the measurement could go.
MiniBooNE will make such a measurement, but it does not have SciBar's
ability to cleanly resolve final states.  Currently MiniBooNE
anticipates being able to make a 10\% measurement of the CC1$\pi^+$
cross-section as a function of neutrino energy, where the limit comes
from the systematic errors associated with the complexity of the final
state.

A more precise CC1$\pi^+$ cross section measurement can be made with a
SciBar/BNB on-axis location. The off-axis location B would be
acceptable as it maintains some of the flux in the energy region of
the T2K beam, but the statistics drop significantly as the threshold
for the process is approached. By the time locations C and D are
reached, the flux is too far from the T2K spectrum to provide useful
measurements.  At the off-axis location H, that maintains the same
mean energy as location A, the rate has dropped by an order of
magnitude. The statistics will still allow for a 5\% measurement of
the integrated CC1$\pi^+$ rate at that position, but any binned
measurements will suffer statistically.

\section{$\numu$ NC$\pi^0$}
\label{sec:nupi0}

\begin{figure}[t]
\begin{center}
\includegraphics[width=3.0in]{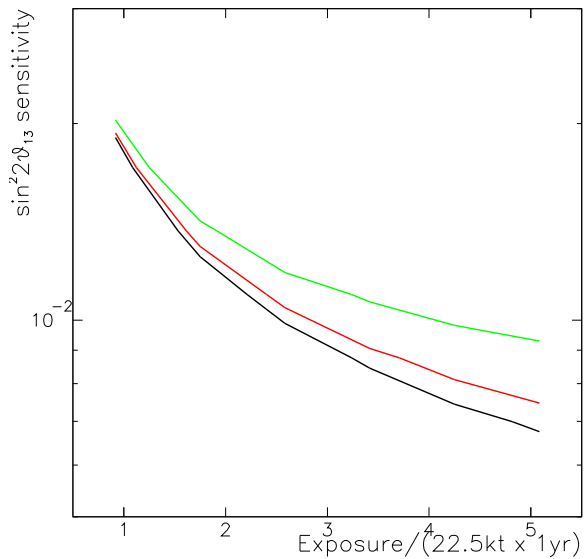}
\end{center}
\caption{\label{fig:nue_sens} \em The expected 90\% CL
  sensitivities for measuring $\sin^2 2\theta_{13}$ for uncertainties
  of 0\% (bottom curve), 10\% (middle curve), and 20\% (top curve) in
  background subtraction.}
\end{figure}

The primary purpose of T2K will be the search for $\nu_\mu$ to $\nu_e$
transitions, and a measurement of the unknown mixing angle
$\theta_{13}$.  This measurement will have significant background
contributions coming from intrinsic $\nu_e$, and $\nu_\mu$ events
misidentified as $\nu_e$ interactions.

As a function of exposure time, Fig.~\ref{fig:nue_sens} shows the
effect on T2K's sensitivity to $\sin^2 2\theta_{13}$,
assuming three different levels of uncertainty in the subtraction of
the $\numu$ misidentified and intrinsic backgrounds. For
these exposures the difference between 10\% and 0\% uncertainty is
minor, but between 10\% and 20\% there is a noticeable change. For
this reason a 10\% uncertainty on the NC$\pi^0$ cross section is
desired.

Currently, the cross-section for NC$\pi^0$ production is poorly known,
with uncertainties well in excess of 10\% and with only one or two
measurements at energies in the few GeV range. Because this is a
neutral current process it is not possible to measure the incoming
neutrino energy on an event by event basis, since the outgoing
neutrino energy is unknown. This means that the higher energy neutrino
beams of K2K, MINOS, and MINER$\nu$A do not allow
these experiments to place useful constraints on the NC$\pi^0$ rate
that might be expected in T2K. That these experiments measure
the NC$\pi^0$ rate at higher energies is very interesting,
however, as this allows the cross-section as a function energy to be
mapped, as described in Sec.~\ref{sec:cohpion}.

Since the neutrino spectrum in the BNB is so well matched to that of
T2K a measurement of the NC$\pi^0$ production rate here is much more
directly applicable to T2K. The difference between these two beams in
the high energy tail does mean, however, that the NC$\pi^0$ production
rate in the BNB will not be exactly the same as that in the T2K beam.
\begin{figure}[h]
  \begin{center}
    \includegraphics[keepaspectratio=true,width=2.75in]{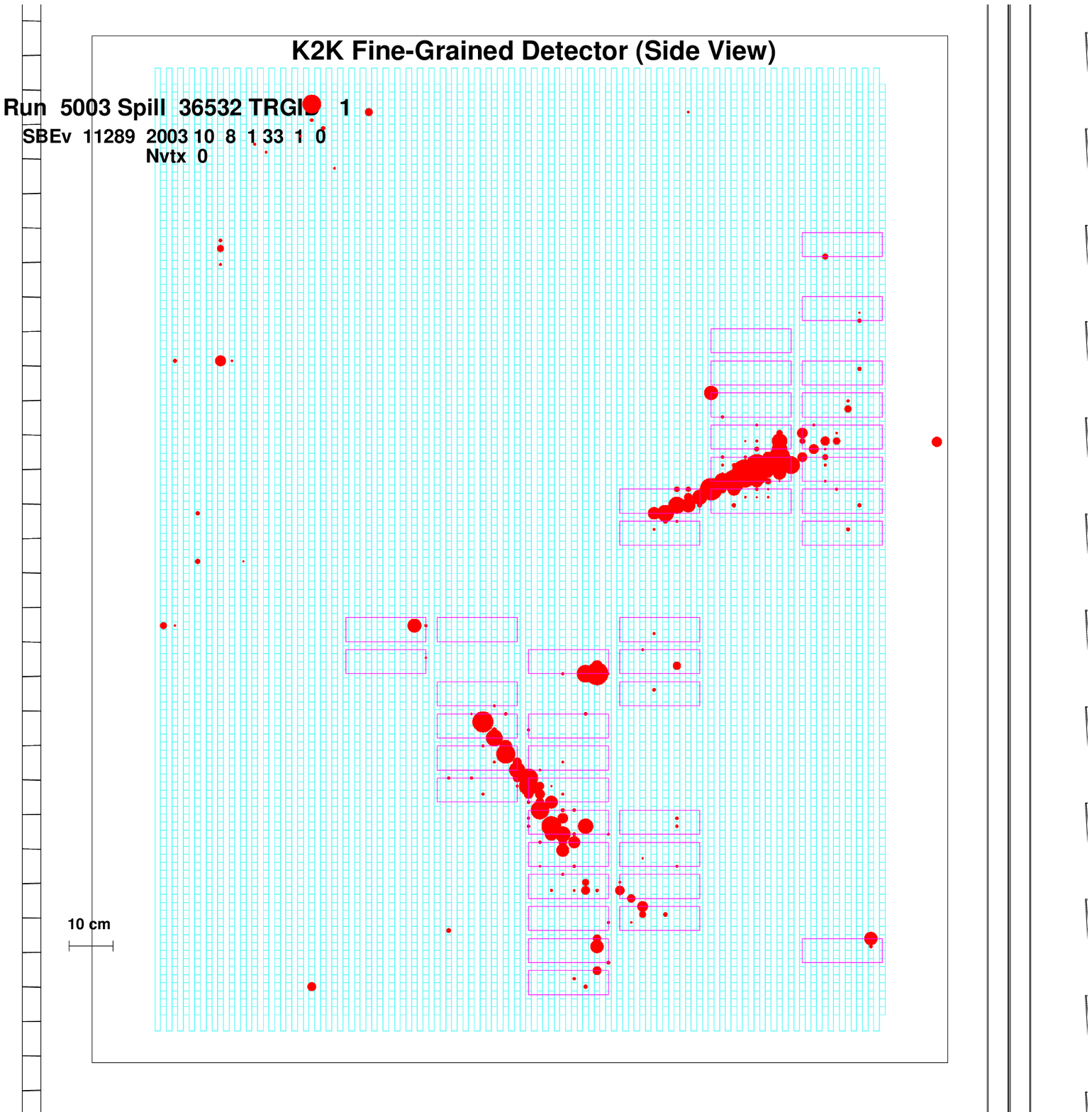}
    \includegraphics[keepaspectratio=true,width=2.75in]{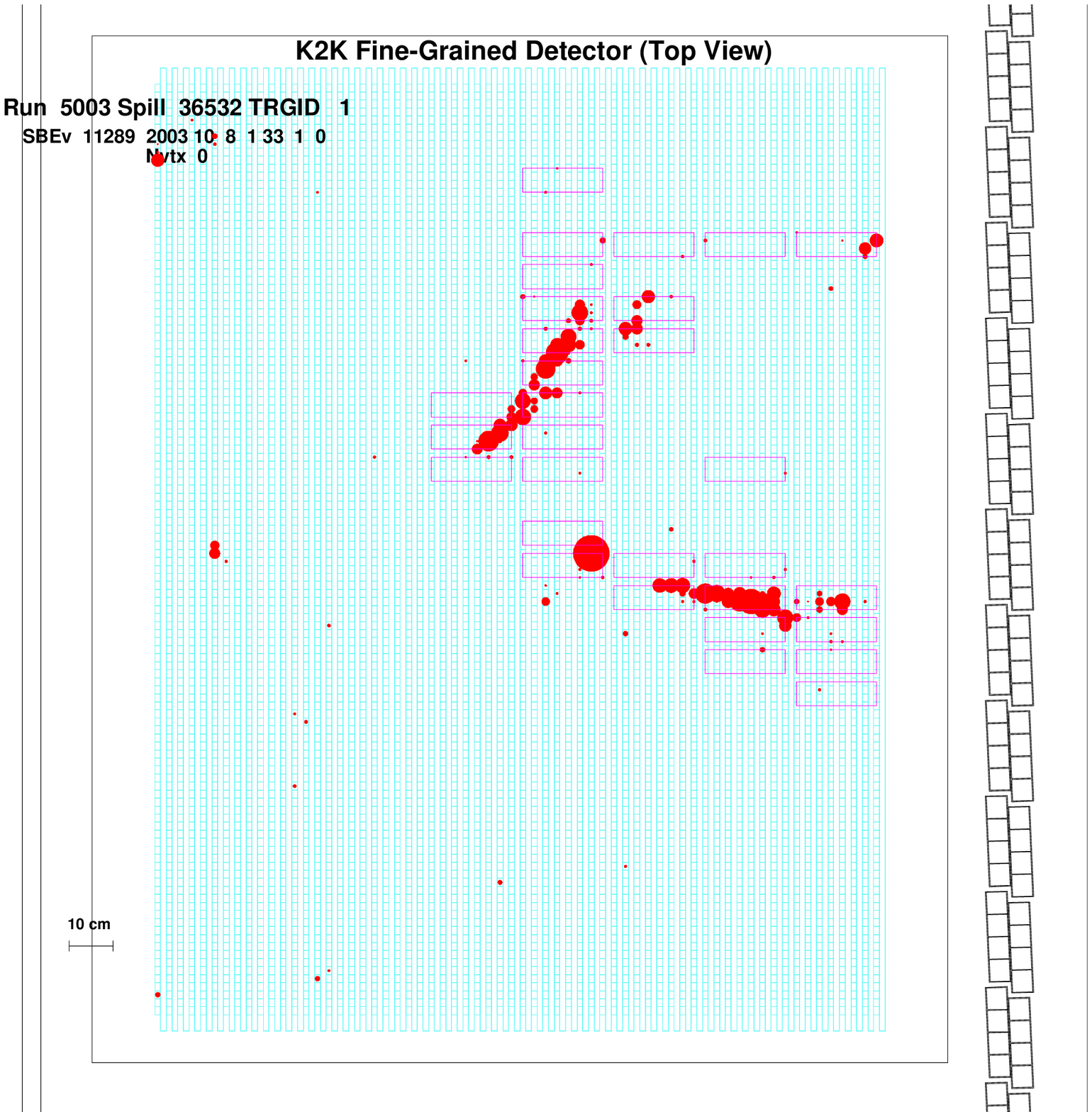}
    \vspace{.25in}
    \includegraphics[keepaspectratio=true,height=2.75in]{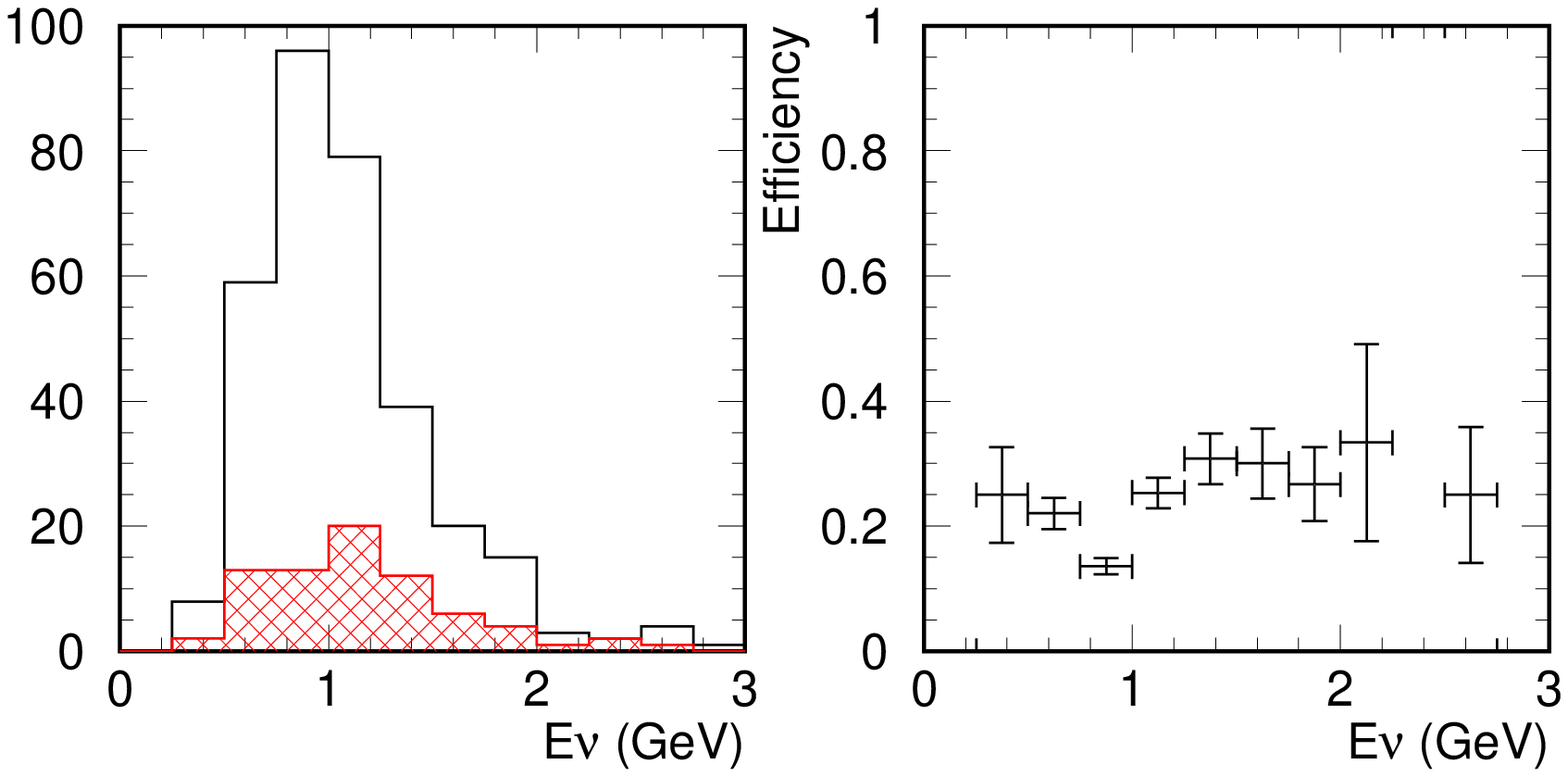}
  \end{center}
  \caption{\label{fig:pi0_display} \em SciBar event displays of a
    NC$\pi^0$ candidate from K2K data (top), and expected SciBar at
    BNB NC$\pi^0$ efficiency (bottom).  In the top panels, the two
    photon tracks point back to a common origin, which is the neutrino
    interaction vertex. In the bottom, the left-hand panel shows the
    energy distribution of NC1$\pi^0$ events that interact in SciBar
    (black curve), including those events that do not emit a $\pi^0$
    in the final state because of interactions within the nucleus, and
    those identified as NC1$\pi^0$ events (red histogram), and the
    right hand panel shows the efficiency for detecting a NC1$\pi^0$
    Monte Carlo event as a function of neutrino energy.}
\end{figure}
\noindent
Table~\ref{table:nu-rates-onaxis} shows that $\sim$3700 NC$\pi^0$
events would be expected from $0.5 \times 10^{20}$ POT with SciBar
on-axis in the BNB and 100~m from the target. We expect a 10\%
uncertainty on the total rate.  The same holds true for MiniBooNE,
which has already about ten times the statistics than expected at
SciBar/BNB.  However, SciBar has one key advantage: it tends to be the
high momentum $\pi^0$s that are most easily confused with electrons,
but it is hard to identify a sample of these in a Cherenkov detector
as it becomes harder to tell the two rings from one another (the same
reason they are misidentified as electrons). SciBar has superior final
state separation capabilities, and an electromagnetic calorimeter in
the forward direction, and hence can distinguish the two EM showers
from the $\pi^0$ decay for higher $\pi^0$ momenta.  Thus,
SciBar will be able to make a better measurement of the NC$\pi^0$
production rate at the critical highest $\pi^0$ momentum than is
achievable at MiniBooNE.

Figure~\ref{fig:pi0_display} shows two views of a SciBar event display
of a NC$\pi^0$ candidate event from the neutrino data run at K2K.  In
the display, two clear photon tracks point back to a common origin,
which is the neutrino interaction vertex. It is possible to
distinguish electron from photon tracks by measuring the average
energy deposited along the track; photon induced tracks will have
twice the deposited energy per track length because they contain two
charged particles, from the photon's pair conversion.  Note that
Figure~\ref{fig:pi0_display} is a display of a real data event.
Figure~\ref{fig:pi0_display} also shows the expected NC$\pi^0$ energy
distribution for all interactions and identified events as well as the
NC1$\pi^0$ efficiency as a function of neutrino energy for SciBooNE.
The plots shown in figure~\ref{fig:pi0_display} show Monte Carlo
events and not data efficiencies, because the NC$\pi^0$ analysis of
K2K SciBar neutrino data is ongoing, and there are not yet public
plots available.

\begin{figure}[h]
  \begin{center}
    \includegraphics[keepaspectratio=true,width=5.5in]{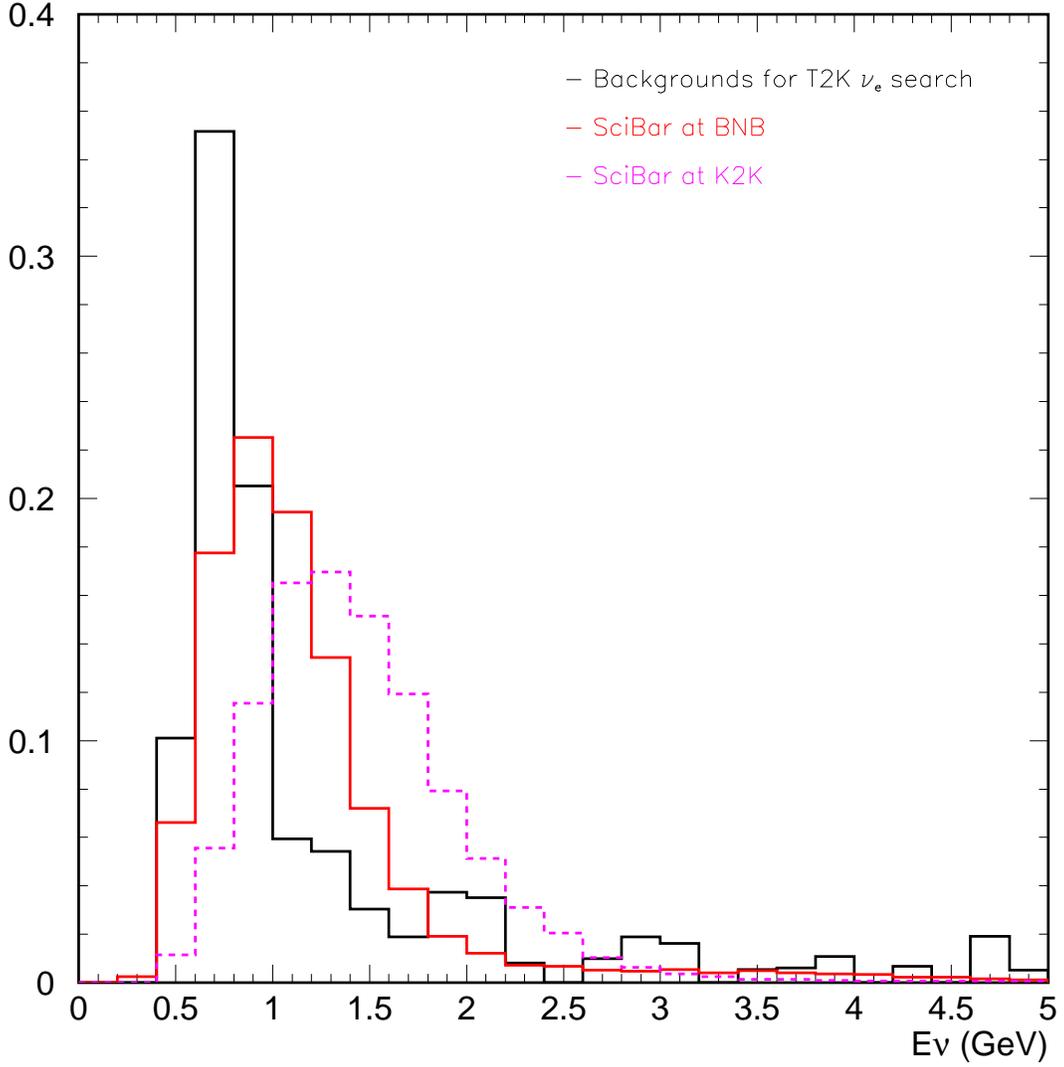}
  \end{center}
  \caption{\em Utility of SciBooNE and K2K NC1$\pi^0$ events for
  constraining T2K misidentification backgrounds. The black curve
  shows the neutrino energy distribution of NC1$\pi^0$ events that are
  misidentified as $\nu_e$ signal events, while the magenta and red
  curves show the energy distribution of identified NC1$\pi^0$ in K2K
  (magenta) and SciBooNE (red). All curves are normalized to unit
  area.  The SciBooNE events span the peak that contains most of the
  T2K background events, indicating that the SciBooNE measurement
  directly samples the energy region responsible for the bulk of the
  T2K misidentification background events.}
  \label{fig:ncpi0_energy}
\end{figure}

Figure~\ref{fig:ncpi0_energy} demonstrates the utility of the SciBooNE
NC1$\pi^0$ measurement in understanding the NC1$\pi^0$
misidentification background for the T2K $\nu_e$ appearance search.
The plot shows the neutrino energy distribution for NC1$\pi^0$ events
that are misidentified as $\nu_e$ events, with the neutrino energy
distributions for events {\it identified} as NC1$\pi^0$ interactions in
SciBar at BNB and K2K.  The figure shows that the SciBooNE measurement
will span the peak of the T2K misidentification events, where the bulk
of the $\nu_e$ background appears, but the K2K measurement does not.
In other words, the existing K2K NC1$\pi^0$ measurement is
insufficient for understanding T2K's NC1$\pi^0$ background.

The on-axis location A is the best position for SciBar to measure
NC$\pi^0$ production as this location maximizes the rate. The off axis
location B is intriguing, however, as its flux has a better match to
the high energy tail of the T2K flux than the on-axis location A.
Many NC$\pi^0$ events come from this tail and so, even though
the mean energy is wrong at location B, it may prove to be a better
location for inferring a T2K NC$\pi^0$ production rate from
SciBar/BNB. The hit in statistics from the farther off-axis locations
C and D render them unusable for this measurement, the same holds true
for location H.

\clearpage

\section{Antineutrino Measurements}
\label{sec:t2k_nubar}

T2K is expected to run in neutrino mode for its first five years of
operation. If there are indications of a finite $\theta_{13}$, T2K
will likely be upgraded, increasing to a 4~MW proton source and a much
larger water Cherenkov detector (Hyper-K). With these upgrades the
experiment would search for CP violation in the neutrino sector,
requiring oscillation measurements of both neutrinos and
antineutrino beams. It will therefore be critical to have good knowledge
of antineutrino cross-sections at this stage.

The state of antineutrino cross-section knowledge in the $\sim$1~GeV
energy range is very poor with only a handful of low statistics
measurements~\cite{fy06_loi}. If MiniBooNE runs in antineutrino mode
in FY06, its primary goal will be to vastly improve this cross-section
knowledge. There are a few ways in which SciBar can further improve these
MiniBooNE measurements. The two channels of the previous two sections
(CC1$\pi^{+/-}$ and NC$\pi^0$) will be important backgrounds to the
disappearance and appearance channels in antineutrino mode and the
advantages of a SciBar measurement described in the previous two
sections for neutrino mode will hold for antineutrino mode as well.

In addition, as was pointed out in Sec~\ref{sec:ws}, SciBar can
measure the spectrum of contaminant neutrinos in antineutrino mode in
the BNB and thus improve an antineutrino CC QE cross-section made with
MiniBooNE tank data. SciBar can also use its antineutrino CC QE events
to measure this cross section. The statistics will be lower than the
data from the MiniBooNE tank (assuming they have the same beam
exposure), but this will be a systematics limited measurement and 
SciBar can benefit from some cancellation of
systematics by virtue of the fact that it measures both the
antineutrino CC QE events and the neutrino CC QE contamination in the
same detector.

The proposed SciBar/BNB antineutrino run of $\sim1.5 \times
10^{20}$ POT will provide healthy numbers
for an antineutrino CC QE measurement and sufficient numbers for the
CC1$\pi^{+/-}$ and NC$\pi^0$ measurements. This will also ensure that
the separation of neutrino CC QE from antineutrino CC QE in
antineutrino mode will be robust.  In any of the other locations there
will probably be insufficient statistics to make SciBar measurements
superior to the ones that will be done using MiniBooNE neutrino mode
tank data.

K2K never ran in antineutrino mode and, since the experiment has been
terminated, will not in future. The NuMI beamline is capable of
switching to antineutrino mode and so MINER$\nu$A and MINOS will
probably make antineutrino measurements at some point in the future,
but NuMI is a shared beamline and the needs of the oscillation
measurements will likely come first. It is therefore unlikely that
these experiments would be able to operate in antineutrino mode for
several years. When they do they will be at higher energy which will
provide an attractive complement to the lower energy SciBar and
MiniBooNE measurements.

\chapter{Leveraging MiniBooNE}
\label{chap:leverage}

MiniBooNE is a neutrino oscillation experiment at Fermilab, whose
primary physics goal is the confirmation or refutation of the LSND
oscillation signal~\cite{lsndfinal}.  A description of MiniBooNE's
detector and analysis methods can be found elsewhere~\cite{runplan}.

We describe three measurements that SciBar can make that will improve
current or planned MiniBooNE measurements.  Only one of these
measurements, $\numu$ disappearance, is aided by concurrent
MiniBooNE/SciBar running.  The results of the other two SciBar
measurements could be applied to MiniBooNE analyses after the fact,
although concurrent running is preferred to ensure that the neutrino
beam conditions are identical.

\section{Wrong-Sign Backgrounds}
\label{sec:ws}

\begin{figure}
\center
{\includegraphics[width=2.5in]{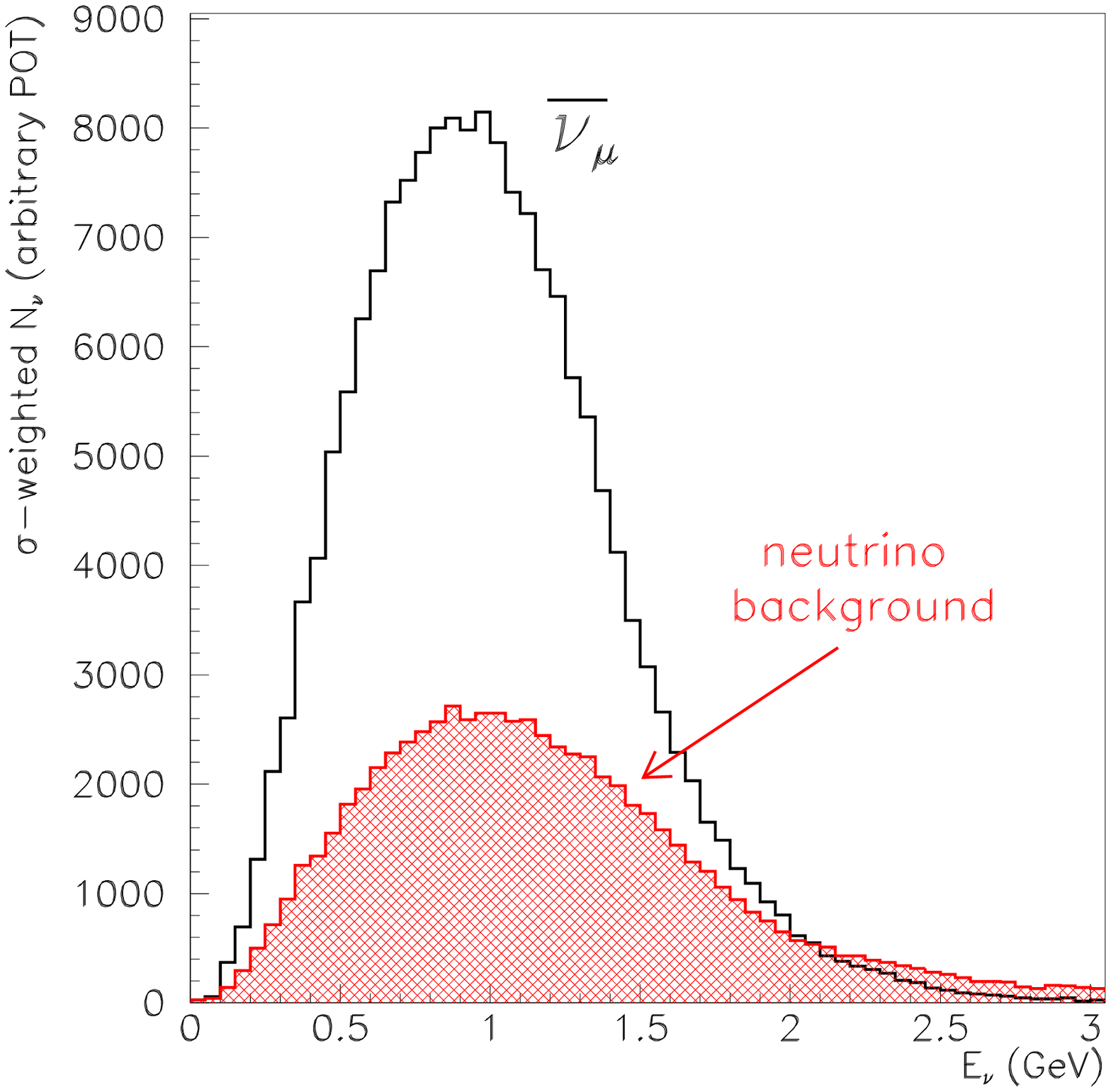}
\includegraphics[width=2.5in]{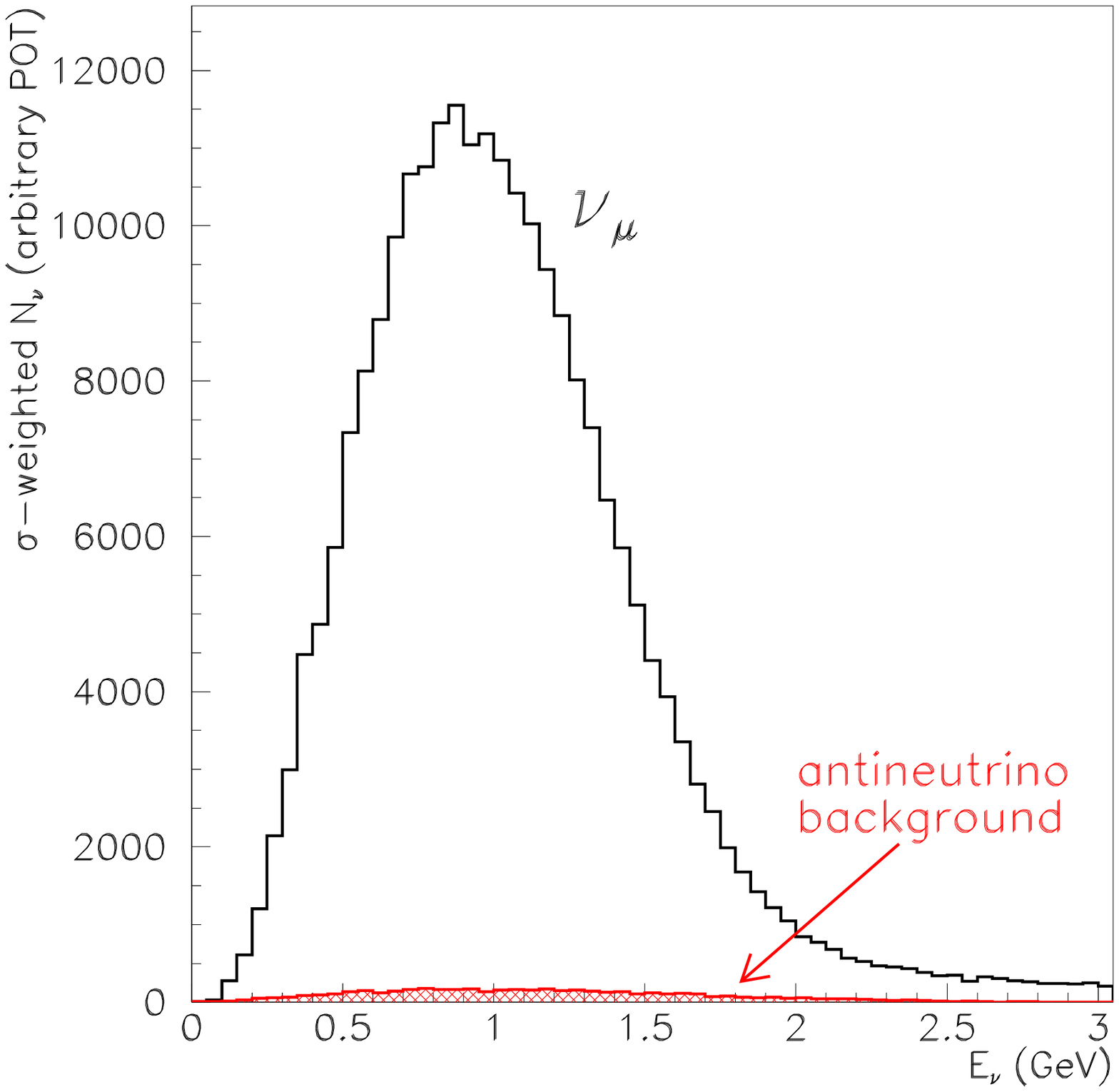}}
\vspace{-0.2in}
\caption{\em Expected energy spectra for right-sign and wrong-sign
  neutrino events (cross-section weighted) in antineutrino (left) and
  neutrino (right) modes for the on-axis (A) SciBar detector location.
  In each plot, the wrong sign events are shown with the cross-hatched
  histogram.}
\label{fig:nubarmode-ws-flux}
\end{figure}

Having precise knowledge of neutrino (``wrong-sign'') backgrounds in
data collected in antineutrino mode running is important for any
antineutrino cross section measurements, including those being planned
with phase II running at MiniBooNE~\cite{fy06_loi}.  At MiniBooNE,
these wrong-sign backgrounds comprise $\sim30\%$ of the anticipated
antineutrino mode CC QE event rate (36\% of the total rate are WS
events, Figure~\ref{fig:nubarmode-ws-flux}), and contribute a direct
source of error on any potential antineutrino cross section
measurements.  Using a combination of several novel techniques for
directly measuring the wrong-sign rates in the MiniBooNE
detector~\cite{fy06_loi}, MiniBooNE has reduced this background
contribution to a few-$\%$ uncertainty on their projected antineutrino
cross sections measurements.

SciBar is uniquely suited to provide an additional measurement of
the wrong-sign contamination in the antineutrino Booster beam by
exploiting the fact that, unlike MiniBooNE, the fine-grained
detector can differentiate between final states with protons
versus neutrons, and hence can distinguish neutrino versus
antineutrino QE interactions on an event-by-event basis:

\begin{eqnarray}
  \numu \, n \rightarrow \mu^- \, p \\ 
  \numubar \, p \rightarrow \mu^+ \, n
\end{eqnarray}

\begin{figure}[h]
\center
{\includegraphics[width=5.5in]{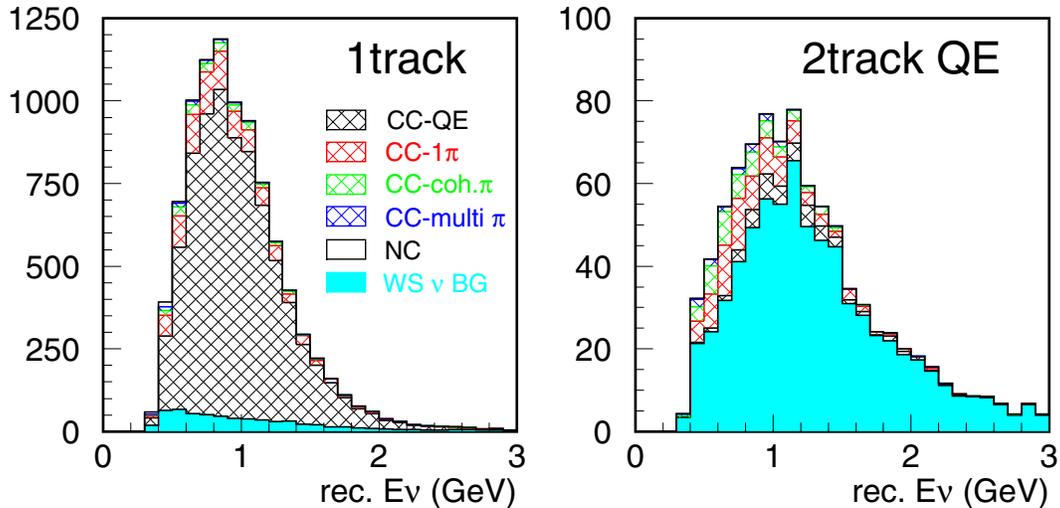}}
\vspace{-0.2in}
\caption{\em Monte Carlo generated reconstructed energy distributions
  for antineutrino mode QE events in the SciBar detector passing 1
  track (left) and 2 track (right) selection requirements. These
  particular plots were generated assuming $1\times10^{20}$ POT in
  $\nubar$ mode, assuming an on-axis location at z=100~m.}
\label{fig:scibar-nubarmode-erecon-ccqe}
\end{figure}

\noindent Based on their differing final state composition, QE
neutrino interactions are expected to have two tracks (one each from
the muon and proton) while antineutrino interactions are expected to
have only one track (from the muon).
Figure~\ref{fig:scibar-nubarmode-erecon-ccqe} shows the reconstructed
energy distributions for QE events passing one and two track selection
in the SciBar detector.  These plots were made with the full analysis
cuts developed for the SciBar CC QE analysis using K2K neutrino beam
data. Assuming a $1.5 \times 10^{20}$ POT antineutrino run on-axis,
the one track requirement yields a sample of $\sim20,000$ events, of
which $59\%$ are $\numubar$ QE interactions, $10\%$ are CC $1\pi$
backgrounds, and $29\%$ are $\numu$ QE wrong-sign backgrounds. Further
requiring less than 10~MeV deposited in the vertex strips reduces the
sample to $\sim$10,000 events, but with only 7\% WS background events
total.  This sample provides a direct measurement of the antineutrino
spectrum that is impossible with MiniBooNE tank data alone.

On the other hand, requiring two tracks in the event isolates a sample
of $\sim1,400$ events that is $80\%$ pure $\numu$ QE wrong-sign
backgrounds. Applying the converse vertex activity cut yields a sample
of $\sim$900 events that are 80\% pure WS.  This yields a direct
measurement of the energy spectrum of the neutrino background
(Figure~\ref{fig:scibar-nubarmode-erecon-ccqe} right panel) superior
to that achievable with MiniBooNE alone.  Using the angular
distributions of the outgoing muons from CC QE events, MiniBooNE
expects to constrain the WS background to 7\% uncertainty for their
full 2006 data run~\cite{fy06_loi}, with no information about energy
dependence.  By splitting the event sample into energy bins, the
energy dependence of the WS contamination can be extracted as a
function of energy.  Using four energy bins between 0 and 1.5~GeV,
MiniBooNE can extract the WS content with $\sim$15\% uncertainty in
each energy bin.  Using the two track sample, SciBar can extract the
WS content with 15\% statistical uncertainty in 100 MeV bins up to
1.5~GeV, a marked improvement over the MiniBooNE-only constraint.

In this way, SciBar can provide a superior constraint on the energy
spectrum of wrong-sign background events in antineutrino running at
MiniBooNE.  Combining this spectral constraint with measurements of
the overall wrong-sign rate obtained in the MiniBooNE detector will
lend further confidence and precision to MiniBooNE antineutrino cross
section measurements, especially those that are binned in energy.

%
%
%
%

\begin{figure}
\center
{\includegraphics[width=4.0in]{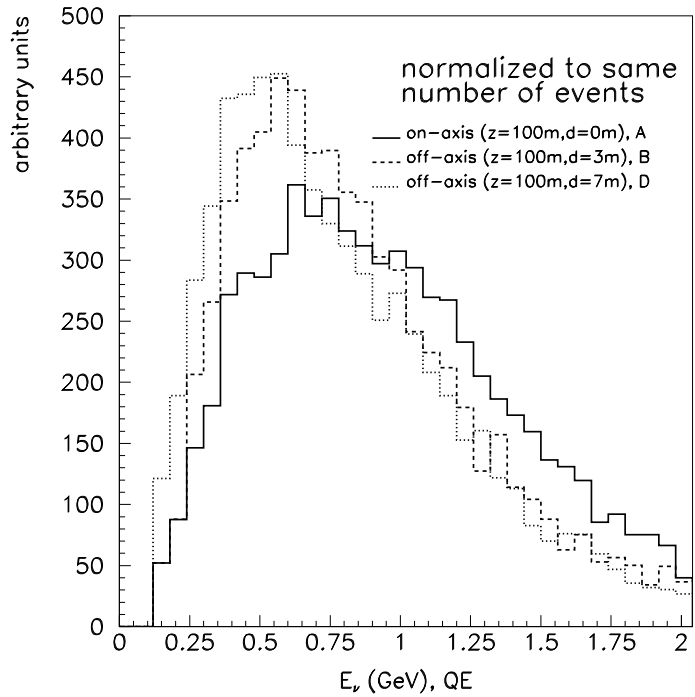}}
\vspace{-0.2in}
\caption{\em Generated neutrino energy distributions for wrong-sign QE
         neutrino events in antineutrino mode for z=100m detector
         locations on-axis (A) and two off-axis locations at 300m cm (B) 
         on the surface (D). The three distributions have been relatively
         normalized so as to compare spectral shapes.}
\label{fig:scibar-nubarmode-ws-enu}
\end{figure}

This wrong-sign event contamination actually increases as the
SciBar detector is moved off-axis because one loses the focusing
benefits of the horn (the wrong-sign fraction increases from
$30\%$ on-axis to $50\%$ by the time one reaches the surface at
z=100m). Despite this, off-axis measurements of the neutrino
energy spectrum in the antineutrino beam are not easily
transportable as constraints on the on-axis MiniBooNE beam. This
is largely due to the fact that the spectrum shifts toward lower
energies as one moves off-axis
(Figure~\ref{fig:scibar-nubarmode-ws-enu}). In addition, for a
detector location at z=100, the 300cm off-axis wrong-sign event
samples are down by a factor of two, and are decreased by a
factor of four at the surface.  This combination of sampling a
different wrong-sign energy distribution than the on-axis
MiniBooNE location and the degradation in the event sample make
it less clear how useful off-axis running is toward constraining
neutrino backgrounds in antineutrino running at MiniBooNE. To
gain full benefit, one really needs to be on-axis to provide a
useful spectral measurement.

\clearpage

\section{$\numu$ Disappearance} 
\label{sec:numudis}

In models with sterile neutrino flavors, the rate of $\numu$ or
$\numubar$ disappearance can be significantly greater than $\nue$ or
$\nuebar$ appearance.  Thus, such searches provide information on
additional mixing parameters beyond confirmation of the LSND signal.

\begin{figure}
\center
{\includegraphics[width=2.8in]{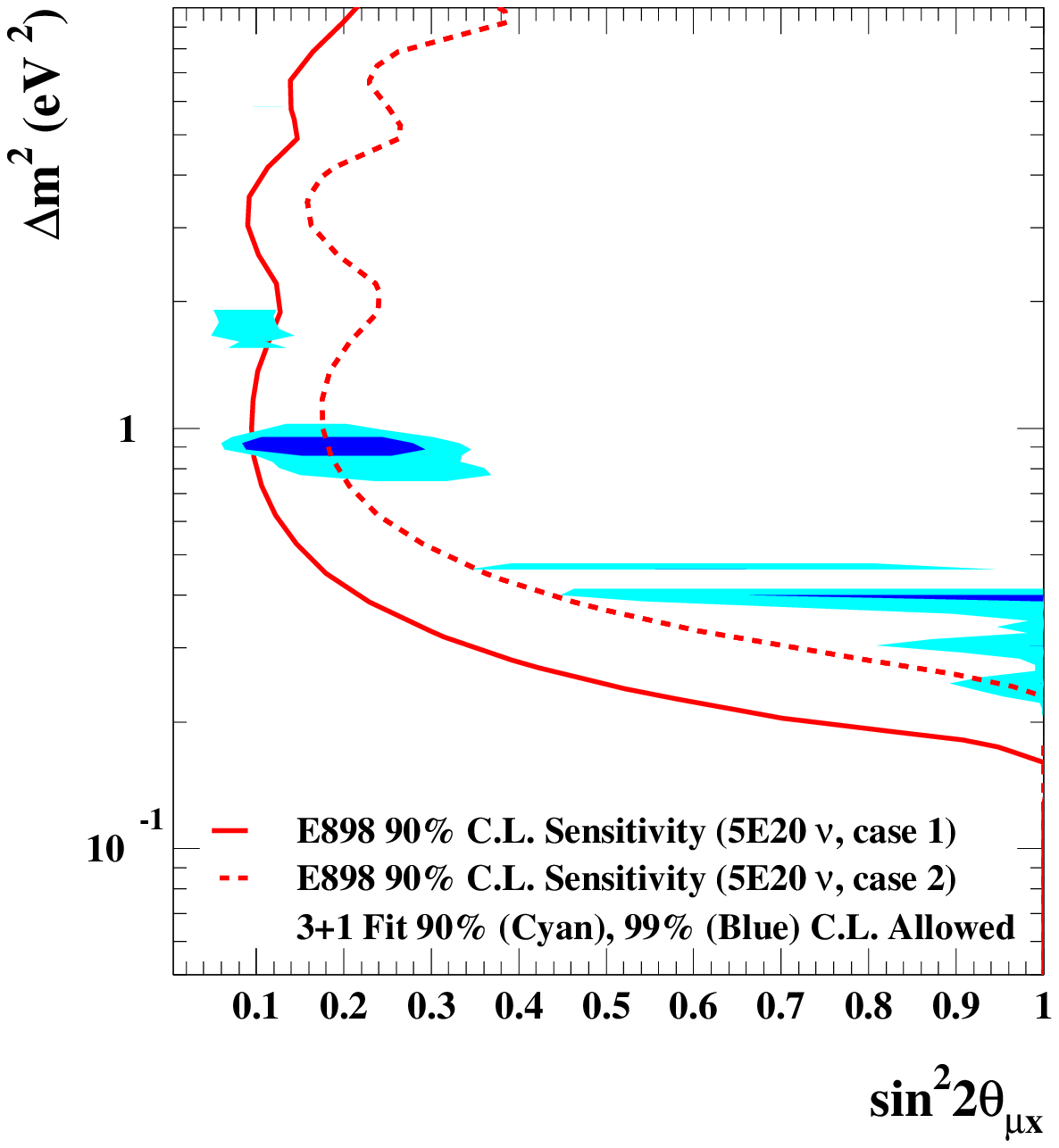}}
{\includegraphics[width=2.8in]{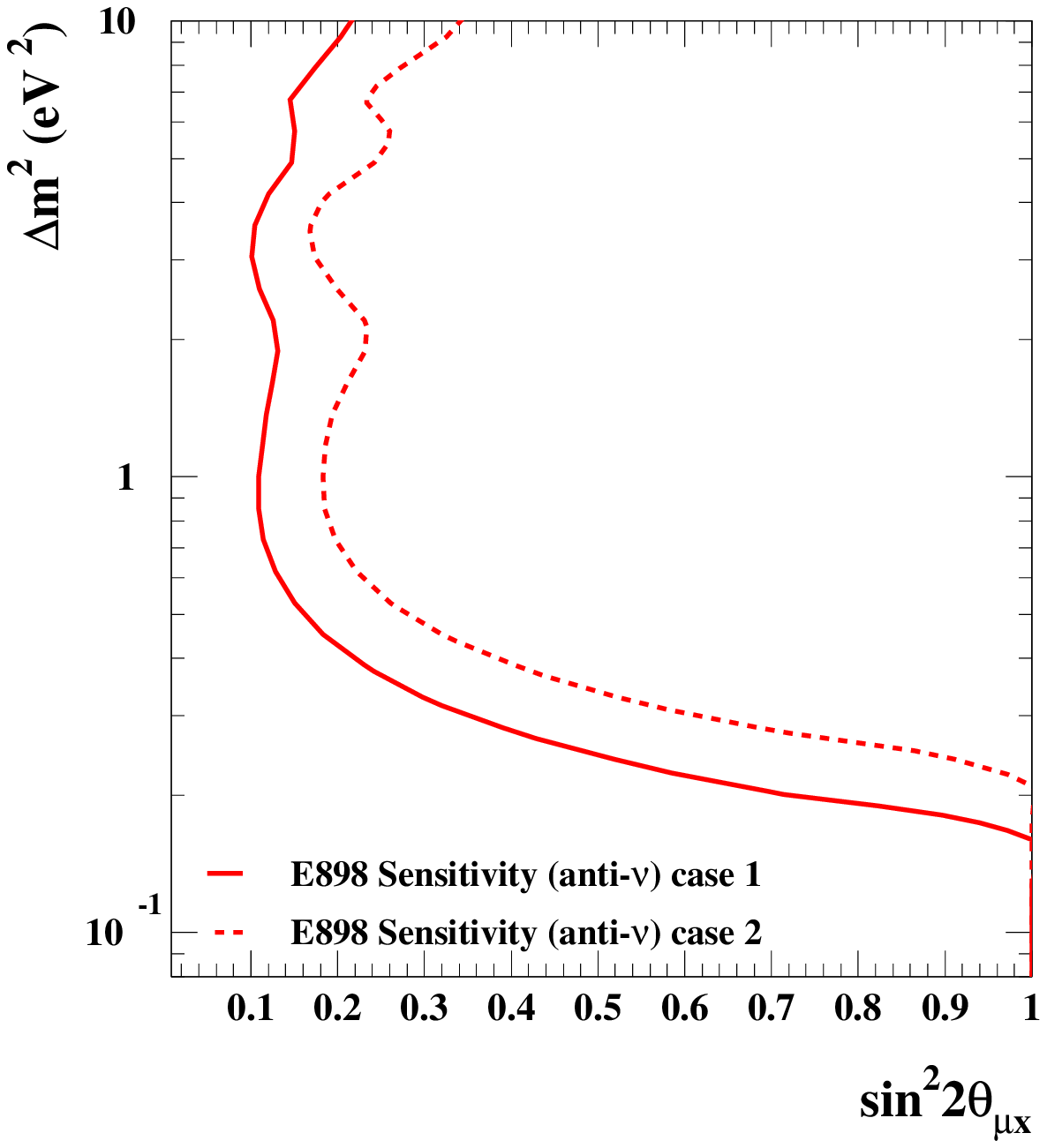}}
\vspace{-0.1in}
\caption{\em The MiniBooNE 90\% confidence level sensitivity to
  $\numu\!\rightarrow\!\nu_x$ (left, 5$\times$10$^{20}$POT) and
  $\numubar\!\rightarrow\!\overline{\nu}_x$ (right,
  1.5$\times$10$^{20}$POT) oscillations.  The projected MiniBooNE
  sensitivity is shown for two cases of systematic uncertainties; the
  solid line indicates case 1: 5\% shape and 10\% normalization
  errors, and the dotted line indicates case 2: 10\% shape and 25\%
  normalization errors.  In the left hand panel, we include the
  allowed regions for 3+1 sterile neutrino models, and note that the
  case 2 sensitivity curve does not cover these.}
\label{fig:numu_dis}
\end{figure}

The availability of a near detector significantly extends MiniBooNE's
$\numu$ disappearance reach by offering a measured constraint on the
un-oscillated $\numu$ flux normalization and energy spectrum of the
BNB.  This benefit is only realized if SciBar is placed in the on-axis
location.

In the following section, we present two $\numu$ disappearance studies
using the MiniBooNE CC QE selection cuts for both $\nu$ and $\nubar$
modes~\cite{runplan},~\cite{jocelyn-ccqe}.  We do not present detailed
near/far event spectrum ratio studies; rather, we show only how
changes in the systematic errors affect the oscillation sensitivities.
More quantitative studies are ongoing.  We note that the event rates
in SciBar and MiniBooNE are dominated by neutrino interactions on
carbon, so the plastic scintillator (CH) of SciBar is quite comparable
to the mineral oil of MiniBooNE (CH$_2$).

\subsubsection{$\nu$ Running}

For neutrino running, the use of a near detector will not improve the
sensitivity to $\numu$ disappearance with only
0.5$\times$10$^{20}$~POT~\cite{finesse}.  It is crucial to use
concurrent data for such analyses, and the short neutrino run will not
provide sufficient statistics to perform a joint $\numu$ disappearance
search with SciBar and MiniBooNE data that will approach the expected
sensitivity of the MiniBooNE neutrino run up to that time.  It will,
however, independently measure the un-oscillated $\numu$ flux, and
thus provide an external constraint on the flux normalization and
spectrum.  We show the expected 90\% confidence level
$\numu\,\rightarrow\,\nux$ sensitivity curves under two different
systematic error assumptions in Figure~\ref{fig:numu_dis}(left).  The
figure demonstrates the effects of increased normalization and shape
systematics, and thus indicates the utility of an external
measurement of the neutrino flux.

\subsubsection{$\nubar$ Running}

A disappearance search in antineutrino mode, when compared with a
disappearance search in neutrino mode, provides a powerful test of CPT
invariance.  While CP violation can only be observed in an {\em
  appearance} experiment --- by observing an asymmetry between the
appearance rates in neutrinos and antineutrinos --- the appearance
mode is unable to distinguish if the asymmetry is the result of CP or
CPT violation.  As a result, one needs to additionally search for an
asymmetry in a {\em disappearance} experiment.  Moreover, the
potential for a larger disappearance rate means that a disappearance
asymmetry may be observable even if an appearance asymmetry is not.

As described in Section~\ref{sec:ws}, the SciBar detector would allow
us to extract the energy spectrum of the wrong-sign backgrounds in
$\nubar$ running.  Exploiting this reduces the systematic error on the
shape of the $\numubar$ flux for $\numubar$ disappearance analyses.
In Figure~\ref{fig:numu_dis}(right), we show the expected sensitivity
to $\numubar\!\rightarrow\!\overline{\nu}_x$ oscillations for two
cases of systematic errors. The sensitivity region is noticeably
curtailed for the case of poorer systematic errors.

\clearpage

\section{Intrinsic $\nue$  Contamination}
\label{sec:nue}

The precision of MiniBooNE's $\nue$ appearance measurement is limited
by knowledge of the flux of intrinsic $\nu_e$s from decays of K$^+$,
K$_L^0$, and $\mu^+$ in the 50~m beam decay pipe. MiniBooNE has a
variety constraints on these different components, and has reported a
goal of $\sim5\%$ uncertainty on the intrinsic $\nu_e$ background, 5\%
on $\nue$ from $\mu^+$ decay, 5\% on K$^+$ decay and 6\% on K$_L^0$
decay~\cite{runplan}.  Even with this level of systematic uncertainty, 
it will be important to have a cross check on the $\nu_e$ backgrounds, 
especially if MiniBooNE sees a signal.

\begin{figure}
  \begin{centering}
    \includegraphics[width=2.5in]{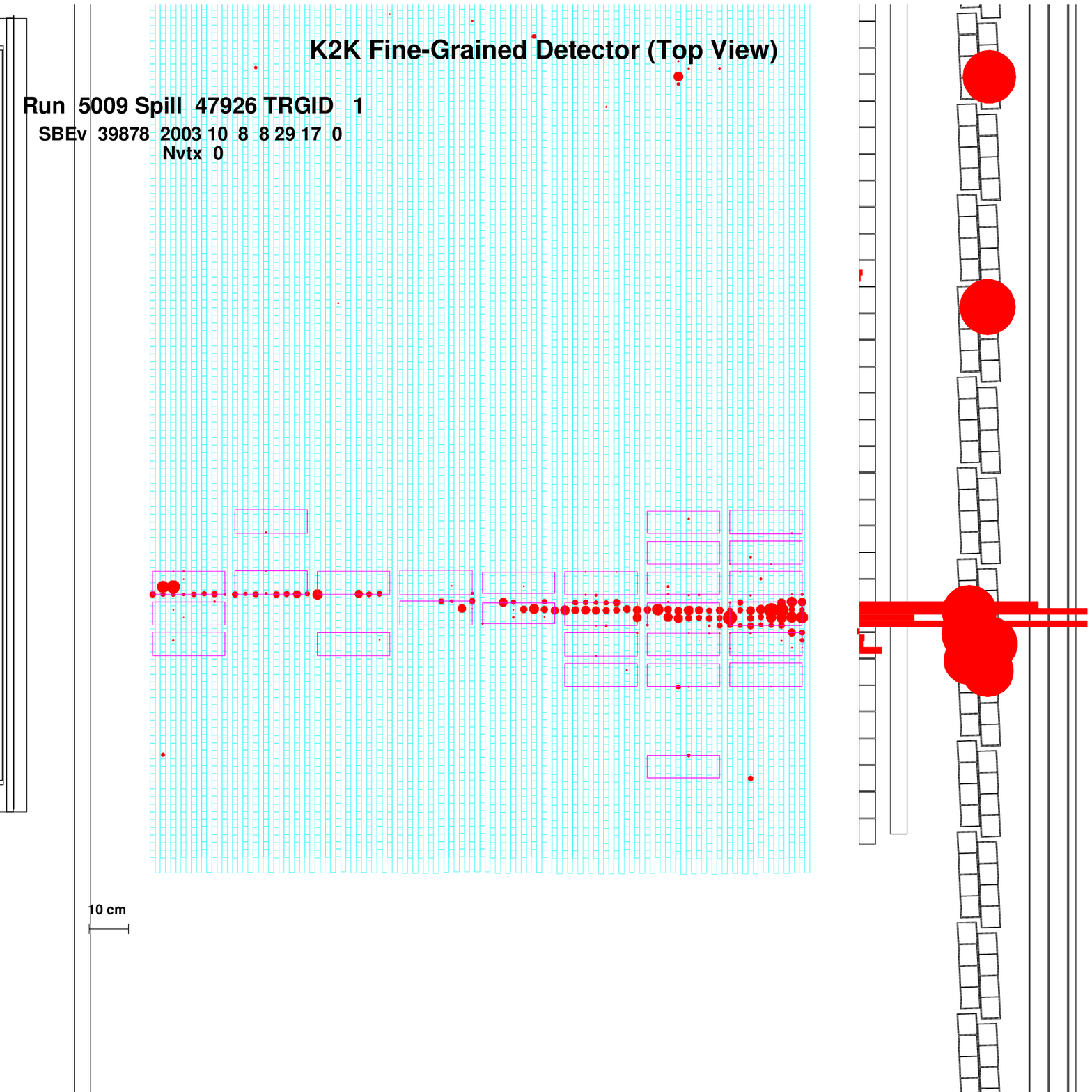}
    \includegraphics[width=2.5in]{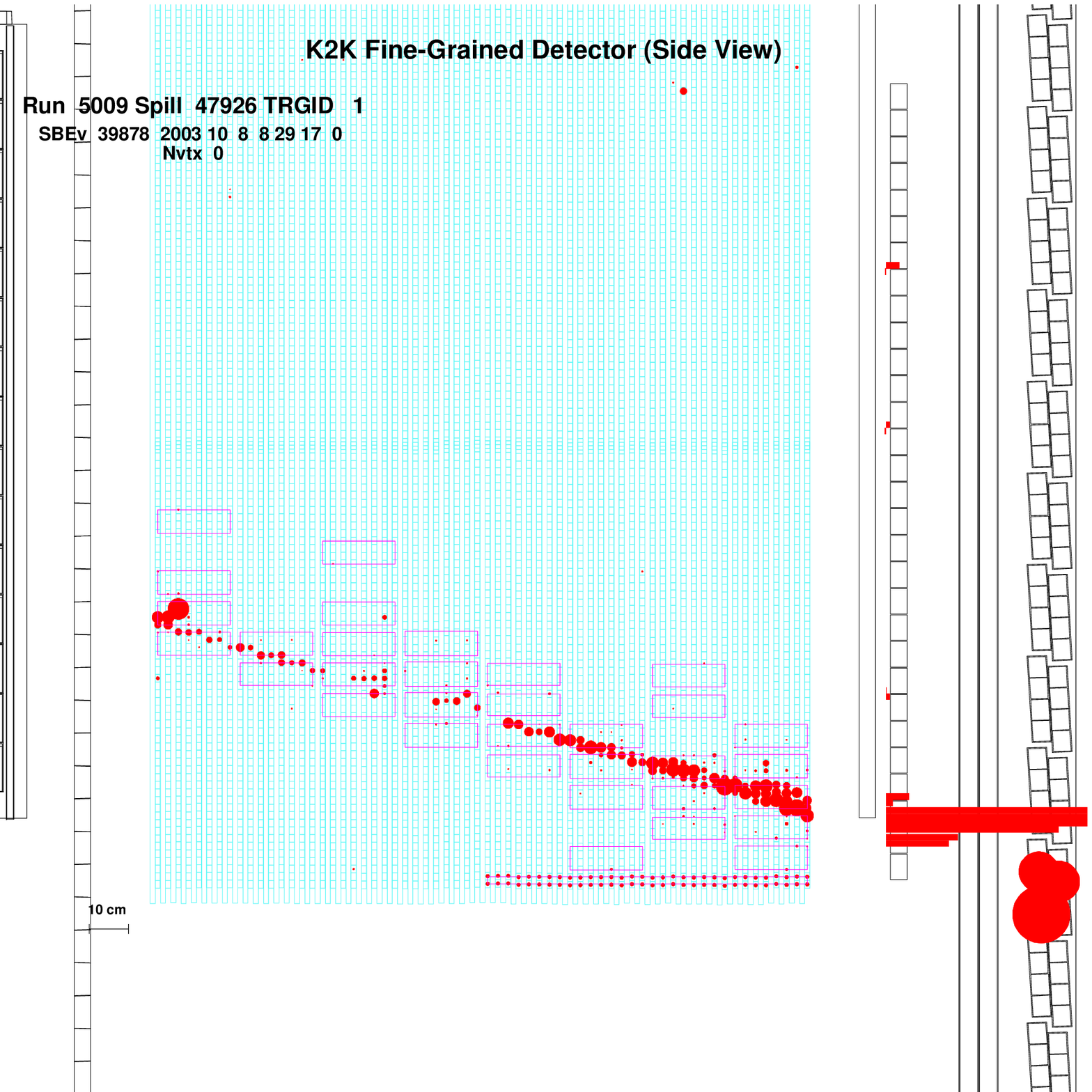}
    \vspace{-0.2in}
    \caption{\em SciBar $\nue$ CC QE candidate event display.  The
      electron's track starts with very low energy deposition but
      grows broader as it starts to shower and culminates in a huge
      energy deposit in the EC and the first layer of the MRD.}
    \label{fig:nue_evt_display}
  \end{centering}
\end{figure}

For $0.5 \times 10^{20}$ POT, there should be $\sim$490 charged
current $\nu_e$ interactions in SciBar. Based on detailed Monte Carlo
simulations, SciBar is expected to have a $\nu_e$ cut efficiency of
21\% and a purity of 88\% for electrons above 0.5 GeV (performance
numbers for lower energy electrons are not available at this time,
because the analysis is still in progress).  Additionally, only a
subset of the $\nu_e$ passing through SciBar will also pass through
the MiniBooNE tank. Considering these uncertainties, we expect to make
a 10-20\% measurement of the intrinsic $\nu_e$ component of the
beam.

Figure~\ref{fig:nue_evt_display} shows two views of an event display
of a $\nue$ CC QE candidate event in SciBar.  The electron's track,
which starts with low energy deposition but grows broader as it starts
to shower and culminates in a huge energy deposit in the EC and the
first layer of the MRD, is quite distinct from the muon track seen in
Figure~\ref{fig:cc_evt_displays}.

Although a 10-20\% measurement of the intrinsic $\nu_e$ flux does not
compete with MiniBooNE's reported goal of $\sim5\%$, it has one very
important feature: it is a \emph{direct} measurement of the $\nu_e$s
in the same beam that goes through the MiniBooNE tank. All the other
ways in which MiniBooNE can determine the $\nu_e$ flux are indirect.
The SciBar $\nue$ measurement is only valuable to MiniBooNE if the
detector is on-axis. At the off-axis locations, the $\nue$ event rates
drop rapidly; more importantly, the flux through SciBar would no
longer be the same flux that passes through the MiniBooNE tank.

\chapter{Cost and Schedule}
\label{chap:cost}

There is a window of opportunity to bring SciBar to Fermilab, but this
window will only remain open as long as the BNB continues to operate.
Fermilab's current schedule closes the BNB by the end of 2006. A study
carried out by Fermilab and KEK indicates that an operating SciBar
could occupy the beam within nine months of approval.

We therefore request approval before the end of the CY2005, to allow
funding grants for participation in SciBar at BNB to be submitted with
a positive statement of approval at FNAL.  This, combined with our 9
month schedule, means that we would expect to be taking cosmic ray
data with SciBar in the on-axis detector enclosure by 1 October,
2006.

There are three sub-detectors in SciBar, described in
Section~\ref{sec:intro_det}. SciBar and the EC will be shipped from
KEK, whereas the MRD can be easily assembled at Fermilab using
materials from retired fixed-target experiments.  The installation of
an enclosure in the BNB, shipping and assembly of detectors at
Fermilab, and construction and assembly of the MRD will take about
nine months.

The schedule depends on successfully decoupling the larger tasks, so
that they can proceed in parallel.  Reconstruction of SciBar and the
EC will take place in the NuMI surface hall (MI-65).  They will each
be mounted on a platform, so that at completion they will be lifted
onto a flatbed truck, and taken, fully constructed, to the detector
enclosure for installation. Each subdetector will be mounted on a
platform so that it can be brought by truck to the detector enclosure.
The MRD will be built in two modules to assure that we keep the weight
of each module below 15 tons.  Placing the sub-detectors on the floor
of the detector enclosure will require rental of a 100 Ton mobile
crane for about one week.

The materials needed for the MRD have already been identified, and
their assembly could be done before the arrival of the detector from
KEK, so that technician and physicist time would be free for the
assembly of SciBar and the EC.

Once the decision to proceed is made, SciBar could arrive at Fermilab
within four months of the decision. This allows four months for
assembly of SciBar and the EC at Fermilab.  In month 1, KEK will
prepare to disassemble the detector, making all of the arrangements to
commit students and technicians to work on the project.  In month 2,
KEK will disassemble cables, front-end electronics, PMTs and fibers.
In month 3, the scintillator in SciBar and the EC will be
disassembled.  Shipping should take approximately one month.
Installation at Fermilab will include about two weeks to install the
scintillator into the frame to reconstruct the SciBar detector. Then a
month will be needed to connect the fibers, PMTs, and front-end
electronics.  At this point the detector can be tested with cosmic
rays.  After the detector is installed in the beam, about two weeks
will be needed to connect cables, back-end electronics, and the DAQ
system.  These time estimates are based on experience from
installation at K2K.

The critical path for occupancy in the beam will be the construction
of the detector enclosure.  A design study was carried out by FESS and
PPD engineers to derive a cost estimate and schedule for the detector
enclosure.  These are given in Appendix A.  The detector enclosure
will be a vertical shaft, twenty feet deep. The shaft will be capped
with a shed made of light materials and with a removable roof.
Installation of the detectors will be done by a mobile crane---the
detectors lowered through the roof onto the floor of the shaft.  After
the detectors are installed, a mezzanine will be placed a few feet
above to provide room for electronics racks.  Cables from the detector
will run directly into the bottom of the relay-racks. One relay-rack
will be required on the enclosure floor next to the SciBar detector.
The Data Acquisition System will come from Japan; on-site data storage
and analysis will be done with Fermilab's Enstore system and local
computing.

Two vertical ladders will provide access to the detector enclosure.
The top ladder starts at grade and terminates at the mezzanine. The
lower ladder leads from the mezzanine to the enclosure floor.  The
shaft will have minimal need for lighting and environmental controls,
since most of the work associated with assembly of the detectors will
be done in the NuMI surface hall (MI-60). In one year and four months
of running at KEK, access to the detector was required only twice.
Dehumidification will be needed only to keep the enclosure air below
the dew-point.  A gas fire protection system will be used to avoid any
need to bring ICW water to the building.  This is currently under
review.  Power will be brought in from the nearby MI12 service
building as a 480~V service, using a small step-down transformer at
the enclosure to convert to 120~V house power. A communication line
will also be run between MI12 and the SciBar enclosure for telephone
and Ethernet connections.  A new, full three dimensional model for the
detector enclosure is being developed and will be done before
December, along with an updated cost estimate and a review of the
detector environmental conditioning requirements and safety
considerations.  These updates will be available before the December
PAC meeting.

The construction schedule of the detector enclosure requires about
nine months between approval and beneficial occupancy; the design
process takes about two months; two months are also required for the
procurement process: placing an ad for an RFP, evaluating and
selecting a bid, etc.  The period of construction is about 4.5 months.
See Appendix A for more details.
   
FESS has prepared a cost estimate for civil construction, which is
given in Appendix A. The anticipated cost for the civil contract is
about \$290,000. Engineering costs at (21\% of contract price) would
be about \$60,000. Contingency and overhead at nearly 50\% add
approximately \$160,000 to the total project cost.

The assembly of the detectors onto platforms, and installation into
the detector enclosure will add $\sim$\$5,000 each for the four
sub-detectors.  Crane rental for a week is $\sim$\$5,000.  A rigging
crew may be needed for about one week.  This adds up to $\sim$\$30,000
in Laboratory M\&S.

KEK will be responsible for the cost to disassemble, package and ship
the detector to Fermilab, and to return it to Japan.

\chapter{Conclusions} 
\label{chap:conclusions}

The marriage of K2K's fine-grained SciBar detector and the Booster
Neutrino Beamline presents a unique, low risk, and low cost
opportunity for low energy neutrino and antineutrino measurements that
are useful to the neutrino community at large.

The present knowledge of neutrino cross sections in the few GeV region
is not commensurate with the physics goals of future oscillation
experiments~\cite{itow}.  Based on recent experience, low energy
neutrino cross sections may still have some surprises in store.  For
example, MiniBooNE realized an important new class of background
events for experiments that seek to identify $\nu_e$'s, from radiative
$\Delta$ decay, that had been previously overlooked.  Further, both
MiniBooNE and K2K observe a deficit of events in data with respect to
Monte Carlo at $Q^2 < 0.2$~GeV$^2$, which is attributed to a lack of
theoretical understanding of the nuclear
environment~\cite{jocelyn-ccqe,k2k_coherent}.  As we consider the
future, with plans for precision oscillation measurements, we must ask
ourselves what new surprises await.  The cross section measurements
proposed here will help to lay the foundation needed for the future
off axis programs, and ensure that any new surprises will be found
soon enough to determine strategies to handle them.

This effort complements the existing and future neutrino programs at
Fermilab, providing important input to MiniBooNE as well as crucial
cross section measurements for off-axis neutrino experiments, most
especially T2K.  This project utilizes a pre-existing detector and an
operating beamline which are both well understood and have both
demonstrated high quality performance.  Additionally, this modest
investment will complement the lab's existing neutrino program by
providing a significant and high quality data set that will be useful
for training students.  It will also draw to Fermilab a significant
number of neutrino physicists who would otherwise concentrate their
efforts in Europe or Japan in 2007.

The window of opportunity to bring SciBar to Fermilab will only remain
open as long as the BNB continues to operate.  We therefore request an
extension of the data run of the BNB through the end of FY2007,
regardless of the result of the MiniBooNE $\nue$ appearance search.
Furthermore, we request approval before the end of calendar year 2005,
to allow our collaborators to request funding to work on SciBar at
BNB.  Prompt approval combined with our 9 month schedule means that we
would expect to be taking cosmic ray data with SciBar in the on-axis
detector enclosure by 1 September, 2006, and neutrino beam data as
soon as the summer accelerator shutdown is over.

\appendix

\chapter{Off-Axis NuMI Locations}
\label{chap:numi}

Positioning the SciBar detector in the NuMI beamline was also
considered as a possibility. We have studied neutrino fluxes in the
NuMI surface hall, as well as several locations in the existing NuMI
off-axis tunnel.  Table~\ref{table:numi-offaxis} shows the locations
and off-axis angles of four of the specific locations considered.  For
sufficiently small angles, one can calculate the expected $\nu_{\mu}$
flux and energy for the two-body decay of a single $\pi^+$ of energy
$E_{\pi}$,

\begin{eqnarray}
  \Phi_\nu = \frac{A}{4\pi r^2} \left( \frac{2\gamma}{1+\gamma^2\theta^2}\right)^2 \label{eqn:offaxis_f}\\
  E_\nu = \frac{(m_{\pi}^2-m_{\mu}^2)}{m_{\pi}^2} \frac{E_{\pi}}{(1+\gamma^2\theta^2)}, \label{eqn:offaxis_e}
\end{eqnarray}

\noindent 
for a detector of cross sectional area $A$, at a distance $r$ from the
decay point of the pion and angle $\theta$ with respect to the pion's
momentum.  Note that $\gamma = E_{\pi}/m_{\pi}$, and that the formulas
can also be used to calculate the flux and energies for neutrinos from
two-body decays of $K^+$, with the appropriate substitutions.
Table~\ref{table:numi-offaxis} also gives the expected peak $\nu_{\mu}$
energy from $\pi^+$ decay for each of the locations considered.

\begin{table}[h]
\centering
   \begin{tabular}{ c c c c c c } \hline
Location               &   x(m)   &   y(m)   &   z(m) & $\theta$(mrad) & peak E$_{\nu_{\mu}}$ (GeV)\\ \hline
Near 2a                &    14    &      0   &    740 &  16   &  1.8  \\ 
Near 3a                &    14    &     -6   &    940 &  19   &  1.6  \\ 
NuMI surface building  &     0    &     71   &    940 &  76   &  0.4  \\ 
MiniBooNE              &    26    &     78   &    745 &  111  &  0.25 \\  
\hline
   \end{tabular}
   \caption{\em Comparison of positions of four off-axis locations
   in the NuMI neutrino beam, and the peak $\nu_{\mu}$ energy from
   pion decays at that location calculated using the ``off-axis
   formulas,''(see equation~\ref{eqn:offaxis_e})}
   \label{table:numi-offaxis}
\end{table} 

\begin{figure}[h]
\center
{\includegraphics[width=6.0in]{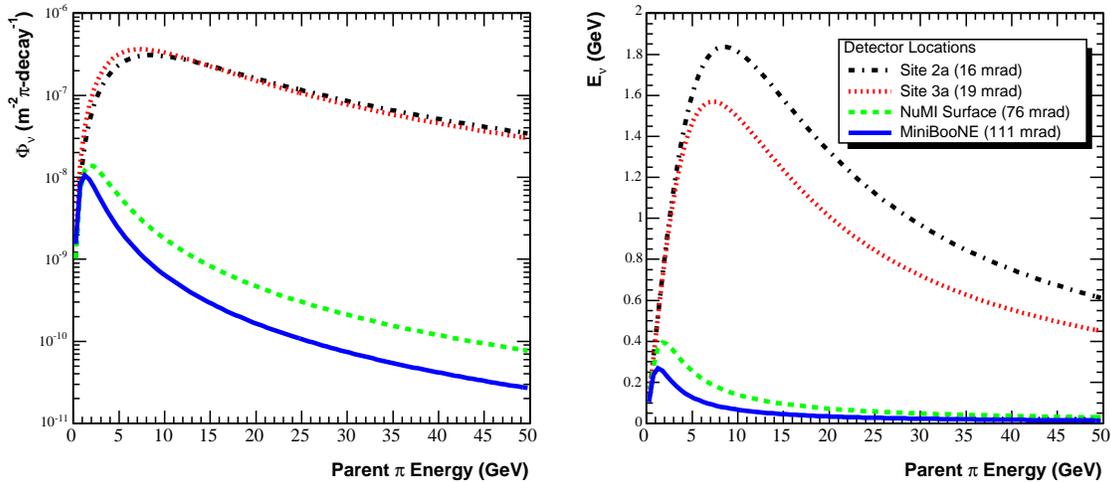}}
\vspace{-0.2in}
\caption{\em Comparison of possible neutrino fluxes (left) and
energies (right) from pion decays as a function of pion energy, at
four off-axis angles for the SciBar detector. The four off axis angles
considered are based on the locations of NuMI Sites 2a and 3a, as well
as the NuMI surface hall and the MiniBooNE detector, which is included
since neutrinos from the NuMI beam have already been observed in the
MiniBooNE detector. Note the left hand plot (flux) is shown on a log 
scale.}
\label{fig:numi-offaxis}
\end{figure}

\begin{figure}
\center
{\includegraphics[width=3.0in]{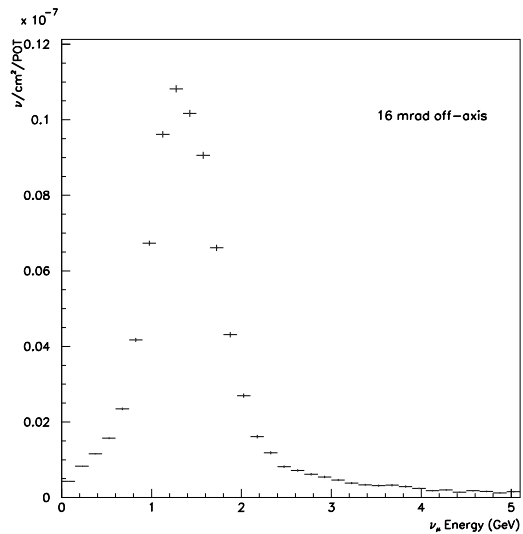}}
\vspace{-0.2in}
\caption{\em Expected $\nu_{\mu}$ flux from $\pi$ decay in the NuMI LE
configuration for the off-axis Site 2a, 16~mrad off axis.  The peak
neutrino energy is $\sim$1.4~GeV, about 200~MeV higher than the peak
energy at K2K and 800~MeV higher than the expected peak energy for
T2K.}
\label{fig:numi_2a}
\end{figure}

Figure~\ref{fig:numi-offaxis} shows the $\nu_{\mu}$ flux and energy
curves as functions of parent pion energy, given by
equations~\ref{eqn:offaxis_f} and~\ref{eqn:offaxis_e}, for the four
locations in Table~\ref{table:numi-offaxis}.  The peak neutrino
energies for sites 2a and 3a are at 1.8~GeV and 1.6~GeV from pions of
energy 7~GeV and 9~GeV, respectively.  However, the neutrino flux for
site 2a (16~mrad) falls relatively slowly as a function of pion
energy, so that the expected neutrino flux from pion decays in flight
for the low energy (LE) NuMI configuration at Site 2a peaks around
1.3~GeV, as shown in Figure~\ref{fig:numi_2a}.  This neutrino flux was
calculated using the full gnumi beam Monte Carlo used by the NuMI
collaboration, with the beam configured in LE mode.  The neutrino flux
for site 3a (19~mrad) is not quite as flat as a function of pion
energy, so the integrated neutrino flux is expected to peak closer to
the peak pion energy than for site 2a.  Flux studies for site 3a using
gnumi are ongoing.

Unfortunately, there are several drawbacks to these locations. Most
obviously, one loses the direct physics benefits to MiniBooNE with a
NuMI site (Chapter~\ref{chap:leverage}).  Additionally, the numbers in
Table~\ref{table:numi-offaxis} indicate that the available locations
(see Table~\ref{table:numi-offaxis}) do not offer a $\nu_{\mu}$ energy
distribution that is suitably close to the expected T2K flux to make
the cross section measurements needed for T2K
(Chapter~\ref{chap:t2k}).  Sites 2a and 3a are too high in energy and
the NuMI surface hall is too low in energy.

\begin{figure}
\center
{\includegraphics[width=3.0in]{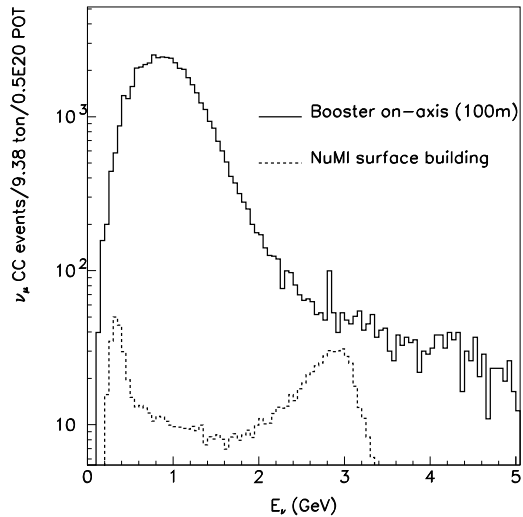}}
\vspace{-0.2in}
\caption{\em Comparison of cross section weighted energy distributions
for CC $\numu$ events in a 9.38 ton carbon detector for $0.5 \times 10^{20}$
POT at Booster and NuMI locations.  Note that these event
distributions do not include the effects of detector acceptance or
cut efficiencies.}
  \label{fig:numi-surface}
\end{figure}

Figure~\ref{fig:numi-surface} compares the predicted energy
distributions for CC $\numu$ events at the NuMI surface and Booster
on-axis SciBar locations. This figure also shows the high energy
neutrino peak from kaon decays.  The NuMI surface hall event rate
peaks below and above the T2K energy peak ({\em cf.}
Figure~\ref{fig:onaxis_comp}), although this figure does not include
the effect of detector acceptance, which would largely cut out the
high energy (kaon) peak.  Studies of cross-section weighted event
rates for site 2a and 3a are ongoing.

While we do have several ongoing studies, all of which will be
completed by December, it seems unlikely that the NuMI off-axis
locations offer a neutrino flux suitable for the physics goals we have
set out to accomplish.

\chapter{Civil Construction Documents}
\label{appendix}

\begin{figure}[htbp]
  \begin{center}
    \includegraphics[keepaspectratio=true,height=90mm]{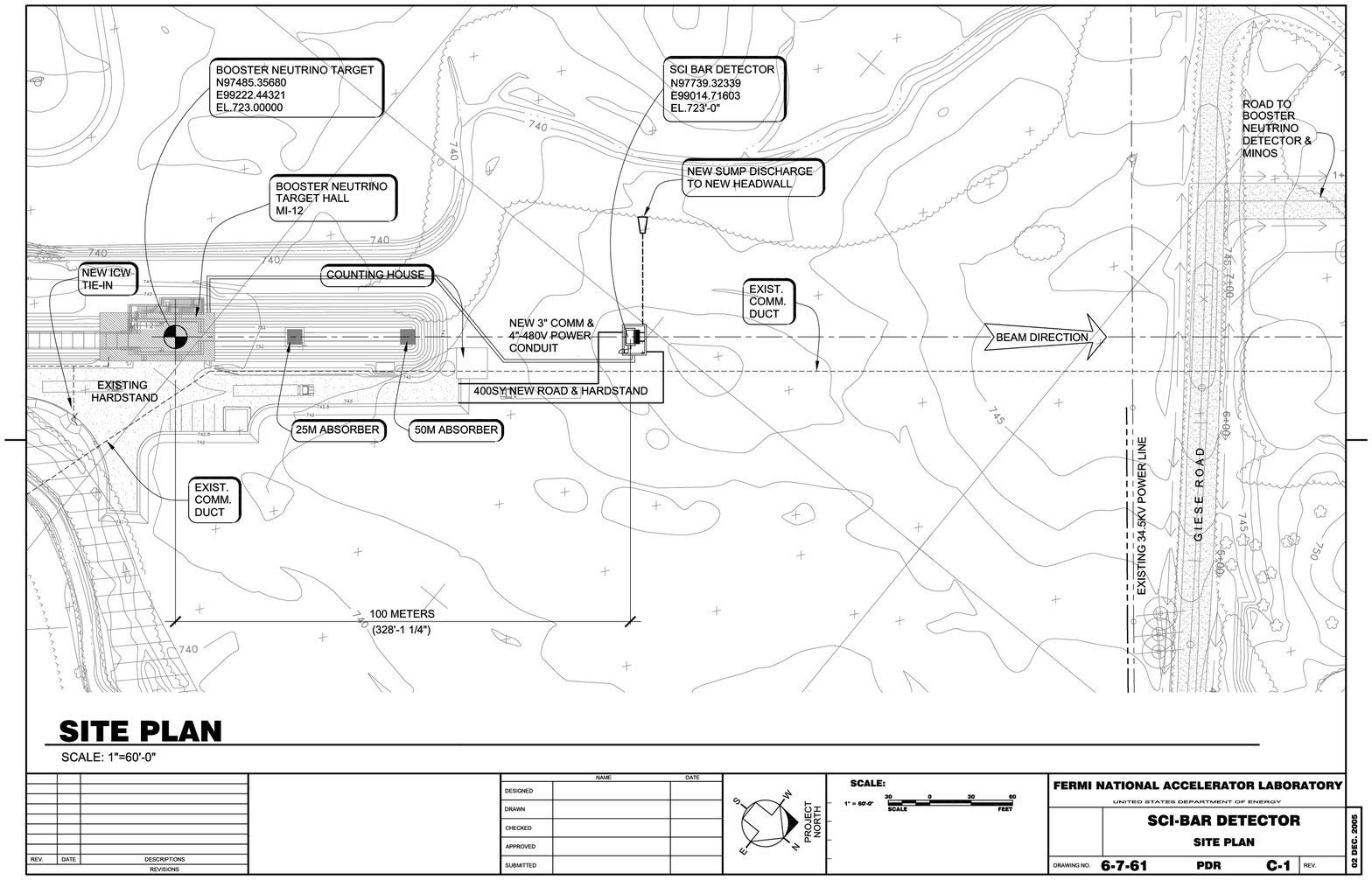}
  \end{center}
  \caption[Schematic view of SciBar]{\em Site Drawing.}
  \label{fig:PDR_1.eps}
\end{figure}

\begin{figure}[htbp]
  \begin{center}
    \includegraphics[keepaspectratio=true,height=90mm]{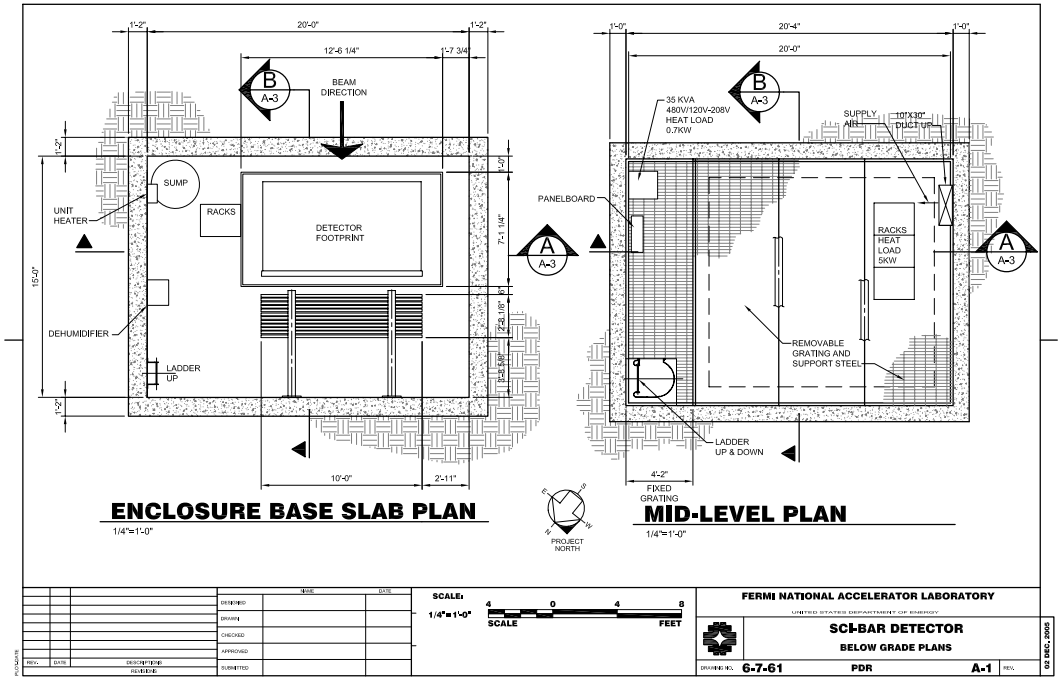}
  \end{center}
  \caption[Floor level of the beam enclosure.]{\em Sketch of the floor level of the enclosure.}
  \label{fig:PDR_2.eps}
\end{figure}

\begin{figure}[htbp]
  \begin{center}
    \includegraphics[keepaspectratio=true,height=90mm]{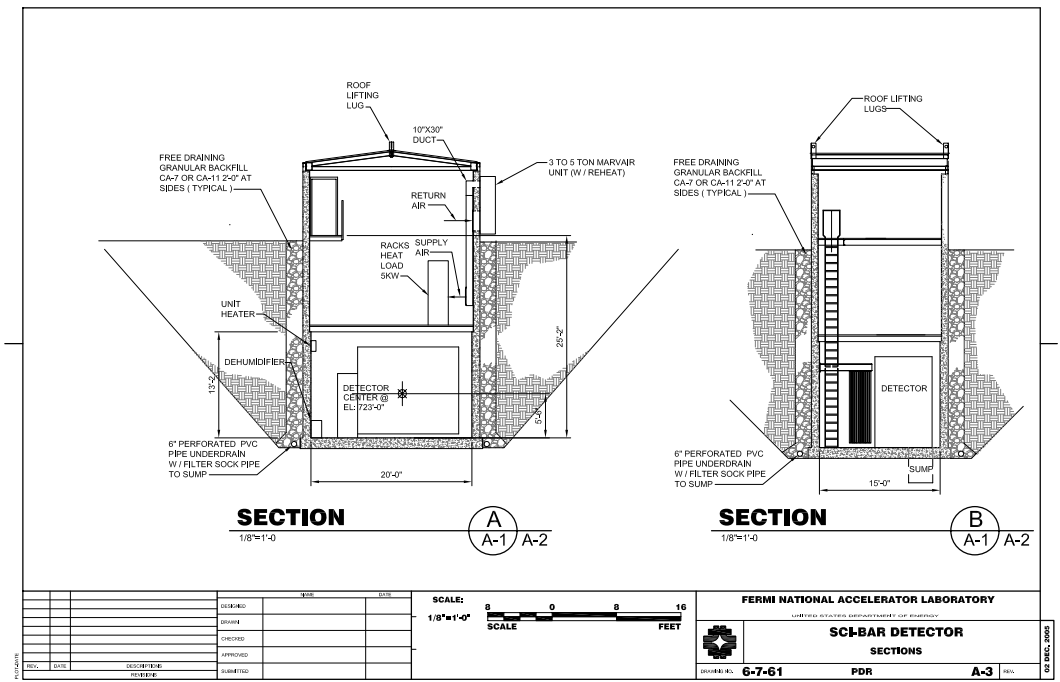}
  \end{center}
  \caption[Beam enclosure elevation view.]{\em 
    Elevation views of the beam enclosure.}
  \label{fig:PDR_4.eps}
\end{figure}

\begin{figure}[htbp]
  \begin{center}
    \includegraphics[keepaspectratio=true,height=90mm]{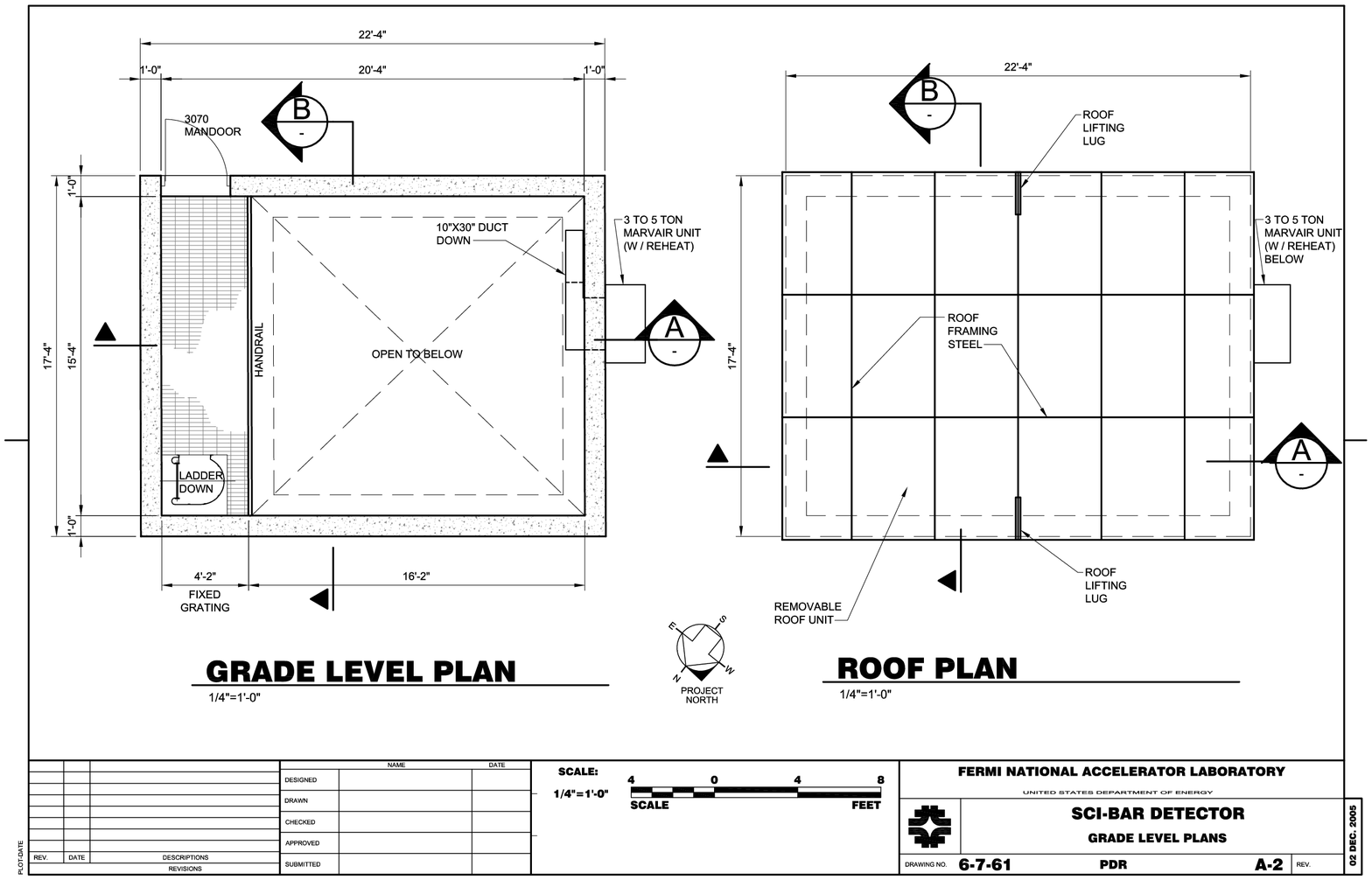}
  \end{center}
  \caption[Beam enclosure view at grade level.]{\em 
    Plan view of the enclosure at grade level.}
  \label{fig:PDR_3.eps}
\end{figure}

\begin{figure}[htbp]
  \begin{center}
    \includegraphics[keepaspectratio=true,width=5.5in]{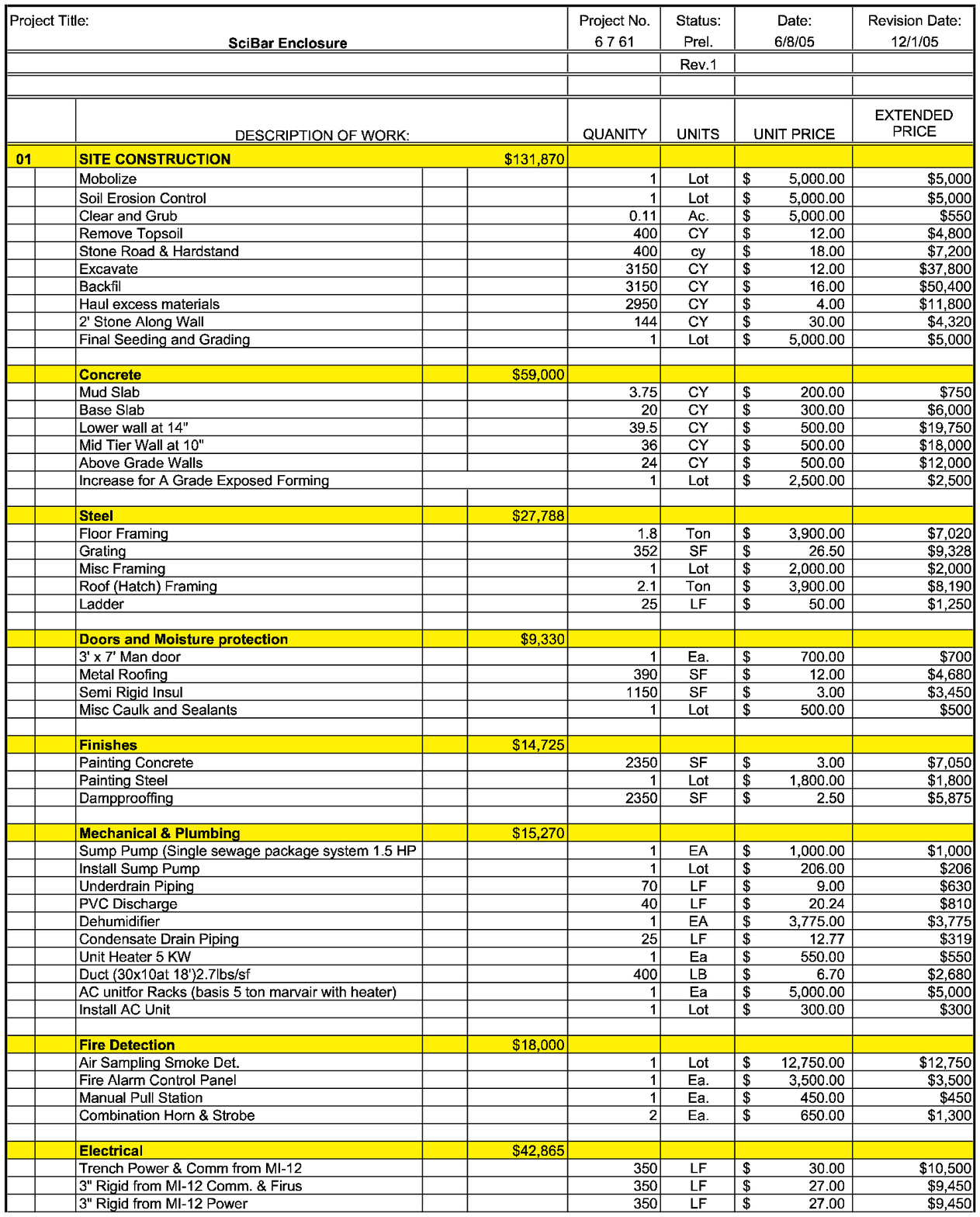}
  \end{center}
  \caption[Cost of Civil Construction]{\em 
    Fully loaded cost estimate developed by Fermilab's FESS department, continued in Figure~\ref{fig:k2k_costsb}.}
  \label{fig:k2k_costsa}
\end{figure}
\begin{figure}[htbp]
  \begin{center}
    \includegraphics[keepaspectratio=true,width=5.5in]{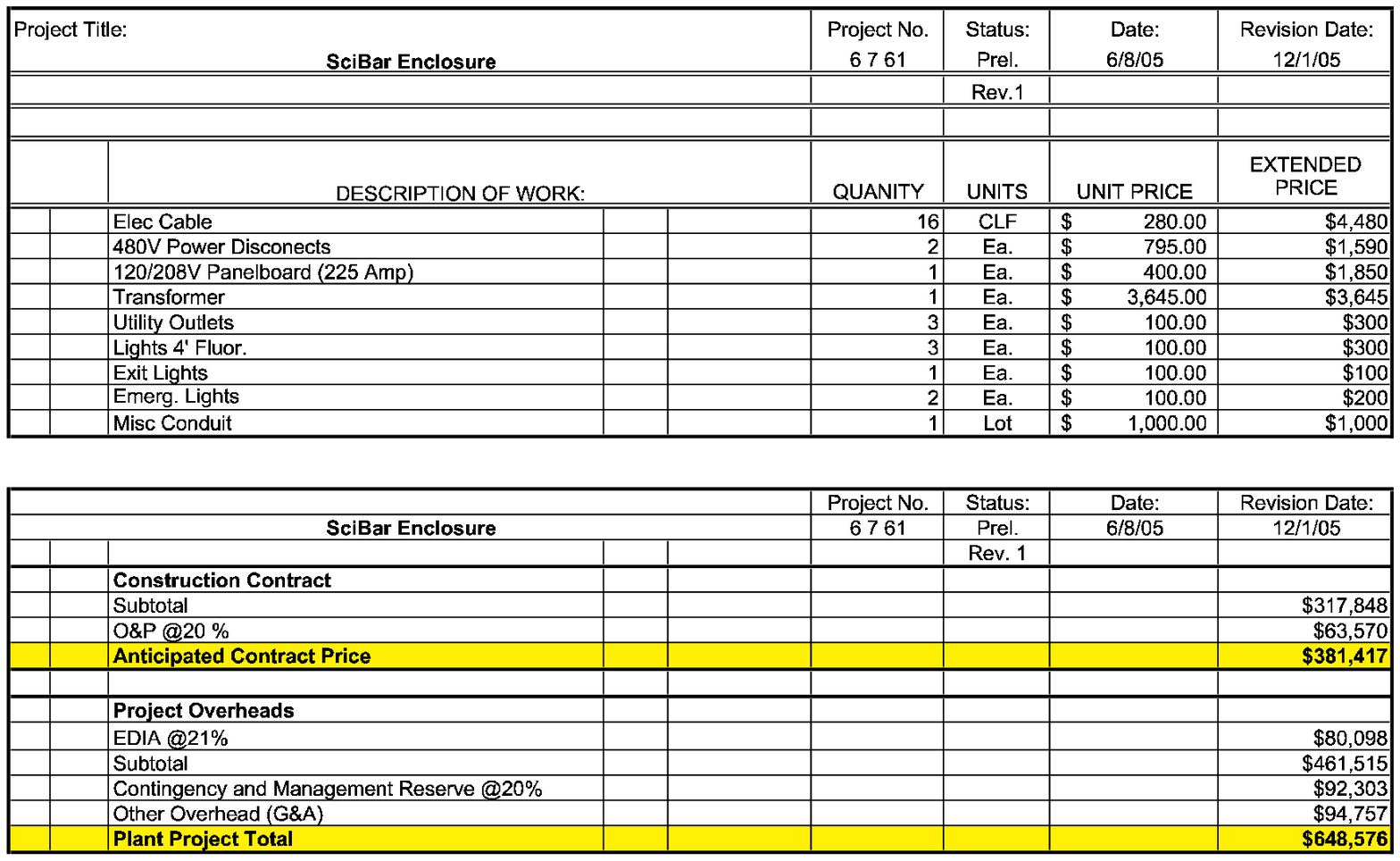}
  \end{center}
  \caption[Cost of Civil Construction]{\em 
    Fully loaded cost estimate developed by Fermilab's FESS department, continued from Figure~\ref{fig:k2k_costsa}.}
  \label{fig:k2k_costsb}
\end{figure}

\clearpage

\begin{figure}[hbp]
  \begin{center}
    \includegraphics[keepaspectratio=true,height=90mm]{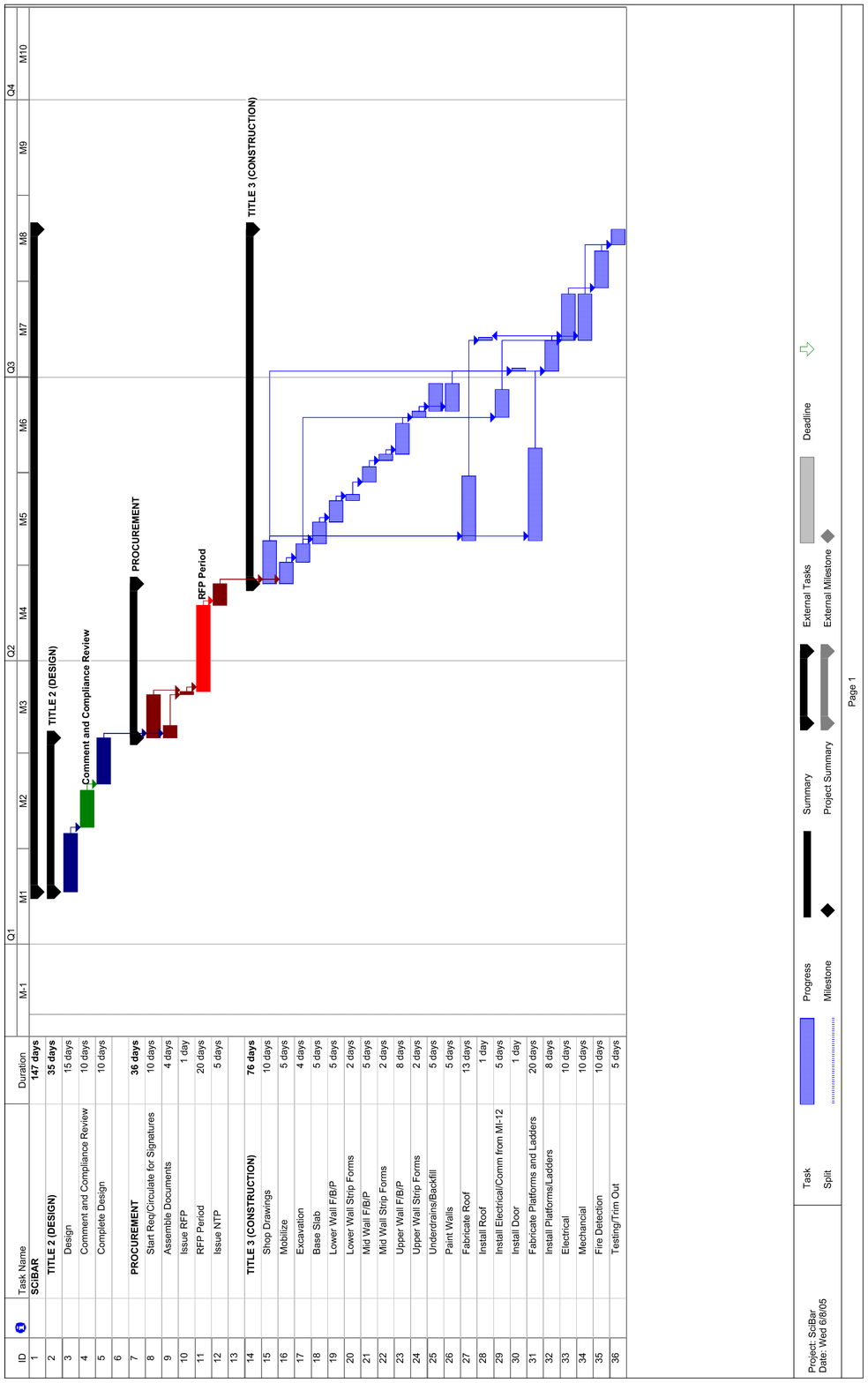}
  \end{center}
  \label{fig:k2k_sched}
\end{figure}


\end{document}